\documentclass[fleqn,usenatbib,useAMS]{mnras}

\usepackage{graphicx}	
\usepackage{amsmath}	
\usepackage{amssymb}	
\usepackage{multicol}        
\usepackage{bm}		
\usepackage{pdflscape}	

\usepackage{verbatim}

\usepackage{savesym}
\savesymbol{comment}
\usepackage{comment}

\newcommand{\kms}{\,km\,s$^{-1}$} 

\usepackage[T1]{fontenc}
\usepackage{ae,aecompl}

\usepackage{newtxtext,newtxmath}

\title[Q0107: Gas around galaxy groups]{Modelling gas around galaxy pairs and groups using the Q0107 quasar triplet}

\author[Beckett et al.]{Alexander Beckett$^{1, 2}$\thanks{Contact e-mail: \href{mailto:abeckett@stsci.edu}{abeckett@stsci.edu}},
Simon L. Morris$^{1}$,
Michele Fumagalli$^{3, 4}$,
Nicolas Tejos$^{5}$,
\newauthor
Buell Jannuzi$^{6}$,
Sebastiano Cantalupo$^{3}$
\\
$^{1}$Centre for Extragalactic Astronomy, Durham University, South Road, Durham DH1 3LE, UK\\
$^{2}$Space Telescope Science Institute, 3700 San Martin Drive, Baltimore, MD 21218, USA\\
$^{3}$Dipartimento di Fisica G. Occhialini, Universit\`{a} degli Studi di Milano Bicocca, Piazza della Scienza 3, 20126 Milano, Italy\\
$^{4}$INAF - Osservatorio Astronomico di Trieste, via G. B. Tiepolo 11, 34143 Trieste, Italy\\
$^{5}$Instituto de F\'{i}sica, Pontificia Universidad Cat\'{o}lica de Valpara\'{i}so, Casilla 4059, Valpara\'{i}so, Chile\\
$^{6}$Steward Observatory, University of Arizona, 933 North Cherry Avenue, Tucson, AZ 85721, USA
}
\date{Last updated 2023 Feb 21; in original form 2023 Feb 21}

\pubyear{2023}

\begin{document}
\label{firstpage}
\pagerange{\pageref{firstpage}--\pageref{lastpage}}
\maketitle

\begin{abstract}
We examine to what extent disk and outflow models can reproduce observations of \ion{H}{i} gas within a few virial radii of galaxies in pairs and groups. Using highly-sensitive HST/COS and FOS spectra of the Q0107 quasar triplet covering Ly$\alpha$ for z$\lesssim$1, as well as a deep galaxy redshift survey including VIMOS, DEIMOS, GMOS and MUSE data, we test simple disk and outflow models against the \ion{H}{i} absorption along three lines-of-sight (separated by 200-500 kpc) through nine galaxy groups in this field. These can be compared with our previous results in which these models can often be fit to the absorption around isolated galaxies. Our models can reproduce $\approx$ 75\% of the 28 identified absorption components within 500 \kms{ }of a group galaxy, so most of the \ion{H}{i} around groups is consistent with a superposition of the CGM of the individual galaxies. Gas stripped in interactions between galaxies may be a plausible explanation for some of the remaining absorption, but neither the galaxy images nor the galaxy and absorber kinematics provide clear evidence of such stripped material, and these unexplained absorbers do not preferentially occur around close pairs of galaxies. We find \ion{H}{i} column densities typically higher than at similar impact parameters around isolated galaxies ($\approx$ 2.5$\sigma$), as well as more frequent detections of \ion{O}{vi} than around isolated galaxies (30\% of sightlines to 7\%).

\end{abstract}





\newpage

\section{Introduction}

The exchange of gas between galaxies and their surroundings plays a vital role in their evolution. Cool gas that provides fuel for star formation needs to be accreted from outside the galaxy in order to explain the observed star-formation rates and gas content of galaxies \citep[e.g.][]{freundlich2013, scoville2017}. Cool gas is also ejected from galaxies by stellar-feedback- or AGN-driven winds, and models suggest that these processes are important in regulating its star formation \citep[e.g.][]{lehnert2013, somerville2015, salcido2020}. This reservoir of gas surrounding galaxies is known as the circumgalactic medium (CGM, e.g. \citealt{tumlinson2017}), and is usually defined as extending to the virial radius.

This material is difficult to observe in emission due in part to its low density (although emission-line maps of gas on CGM scales have become more common in recent years, e.g. \citealt{chen2019a, fossati2019a, zabl2021a, leclercq2022}), so is most commonly detected using absorption features along the lines-of-sight to background sources, often quasars \citep[e.g.][]{bahcall1969, bergeron1986, bergeron1994, weymann1998, adelberger2005, chen2010a, prochaska2011, rubin2018, pointon2019, wilde2021, lehner2021}. These absorption-based studies often discuss large numbers of galaxy-sightline pairs, whether as targeted surveys \citep[e.g.][]{tumlinson2013, bielby2019} or by utilizing other surveys and/or archival data \citep[e.g.][]{wild2008}, but are usually limited to a single line-of-sight through the gas around any galaxy or group. Some studies are able to utlilize quasar pairs or triplets \citep[lensed or projected, e.g.][]{fossati2019, maitra2019a}, bright background galaxies \citep[e.g.][]{zahedy2016, peroux2018, chen2020, okoshi2021} or gravitationally-lensed arcs \citep[e.g.][]{lopez2018, lopez2020, tejos2021, mortensen2021}, but these are not common enough to produce statistically meaningful samples of galaxy-absorber pairs.


Many observations find evidence for disk-like accreting and rotating structures along the major axis of galaxies \citep[e.g.][]{charlton1998, steidel2002, bouche2016, ho2017, zabl2019, french2020},
and this is generally reproduced by simulations, with simulated CGM structures aligned with the galaxy major axis \citep[e.g][]{ho2019, mitchell2020, defelippis2020, hafen2022b}. Similarly, evidence for outflowing material is often found along the minor axis of galaxies \citep[e.g.][]{bland1988, heckman1990, finley2017, lan2018, schroetter2019, burchett2021, zabl2021a}, which are also reproduced by simulations \citep[e.g][]{nelson2019, mitchell2020a, pandya2021}. This leads to some observations finding a bimodality in the azimuthal angles of detected absorption \citep[absorption is found primarily along the major and minor axes, e.g.][]{kacprzak2012, bouche2012, kacprzak2015}, as well as differences in flow rates and metallicities along galaxy major and minor axes in simulations \citep[e.g.][]{peroux2020}.

However, gas flows around galaxy groups are likely far more complex. For example, in the nearby M81/M82 group, \ion{H}{i} observations reveal a conical outflow structure on small scales, but clear distortion on larger scales due to the other group galaxies \citep[e.g.][]{sorgho2019}. Tidal interactions are expected to produce a significant fraction of Ly$\alpha$ absorbers in groups \citep[e.g.][]{morris1994}, and likely contribute to other observed ions \citep[e.g.][]{chen2009}. Some studies have identified possible tidal material in absorption, but this is usually difficult to distinguish from other origins \citep[e.g.][]{chen2014, guber2018}. Ram-pressure stripping is also likely to affect the state of the CGM and any intra-group gas during interactions \citep[e.g.][]{fumagalli2014, fossati2019a}. This removal of gas from galaxies and their CGM affects not only the gas itself, but also leads to `quenching' of star formation in group galaxies \citep[e.g.][]{peng2010, wetzel2013, jian2017, kuschel2022}

There have been numerous case studies focusing on the gas in individual galaxy groups, using a variety of absorption features, but also emission lines in more recent cases. 
Some of these find material that can be associated with a particular galaxy, including some likely outflows and accretion \citep[e.g.][]{peroux2017, johnson2018}, but also material that cannot be associated with a galaxy, and therefore forms an intra-group medium \citep[e.g][]{bielby2017a, epinat2018}, as well as suggestions of tidal material \citep[e.g.][]{kacprzak2010, chen2019a}. Where multiple absorption/emission components can be identified, material consistent with both disk/outflow structures and tidal/intra-group material can sometimes be seen in the same galaxy group \citep[e.g.][]{nateghi2021, leclercq2022}.

Statistical studies utilizing large samples of galaxy groups are also used to determine the dominant processes in these systems. \citet{bordoloi2011} found that \ion{Mg}{ii} absorption near galaxy groups could be due to a superposition of the halos around individual galaxies. In contrast, \citet{nielsen2018} found that a superposition model could not match the absorber kinematics, and preferred a model in which most cool gas was associated with the group itself rather than any member galaxy. More recently, \citet{dutta2020} found \ion{Mg}{ii} absorption more extended around galaxy groups than isolated galaxies, but also a clear dependence on galaxy properties, suggesting that both the galaxy halos themselves and the interactions between galaxies contribute to the absorption. \

If material stripped from galaxies contributes substantially to the gas content in groups, then we may expect gas in these groups to have a higher metal content (having passed through the galaxy and been enriched by cycles of star formation, e.g. \citealt{oppenheimer2016}). However, this is not found by \citet{pointon2019}.

The distribution of gas in groups therefore remains unclear. Emission from the CGM remains difficult to detect due to a low density, which usually limits such studies to scales of $\lesssim$100 kpc \citep[e.g.][]{burchett2021, leclercq2022}. There remains inconsistency in defining galaxy groups, which are subject to different algorithms, linking lengths, and detection limits in different studies. This makes direct comparisons between observations, or with simulations, difficult \citep[e.g.][]{oppenheimer2021}. The power of absorption studies is limited by the presence of sufficiently bright background sources, so in most cases the gas around any galaxy group is probed by only a single line-of-sight and we cannot measure gas properties elsewhere in the group.

It is this final difficulty which we seek to address in this paper. We study the Q0107 field, a quasar triplet at z $\sim$ 1 with separations of $\sim$ 1 arcminute ($\approx$ 400 kpc at z = 0.5), with basic quasar properties given by Table \ref{table:past_papers}. The use of multiple lines-of-sight through a densely-surveyed field helps to constrain the gas structures and properties on CGM scales. 

This field has been utilized in several previous studies, including early studies of the size scales of Ly$\alpha$ absorbers \citep[e.g.][]{dinshaw1997, dodorico1998, young2001}, analysis of the coincidences between absorption in the different lines-of-sight \citep[e.g.][]{petry2006, crighton2010}, construction of the 2D 2-point correlation functions of absorption in both \ion{H}{i} and \ion{O}{vi} (\citealt{tejos2014}, hereafter T14, \citealt{finn2016}), and detailed radiative transfer modelling of a small number of absorbers \citep[e.g.][]{muzahid2014, anshul2021}. Improvements to the available data, such as higher-resolution spectra of the three quasars, high-resolution imaging from the Hubble Space Telescope, and IFU observations of the field, have enabled our recent works extending these results.

In \citet{beckett2021}, hereafter denoted as Paper 1, we examined some statistical properties of the relationship between gas and galaxies in this field. We found a bimodality in the azimuthal angle distribution of detected galaxy-absorber pairs, likely evidence for the existence of disk and outflow structures along the projected major and minor axes of galaxies, extending to $\approx$ 300 kpc. A higher incidence of \ion{O}{vi} absorption along the minor axis, and a preference for the velocity of \ion{H}{i} absorption near the major axis to be aligned with the galaxy kinematics, support this hypothesis. We investigated this further in Paper 2 \citep{beckett2022}, in which we attempted to fit simple disk and outflow models to isolated galaxies in our sample, finding that many absorbers (13 of 26 within 500 \kms, or 12 of 20 within 300 \kms, and impact parameters up to 600 kpc) could be fit using such models, whilst remaining consistent with the observations of the other sightlines.

This work continues to use the disk and outflow models from Paper 2, but extends coverage to galaxy pairs and groups in our sample. By combining disk/outflow models, we can examine the hypothesis in which most absorption in groups results from the CGM of the individual galaxies. We also search for signs of tidal material based on the kinematics of the gas relative to nearby galaxies.

In Section \ref{sec:data} we summarize the galaxy and quasar data used in this study, discussed more extensively in Paper 1 and references therein, as well as selection of the sub-sample of galaxy groups considered in this work. Section \ref{sec:model_summary} summarizes the toy models used in our attempts to reproduce the observed absorption (covered in more detail in Paper 2), whilst Section \ref{sec:group_details} describes in detail the absorption around each galaxy group, and the process of attempting to fit our models to that absorption. We then discuss the overall results in Section \ref{sec:discussion}, and finally summarize in Section \ref{sec:conclusions}. 

Throughout this work we quote physical sizes and distances unless otherwise stated, and use the Planck 2018 flat $\Lambda$CDM cosmology \citep{planckcollaboration2020}, with $\Omega_{m}$ = 0.315 and $H_{0}$ = 67.4 \,km\,s$^{-1}$\,Mpc$^{-1}$.

\begin{table}\label{table:past_papers}
\begin{center}
\begin{tabular}{| c| c| c| c| c|}
 \hline Object & RA (J2000) & Dec (J2000) & Redshift & R-mag \\ \hline
 Q0107-025 A & 01:10:13.14 & -2:19:52.9 & 0.960  & 18.1\\
 Q0107-025 B & 01:10:16.25  & -2:18:51.0 & 0.956 & 17.4\\
 Q0107-0232 (C) & 01:10:14.43 & -2:16:57.6 & 0.726 & 18.4\\ \hline

\end{tabular}
\caption{Co-ordinates, redshifts and R-band magnitudes of the three quasars, taken from \citet{crighton2010}}
\end{center}
\end{table}

\section{Data}\label{sec:data}

This paper uses the same absorber and galaxy catalogues as our previous two papers covering this field, which utilise the galaxy catalogues compiled by T14. We therefore only briefly describe the observations here; these are described in more detail in Paper 1 and T14.

\subsection{IGM data}\label{sec:igm_data}

Spectra of the three quasars were taken to cover the Ly$\alpha$ transition of \ion{H}{i} along each line-of-sight from $z = 0$ to the redshift of the quasar. These use the spectrographs on the Hubble Space Telescope, with the G130M and G160M gratings on the Cosmic Origins Spectrograph \citep[COS, ][program GO-11585, PI: Neil Crighton]{green2012} and the G190H and G270H gratings on the Faint Object Spectrograph (FOS, observations detailed in \citealt{young2001}), covering a wavelength range of 1135-3277 \AA{}. This includes a range of metal ions in addition to Ly$\alpha$. QSO-C was not observed using the G130M and G270H gratings, with these made unnecessary by a sub-damped Lyman-$\alpha$ system at z $\approx$ 0.56 obscuring any flux below 1420 \AA, and a lower QSO redshift respectively. Details of the observations and properties of the observed spectra are listed in Table 2 of Paper 1.

Our absorption line catalogue was produced by T14, and also used by \citet{finn2016} and our earlier papers. It contains 430 absorption systems, of which 272 are \ion{H}{i}. The process of producing this catalogue is detailed in T14.

The signal-to-noise ratio of the COS spectra (using results from \citealt{keeney2012}), as well as the \ion{H}{i} column density distribution of our sample, imply a 3$\sigma$ detection limit of $\sim10^{13} \textrm{cm}^{-2}$. The FOS detection limit is slightly higher, at $\sim10^{13.5} \textrm{cm}^{-2}$.

We note that Ly$\alpha$ absorbers in the IGM are usually found with Doppler widths $>$ 20 \kms \citep[e.g][]{dave2010}, so most are resolved in the COS spectra (Ly$\alpha$ below z $\approx$ 0.45). Absorption in the FOS spectra is often unresolved and more likely to be blended with other features, but for Ly$\alpha$ these issues can often be mitigated using Ly$\beta$ or higher-order Lyman transitions that remain in the COS wavelength range. Where these mitigations are not possible, absorbers may have a large Doppler parameter that appears as a broad-Ly$\alpha$ absorber (BLA). Whether such a BLA appearing around our galaxies may indicate high temperatures or is unresolved narrower absorption is discussed in each case (Section \ref{sec:group_details} and Appendix \ref{sec:more_groups}).

\subsection{Galaxies}\label{sec:gal_data}

Our sample of galaxy groups is drawn from the spectroscopic galaxy catalogue described in Paper 1, which builds on that used in T14. This forms a highly heterogeneous galaxy survey, with VIMOS, DEIMOS, GMOS and CFHT-MOS observations of differing area, depth and completeness (referred to as MOS observations throughout), supplemented by MUSE fields centered on the A and B lines-of-sight. Additionally, high-resolution imaging from the Hubble Space Telescope is used to provide position angle and inclination estimates for some galaxies. These observations, the methods used to combine them and ensure consistency, and the resulting survey properties, are discussed in detail in Paper 1 (their Section 2). In this work we only provide a brief summary of this information.

\subsubsection{MOS} \label{sec:mos}

The multi-object spectrograph on the Canada-France-Hawaii Telescope (CFHT-MOS) \citep{lefevre1994} was used to obtain spectra for 29 galaxies in the Q0107 field, described in \citet{morris2006}, whilst VIMOS \citep{lefevre2003} took 935 spectra (ESO programs 086.A-0970, PI:Crighton; and 087.A-0857, PI: Tejos). 642 galaxies in this field were observed using DEIMOS (\citealt{faber2003}, program A290D, PIs: Bechtold and Jannuzi), with 210 spectra added by GMOS (\citealt{davies1997}, program GS-2008B-Q-50, PI: Crighton).

Galaxies appearing in more than one of these surveys were used to ensure consistent wavelength calibration across all MOS surveys (and therefore consistent redshifts in the resulting catalogue). As DEIMOS has the best resolution, the redshifts of VIMOS and GMOS objects were adjusted to match the DEIMOS frame, requiring a systematic shift of $\Delta z \sim$ 0.0008 for the VIMOS data, and 0.0004 for the GMOS data see Section 3 of T14).

\subsubsection{MUSE} \label{sec:muse}

GTO observations (ESO program ID 094.A-0131, PI Schaye) cover $1' \times 1'$ fields of view around QSOs A and B. These provide galaxy kinematics for galaxies near to the lines-of-sight, and are also slightly deeper than the MOS surveys. The MUSE data cover the wavelength range from 4750 to 9350 \AA{ }with FWHM of $\approx$2.7 \AA, and offer seeing of $0.96''$ for QSO-A, and $0.82''$ for QSO-B.

We reduced the data using a pipeline similar to many recent papers \citep[e.g][]{fumagalli2016, fumagalli2017, fossati2019, lofthouse2020, bielby2020}, primarily using ESO routines, but also utilizing the CUBEX package \citep{cantalupo2019} to apply a renormalization between the different IFUs, stacks and slices, and an improved, flux-conserving sky subtraction.

Objects were detected using the SExtractor software \citep{bertin1996}, applied to the white-light image. Summing the total flux within the SExtractor aperture produced the 1D spectra, for which redshifts were estimated using MARZ \citep{hinton2016}.

\subsubsection{Combined Galaxy Catalogue}\label{sec:catalogues}

Combining the spectra from multiple instruments into a single galaxy catalogue for this field requires matching the astrometry of galaxies observed by multiple instruments, so that duplicates can be removed. We then checked other galaxy properties to ensure these were consistent across the different observations. 

We confirmed accurate astrometry for each instrument by comparing the brightest objects in the MOS and MUSE catalogues with the SDSS source catalogue \citep{albareti2017}. Coordinates from the MOS catalogues matched the SDSS results within 0.5$"$, whilst the MUSE results required a shift of $\approx$1$"$. After this shift, we matched objects appearing within 1$"$ in both catalogues, finding no additional matches within 2$"$. Magnitudes obtained by integrating the MUSE spectra were compared with the results from the MOS surveys, and found to be consistent within the estimated uncertainties. 

We also compared redshifts to ensure that all of our galaxy samples are in the same frame as the absorption features. Across a large sample of galaxy--absorber pairs, any velocity offsets should average to zero, with no systematic offset. We found that the corrections applied when producing the T14 catalogue (described in Section \ref{sec:mos}) had already removed any systematic shift, with precision better than 10 \kms, whilst the MUSE data required a shift of $\approx$ 30 \kms to match the MOS redshifts.

We utilize the duplicate observations, where a single galaxy was observed by the same instrument on multiple occasions, to estimate the uncertainties in our redshift measurements. We calculated the velocity differences between each pair of duplicate spectra, and used the standard deviation of the velocity differences to estimate the uncertainty in the redshifts estimated from each instrument. These are given in Table 5 of Paper 1, and vary between 30 \kms{ }(for well-determined redshifts in DEIMOS) and 190 \kms{ }(for single-feature detections in VIMOS).

Our MUSE observations contain no duplicates. As MUSE has a higher resolution than GMOS, but lower than DEIMOS, we take the GMOS values as conservative estimates of the velocity uncertainties for MUSE galaxies with the same confidence flags.

\subsubsection{HST imaging} \label{sec:HST}

High-resolution imaging obtained using the ACS instrument on the Hubble Space Telescope \citep{ryon2019} is available for this field in the F814W band (Program GO-14660, PI Straka). This imaging enables a much clearer view of any distortions of the galaxies due to interactions, as well as much improved measurements of position angles and inclinations than are possible with the lower-resolution ground-based imaging available previously. This image does not cover the full field of our MOS surveys, so these improved position angle and inclination estimates are only available for a subset of galaxies in our catalogue, as shown in Figure \ref{fig:survey_layout}.

We modelled galaxies in both our redshift catalogue and the HST image using GALFIT \citep{peng2002}. This uses chi-squared minimization to produce a best-fitting model of a galaxy's morphology. We used the SExtractor results as initial guesses in attempting to fit a Sersic disk to each galaxy. Where necessary, additional model components were introduced until a reasonable fit was found. This more advanced modelling, taking account of the point-spread function of the image,  reduced the average uncertainty on both inclination and position angle measurements by a factor of $\approx$ 3 relative to the SExtractor results. 

We excluded galaxies for which the fit clearly failed to converge to a reasonable result, but included galaxies that have large uncertainties on position angle due to being near to face-on. This produced a list of 109 galaxies for which position angle, inclination and redshift measurements were found.

\subsection{Sample of galaxy groups}\label{sec:sample}

Paper 2 considered only isolated galaxies, defined as those with no detected companion within 500 kpc and 500 \kms, making it unlikely that the region within a galaxy's virial radius overlaps with that of another detected galaxy. We also additionally imposed the constraint that no other galaxy must lie within 1 Mpc of any of the lines-of-sight within the 500 \kms{ }window. In this work we consider galaxies that are not isolated, and therefore for the purposes of this work a group is any set of galaxies with pairwise separations smaller than 500 kpc and 500 \kms, or any set of multiple galaxies within 1 Mpc of at least one of the three lines-of-sight in a 500 \kms{ }window.

This definition of `group' requires only two galaxies, and also has a larger `linking length' (maximum distance between galaxies for them to be considered in the same group) than many similar studies that define group galaxies \citep[e.g.][]{bordoloi2011, nielsen2018, fossati2019}. However, we do use a smaller window than some other studies considering isolated galaxies \citep[e.g. COS-halos,][]{tumlinson2013}. Therefore all galaxies in our sample are considered as 'group' or 'isolated', but we include groups with larger galaxy separations than other works. 


These `group' galaxies make up the majority of our full sample, so it is impractical to model the absorption around all of these (more than 50 groups). As our models rely on position angle and inclination measurements obtained from the HST imaging, we selected our sample from groups that have at least two galaxies lying in the HST field (of which there are 18). Because of the time-consuming nature of modelling group absorption, we manually select a sample of nine groups that meet this restriction, and intentionally span a large range in group properties including redshift as well as number and mass of group galaxies. With the difficulty in determining our uncertainties described in Section \ref{sec:model_summary}, the expected gains from modelling the remaining groups would not substantially improve the strength of our conclusions.

Our range of redshifts and stellar masses is intended to be comparable to that seen in our isolated sample from Paper 2, and therefore includes some groups beyond z $\approx$ 0.73, where only the A and B sightlines are available. The selection was made `blindly' with respect to the nearby absorption, in order to avoid biasing our results. Similarly, we do not preferentially select groups containing galaxies in the MUSE fields, as these are biased towards low-mass, star-forming galaxies and only a small subset of these have useful kinematics.

We also check for any additional bias introduced through this selection, by comparing our nine selected groups with the nine groups not selected for this work. If the largest group in the field (G-202) is removed, the selected and not-selected samples have similar average properties in terms of galaxy number (selected and non-selected samples with a median of 4 galaxies within 1 Mpc and 500 \kms), total galaxy mass ($10^{11.1}$ and $10^{11.2}$ $M_{\odot}$ respectively), and closest impact parameter (120 and 135 kpc). However, our selected sample is slightly biased towards lower redshifts (0.52 vs 0.65). Any alternative sample of groups selected to cover the redshift range seen in Paper 2 would therefore tend to have similar galaxy properties.

We note that the definition of `group' does vary substantially across our large redshift range, such that a group at low redshift would be seen as an isolated galaxy is it were at a higher redshift, with any satellite galaxies going undetected. However, this does not strongly affect our sample, as most of our groups have at least two members that could be detected up to z $\approx$ 1. G-517 is the only group likely to be classified as isolated if it lay at a different redshift, whilst G-383 (consisting of two faint galaxies) could appear isolated in a small redshift range but would more likely go undetected entirely.

\begin{table*}
\begin{center}
\caption{Summary of group properties for our selected sample of galaxy groups. Columns are as follows: (1) Group reference used throughout the paper; (2) group central redshift; (3) number of detected galaxies within 500 \kms of the central redshift and 1 Mpc of at least one QSO line-of-sight; (4) total stellar mass of detected galaxies ; (5) closest impact parameter of any galaxy to any line-of-sight; (6) minimum separation between two detected galaxies; (7) r-band luminosity ratio between brightest two galaxies; (8)-(10) estimated galaxy detection limits in the MOS and MUSE surveys, showing continuum estimates in terms of $L_{\star}$ and SFR limits on emission-line-only detections.}
\label{table:sample_groups}
\begin{tabular}{| c| c| c| c| c| c| c| c| c| c| }
\hline 
Group Ref & z & N gals & total $\textrm{M}_{\star}$  & $\textrm{b}_{min}$ & Min Pair & Lum Ratio & Det Lim (MOS) & Det Lim (MUSE) & Det Lim (Line-only)  \\
 & & & $\textrm{log}_{10}(\textrm{M}_{\odot})$ & (kpc) & (kpc) & & ($L_{\star}$) &  ($L_{\star}$) & ($\textrm{M}_{\odot}$/yr)\\
 (1) & (2) & (3) & (4) & (5) & (6) & (7) & (8) & (9) & (10)  \\ \hline

G-202 & 0.202 & 25 & 11.9 $\pm$ 0.2 & 103 & 17 & 2.6 & 0.004 & 0.003 & 0.004 \\
G-238 & 0.238 & 4 & 11.5 $\pm$ 0.2 & 236 & 42 & 1.5 & 0.006 & 0.004 & 0.005 \\
G-383 & 0.383 & 2 & 9.7 $\pm$ 0.2 & 38 & 5 & 1.5 & 0.018 & 0.011 & 0.016 \\
G-399 & 0.399 & 5 & 11.1 $\pm$ 0.2 & 122 & 71 & 3.3 & 0.020 & 0.013 & 0.018 \\
G-517 & 0.517 & 2 & 10.3 $\pm$ 0.2 & 95 & 320 & 29.5 & 0.04 & 0.02 & 0.03 \\
G-536 & 0.536 & 6 & 11.5 $\pm$ 0.4 & 120 & 4 & 1.4 & 0.04 & 0.03 & 0.04 \\
G-558 & 0.558 & 7 & 11.1 $\pm$ 0.5 & 148 & 53 & 1.2 & 0.04 & 0.03 & 0.04 \\
G-876 & 0.876 & 4 & 11.1 $\pm$ 0.4 & 75 & 26 & 2.1 & 0.13 & 0.08 & 0.12 \\
G-907 & 0.907 & 2 & 10.7 $\pm$ 0.5 & 170 & 594 & 1.7 & 0.15 & 0.09 & 0.13 \\
\hline

\end{tabular}

\end{center}
\end{table*}

\begin{table*}
\begin{center}
\caption{Summary of galaxy properties for our selected sample of group galaxies. For each group only galaxies with position angle and inclination measurements are shown. Derivation of these properties is described in more detail in Paper 1. Column descriptions: (1) Group ID used in this work;  (2) Galaxy ID (MUSE and MOS IDs were collated separately; MUSE ID is used for galaxies featuring in both MOS and MUSE catalogues); (3, 4) On-sky coordinates of galaxy; (5) Observed magnitude in the SDSS r-band; (6) Galaxy luminosity in SDSS r-band as a multiple of $L_{\star}$ ($L_{\star}$ estimate from \citet{montero-dorta2009}, uncertainties smaller than 0.005 $L_{\star}$ are omitted); (7) Stellar mass estimated as in \citet{johnson2015} (for most galaxies the largest uncertainty is a scatter of 0.15 dex in their relation); (8) Halo mass estimated using the abundance matching technique from \citet{behroozi2010}; (9) Star-formation flag denoting a star-forming or non-star-forming galaxy; (10) Star-formation rate estimated from galaxy emission lines, using the \citet{kennicutt1998} and \citet{kewley2004} calibrations for H$\alpha$ and [\ion{O}{ii}] respectively (uncertainties are a combination of scatter in these relationships and uncertainty in the line fit); (11) Line used to estimate SFR (SFRs estimated from H$\beta$ using the correlation between SFR estimated from H$\alpha$ and H$\beta$ line luminosity, generating a substantially larger uncertainty); (12) Note of whether emission-line kinematics from the MUSE data are available for this galaxy. }
\label{table:summary_galaxies}
\begin{tabular}{| c| c| c| c| c| c| c| c| c| c| c| c|}
\hline 
Group & Galaxy & RA & Dec & r-band & Luminosity  & $\textrm{M}_{\star}$ & $\textrm{M}_{h}$ & SF Flag & SFR & Line & Kinematics \\
 & & $^{\circ}$ & $^{\circ}$ & & ($L_{\star}$)& $\textrm{log}_{10}(\textrm{M}_{\odot})$ & $\textrm{log}_{10}(\textrm{M}_{\odot})$ & & ($\textrm{M}_{\odot}$/yr) &  \\
 (1) & (2) & (3) & (4) & (5) & (6) & (7) & (8) & (9) & (10) & (11) & (12) \\ \hline
G-202 & B-22 & 17.5600 & -2.3173 & 22.86 $\pm$ 0.03 & 0.02 $\pm$ 0.01 & 8.4 $\pm$ 0.2  & 10.9 $\pm$ 0.3 & non-SF & 0.06 $\pm$ 0.04 & H$\alpha$ & Poor   \\
& 25962 & 17.5589 & -2.3548 & 21.46 $\pm$ 0.01 & 0.07 $\pm$ 0.01 & 8.7 $\pm$ 0.1  & 11.0 $\pm$ 0.3 & SF & 0.5 $\pm$ 0.3 & H$\alpha$ & No   \\
& 31704 & 17.5453 & -2.2958 & 22.58 $\pm$ 0.02 & 0.03 $\pm$ 0.01 & 8.2 $\pm$ 0.1  & 10.8 $\pm$ 0.3 & SF & 0.08 $\pm$ 0.05 & H$\alpha$ & No   \\
& 31787 & 17.5556 & -2.3002 & 21.35 $\pm$ 0.01 & 0.08 $\pm$ 0.01 & 9.7 $\pm$ 0.1  & 11.5 $\pm$ 0.2 & SF & 0.5 $\pm$ 0.2 & H$\alpha$ & No   \\
& 32497 & 17.5567 & -2.3000 & 18.46 $\pm$ 0.01 & 1.10 $\pm$ 0.01 & 11.0 $\pm$ 0.1  & 13.1 $\pm$ 0.5 & non-SF & 2.3 $\pm$ 1.9 & H$\alpha$ & No   \\
& 32778 & 17.5458 & -2.2893 & 19.59 $\pm$ 0.01 & 0.39 $\pm$ 0.01 & 10.2 $\pm$ 0.1  & 11.8 $\pm$ 0.3 & SF & 1.5 $\pm$ 0.5 & H$\alpha$ & No   \\
\\
G-238 & 26677 & 17.5604 & -2.3491 & 20.59 $\pm$ 0.01 & 0.22 $\pm$ 0.01 & 9.7 $\pm$ 0.2  & 11.4 $\pm$ 0.3 & SF & 1.3 $\pm$ 0.4 & H$\alpha$ & No   \\
& 29214 & 17.5395 & -2.3244 & 21.38 $\pm$ 0.01 & 0.11 $\pm$ 0.01 & 8.9 $\pm$ 0.1  & 11.1 $\pm$ 0.3 & SF & 0.4 $\pm$0.1 & H$\alpha$ & No   \\
\\
G-383 & A-48 & 17.5535 & -2.3299 & 23.96 $\pm$ 0.09 & 0.03 $\pm$ 0.01 & 9.6 $\pm$ 0.2  & 11.5 $\pm$ 0.3 & non-SF & <0.04 & H$\alpha$ & No   \\
& A-49 & 17.5534 & -2.3296 & 25.90 $\pm$ 0.51 & 0.02 $\pm$ 0.01 & 8.7 $\pm$ 0.6  & 11.1 $\pm$ 0.4 & non-SF & <0.02 & H$\alpha$ & No   \\
\\
G-399 & A-56 & 17.5605 & -2.3287 & 25.17 $\pm$ 0.27 & 0.01 $\pm$ 0.01 & 8.5 $\pm$ 0.5  & 11.0 $\pm$ 0.3 & SF & 0.05 $\pm$ 0.01 & H$\alpha$ & No   \\
& A-69 & 17.5620 & -2.3255 & 22.06 $\pm$ 0.01 & 0.19 $\pm$ 0.01 & 9.8 $\pm$ 0.2  & 11.6 $\pm$ 0.3 & SF & 0.4 $\pm$ 0.2 & H$\alpha$ & Poor   \\
& B-7 & 17.5672 & -2.3244 & 21.37 $\pm$ 0.01 & 0.36 $\pm$ 0.01 & 10.6 $\pm$ 0.2  & 12.3 $\pm$ 0.5 & non-SF & <0.1 & H$\alpha$ & No   \\
\\
G-517 & B-34 & 17.5780 & -2.3143 & 25.39 $\pm$ 0.26 & 0.02 $\pm$ 0.01 & 7.6 $\pm$ 0.6  & 10.7 $\pm$ 0.4 & SF & <0.04 & [\ion{O}{ii}] & No   \\
& B-43 & 17.5643 & -2.3119 & 21.50 $\pm$ 0.01 & 0.59 $\pm$ 0.01 & 10.3 $\pm$ 0.2  & 12.0 $\pm$ 0.4 & SF & 3.8 $\pm$ 1.5 & [\ion{O}{ii}] & Yes   \\
\\
G-536 & A-36 & 17.5598 & -2.3322 & 21.30 $\pm$ 0.01 & 0.77 $\pm$ 0.01 & 10.9 $\pm$ 0.2  & 12.9 $\pm$ 0.7 & SF & 6 $\pm$ 3 & [\ion{O}{ii}] & Yes   \\
& A-37 & 17.5599 & -2.3320 & 22.30 $\pm$ 0.06 & 0.31 $\pm$ 0.02 & 10.6 $\pm$ 0.2  & 12.4 $\pm$ 0.6 & SF & 1.3 $\pm$ 0.4 & [\ion{O}{ii}] & Poor   \\
& A-40 & 17.5649 & -2.3308 & 25.61 $\pm$ 0.27 & 0.02 $\pm$ 0.01 & 9.4 $\pm$ 0.4  & 11.4 $\pm$ 0.3 & SF & 0.03 $\pm$ 0.02 & [\ion{O}{ii}] & No   \\
\\
G-558 & A-72 & 17.5538 & -2.3253 & 23.66 $\pm$ 0.13 & 0.10 $\pm$ 0.01 & 9.7 $\pm$ 0.5  & 11.6 $\pm$ 0.4 & SF & 0.7 $\pm$ 0.2 & [\ion{O}{ii}] & Yes   \\
& A-75 & 17.5555 & -2.3238 & 25.02 $\pm$ 0.29 & 0.03 $\pm$ 0.01 & 8.3 $\pm$ 0.4  & 10.9 $\pm$ 0.4 & SF & 0.5 $\pm$ 0.1 & [\ion{O}{ii}] & Yes   \\
\\
G-876 & A-32 & 17.5580 & -2.3325 & 23.61 $\pm$ 0.05 & 0.30 $\pm$ 0.01 & 10.6 $\pm$ 0.2  & 12.4 $\pm$ 0.6 & SF & 7 $\pm$ 2 & [\ion{O}{ii}] & Yes   \\
& A-38 & 17.5573 & -2.3318 & 23.88 $\pm$ 0.13 & 0.24 $\pm$ 0.03 & 10.8 $\pm$ 0.3  & 12.6 $\pm$ 0.8 & SF & 4.7 $\pm$ 1.1 & [\ion{O}{ii}] & Yes   \\
\\
G-907 & A-16 & 17.5600 & -2.3347 & 24.34 $\pm$ 0.18 & 0.17 $\pm$ 0.03 & 10.4 $\pm$ 0.5  & 12.2 $\pm$ 0.9 & SF & 2.3 $\pm$ 0.6 & [\ion{O}{ii}] & Poor   \\
& B-19 & 17.5721 & -2.3181 & 24.88 $\pm$ 0.25 & 0.10 $\pm$ 0.03 & 10.4 $\pm$ 0.5  & 12.2 $\pm$ 0.8 & SF & 1.1 $\pm$ 0.3 & [\ion{O}{ii}] & Poor   \\
\\

\hline

\end{tabular}

\end{center}
\end{table*}

\begin{figure*}
\includegraphics[width=\textwidth]{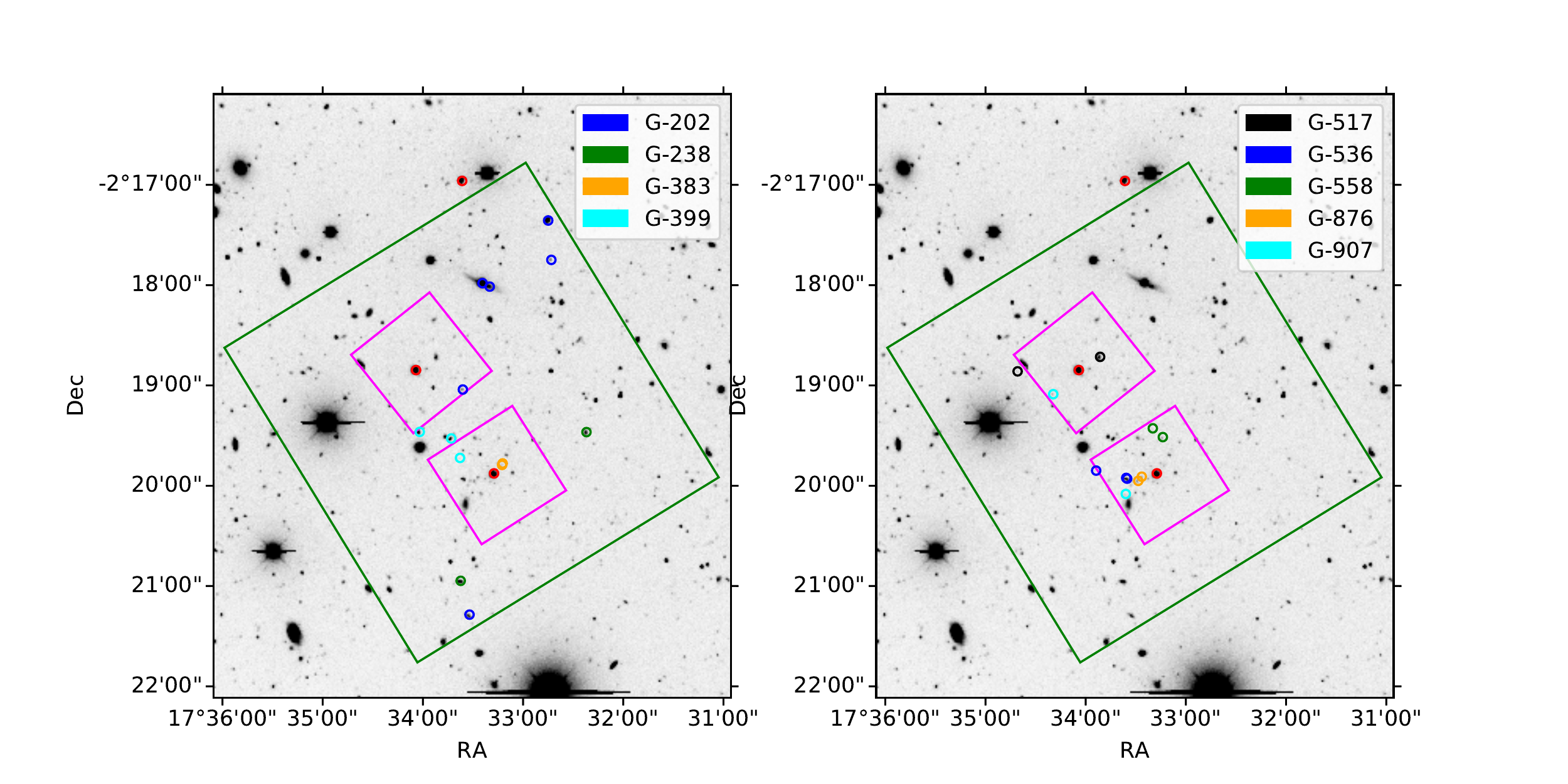}
\caption{The layout of the surveys used in this study. The background image was taken with the Kitt Peak 4-metre Telescope. The solid green square shows the region covered by HST imaging, whilst the smaller magenta squares show the MUSE fields centered on QSOs A and B. The quasars are shown by red circles, with A the southernmost and C the northernmost. The galaxies within the HST field that are analysed in this work are spread across the two panels, with z $<$ 0.5 galaxies in the left panel, and higher-redshift galaxies in the right panel. These are coloured by group and labelled by the group identifier listed in Table \ref{table:sample_groups}. \label{fig:survey_layout}}
\end{figure*}

The properties of the resulting group sample are listed in Table \ref{table:sample_groups}, with individual galaxy properties shown in Table \ref{table:summary_galaxies}, and their locations on the sky illustrated in Figure \ref{fig:survey_layout}. Only galaxies within the HST field are shown (although we discuss the presence of more distant galaxies when modelling the absorption within these groups).

We note that star-formation rates of galaxies in groups and clusters tend to be lower than field galaxies for similar stellar masses \citep[e.g.][]{larson1980, wetzel2013}. This is found for our full sample and group definitions used in Paper 1, with a K-S test on specific star-formation rates yielding a difference of $\approx$ 2.5$\sigma$. The  sample of group galaxies used in this work and the sample of isolated galaxies from Paper 2 are not large enough for a statistical comparison to produce meaningful results, but we do find our group sample to have slightly lower average sSFRs and lower proportion of SF galaxies (using our template classification detailed in Paper 1). 

\begin{figure}
\includegraphics[width=\columnwidth]{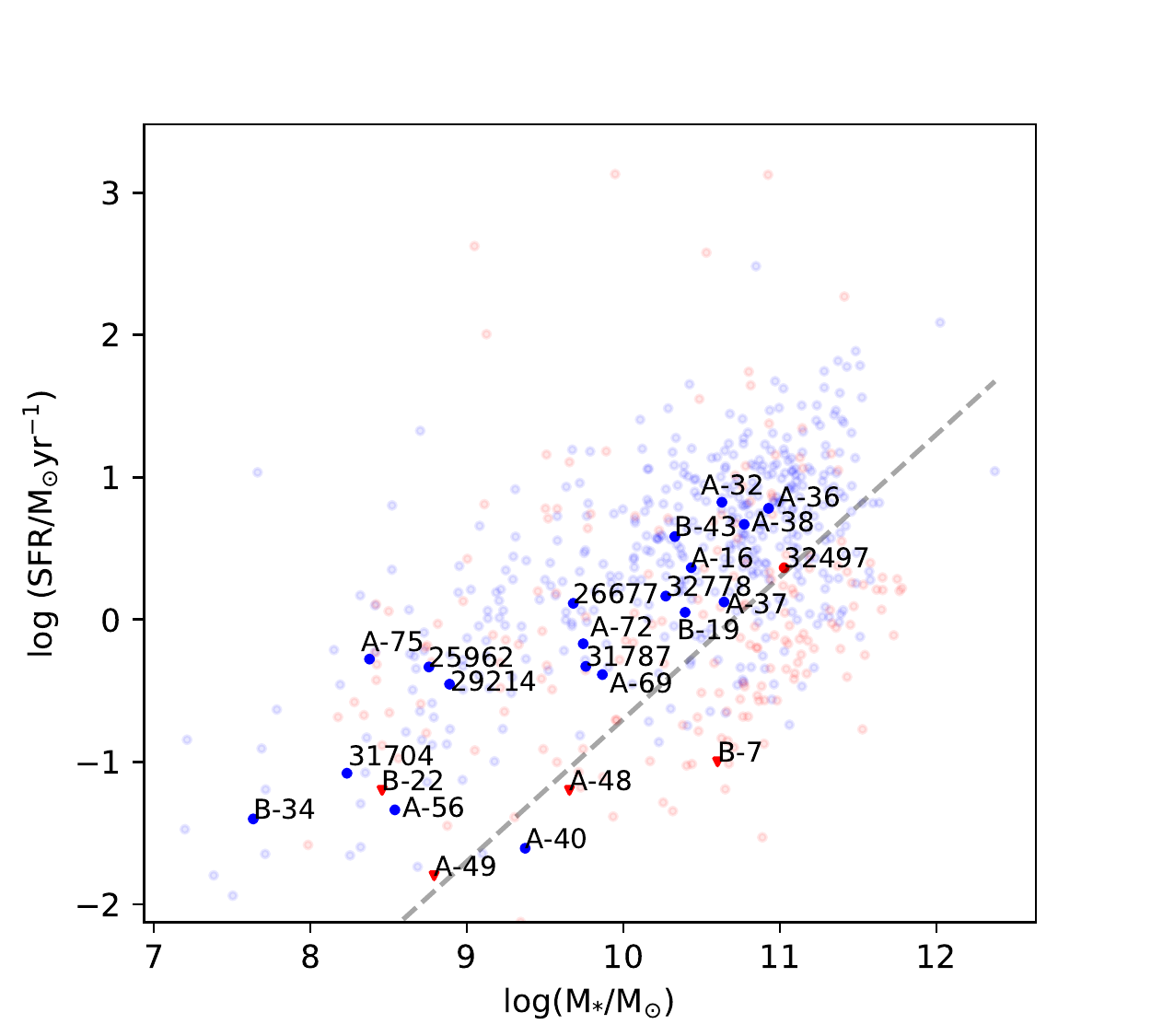}
\caption{Stellar mass vs star-formation rate for galaxies in our sample. Faded points show the overall galaxy sample (identical to Figure 5 from Paper 1), whilst galaxies detailed in this work are bold and labelled with the galaxy MOS or MUSE ID as given in Table \ref{table:summary_groups}. Galaxies identified as star-forming are shown in blue, with non-star-forming galaxies in red. The grey dashed line indicates an sSFR of 0.02 $\textrm{Gyr}^{-1}$, an approximate match to the SF/non-SF designations that were made using template fitting. Masses are estimated using Equations 1 and 2 from \citet{johnson2015} and star-formation rates estimated using the H$\alpha$ or [OII] (3727 \AA) luminosities. These measurements are detailed in Paper 1. The objects marked with triangles are upper limits where no clear emission line is detected. Note that the small number of non-star-forming galaxies with apparent extremely high SFRs are due to fringing effects in the VIMOS data (see T14 Section 3.1) being fit as emission lines. These erroneous measurements do not affect our results.} \label{fig:sfr_mass}
\end{figure}

The stellar masses and SFRs of galaxies in our sample are shown in Figure \ref{fig:sfr_mass}, which can be directly compared to the similar figure in Paper 2. These two subsamples span a similar range in mass and SFR. Both our Paper 2 and Paper 3 subsamples are biased towards low-mass and star-forming galaxies, as we focus on galaxies near to the sightlines, where the increased depth and easier detection of emission-line galaxies using MUSE has a larger impact on the sample.

\section{Models}\label{sec:model_summary}

We use the same three basic models as in Paper 2, namely a power-law halo, bi-conical outflow, and rotating disk. In that work we include a full description of the model parameters, the process of using the galaxy observations and model parameters to generate synthetic spectra, and the process of optimizing for the best-fit models (Section 3 and Appendix A therein).

To briefly summarize, the strength of absorption caused by a spherical halo with power-law density profile is determined by its index $\alpha$ and the distance between the galaxy and sightline, whilst its velocity is constant and may be offset from the galaxy by $v_{\delta}$. An outflow has a constant radial velocity $v_{out}$ and has non-zero density only within a polar angle $\theta_{out}$. We assume a constant outflowing flux, leading to a density profile of $r^{-2}$ within the cone of the outflow. A hollow cone, such that density is zero close to the galaxy minor axis (within $\theta_{in}$), is allowed for, but in most cases is not required. Disks have an exponential profile with scale heights $h_{r}$ and $h_{z}$, alongside radial and azimuthal velocity components $v_{r}$ and $v_{\phi}$ respectively. All three model types also have an allowed thermal or turbulent velocity $v_{t}$ that broadens the absorption profile (we do not attempt to distinguish between thermal and turbulent velocities).  

In addition to these free parameters, the measurements of galaxy position angle and inclination (from the GALFIT models applied to the HST image) are required as inputs to these models, alongside the chosen galaxy orientation ($S_{Nr}$ and $S_{Wr}$, described in Paper 2). Both the direction of galaxy rotation and the direction along the minor axis that points away from the observer may be unconstrained, although the MUSE kinematics and the direction of any spiral arms can be used, if visible, to constrain these before modelling the absorption. We note that none of the models in this paper are constrained by the direction of galaxy spiral arms, and that only in the case of G-536 do we rule out any models based only on the galaxy kinematics.

Model spectra are generated by combining 10pc segments along each line-of-sight. Each segment has a density determined by the distance and orientation between the sightline segment and the galaxy as well as the density profile of the model, and a line-of-sight velocity determined by the model outflow/infall/rotation velocities projected into the direction of the line-of-sight. Combining these yields a column density, and therefore optical thickness, as a function of velocity or wavelength. The model absorption profile is calculated by combining the optical thickness of all model components included (i.e. the total result from all disk and outflow models around galaxies in the group), converting this to a transmission spectrum, and then convolving with the instrumental line-spread function.

As we discuss in Paper 2, creating an automated routine to find the best-fit parameters of these models is complex and time-consuming, so is not considered for a small sample within a single field. Instead we consider each possible model (halo, disk or outflow for each galaxy with inclination and position angle measurements), and determine which absorption components could be fit by such a model without producing excess absorption in the synthetic spectrum over that seen in the observations, for any of the three lines-of-sight. This primarily concerns eliminating model combinations that do not produce velocity offsets in the correct direction, or match the relative column densities that would be produced in the different sightlines. We then iteratively adjust the model parameters to produce a reasonable fit, and finally find the combination of models that reproduces the maximum number of observed absorption components. 

This method identifies the various models capable of reproducing each observed absorption feature, although it does not provide a quantitative measure of the best fit (which would likely depend substantially on the priors chosen on the parameter space). Throughout this work we therefore list the different model combinations found to produce a reasonable approximation of the observed spectra.

\section{Absorption in Galaxy Groups} \label{sec:group_details}

We now apply these models to the absorption in galaxy groups, and attempt to reproduce the \ion{H}{i} absorption components visible in the QSO spectra at the group redshift. Although Ly$\alpha$ is usually preferred, in some cases it is saturated, blended, or lies in the lower-resolution FOS spectra, so Ly$\beta$ provides better constraints on the models. Below we discuss our preferred combination of models for three of these groups, including the model parameters and the reasons for rejecting alternative combinations. The remaining six groups are discussed in Appendix \ref{sec:more_groups}.

\subsection{G-238}

\begin{figure*}
\includegraphics[width=\textwidth]{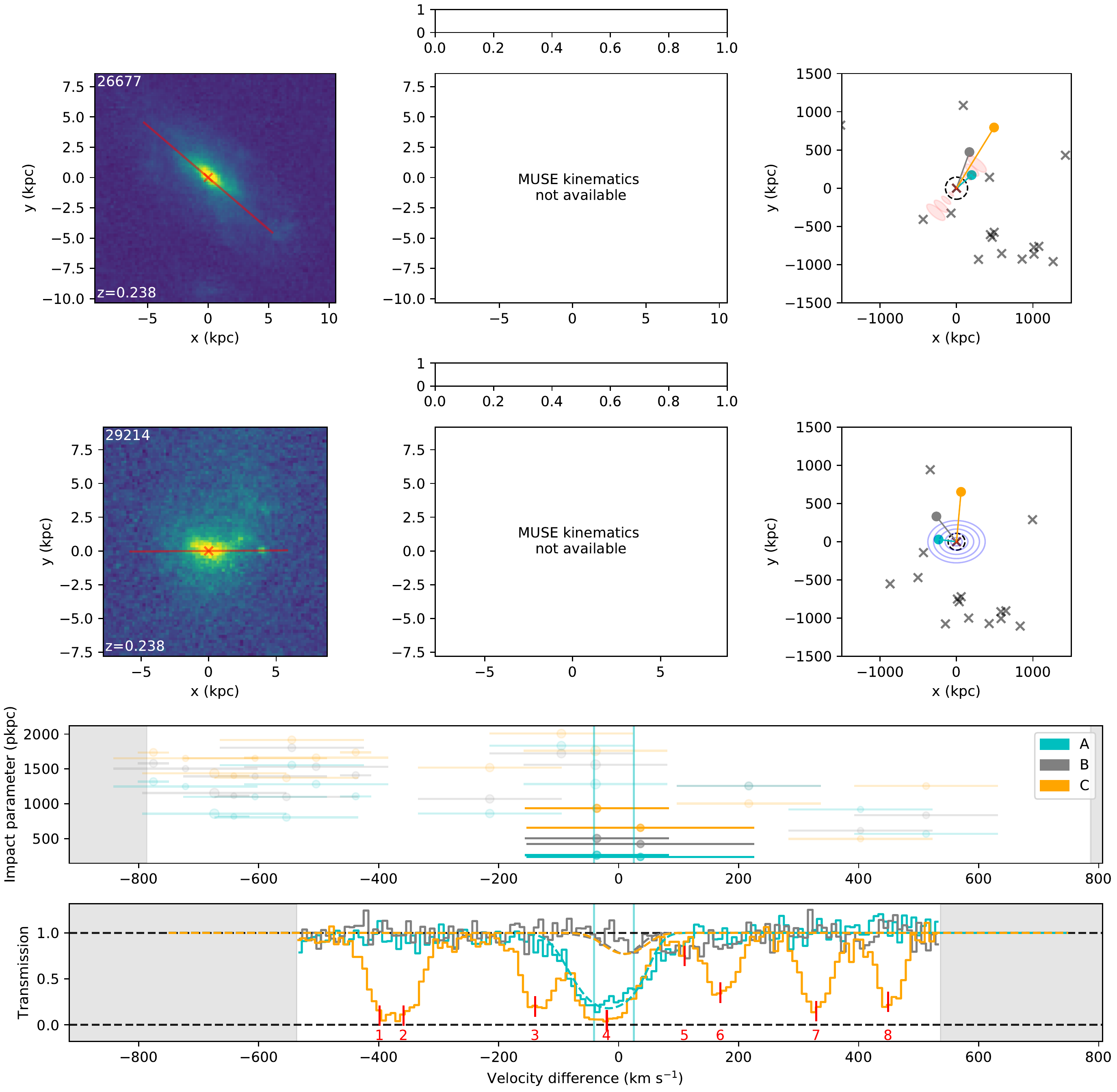}
\caption{Details of the absorption and galaxy environment around group G-238. The upper two rows of panels illustrate the galaxies and the models used, as follows. \textbf{Upper-Left:} HST image of the galaxy, with projected major axis shown in red. \textbf{Upper-Middle:} Velocity map based on emission lines detected in MUSE, scaled and rotated to match the HST image. Only spaxels with a 3$\sigma$ line detection are shown. (Note that both galaxies in this group lie outside the MUSE fields.) \textbf{Upper-Right:} Wide view showing the location of the three QSOs (cyan, grey and orange connected dots representing QSOs A, B and C). The locations of other galaxies at this redshift are indicated by grey crosses; the estimated galaxy virial radius by a dashed black circle; and schematics of models proposed to fit the absorption by red and blue ellipses for outflows and disks respectively. \textbf{Central Panel:} The galaxies at this redshift, with the galaxies shown in the upper panels bolded, and others faded. Each galaxy is shown three times, showing the impact parameter between that galaxy and each of the three QSOs. The horizontal bar denotes the velocity uncertainty derived from the galaxy redshift measurement. \textbf{Bottom panel:} The transmission of the quasar spectra at the wavelength of Ly$\alpha$ at the group redshift. The observed QSO spectra are shown as solid lines, with dashed lines giving the synthetic spectra produced by our models. Vertical lines passing through both panels show detected \ion{H}{i} absorption components. The model shown in the lower panel is a disk with rotation velocity $\sim$ 100 \kms{ }around galaxy 29214, and an outflow with opening angle $20^{\circ}$ and velocity 120 \kms. Additional absorbers identified by red ticks are mostly molecular lines at the redshift of the sub-DLA at z $\approx$ 0.56, except for the weak absorption in LOS-B (5), which is Ly$\beta$ from z $\approx$ 0.47. \label{fig:26677_detailed}}
\end{figure*}

Group G-238 consists of two star-forming galaxies (29214 and 26677) appearing in the HST field at z $\approx$ 0.24. These are the two nearest galaxies to the lines-of-sight at this redshift, although there are several others at larger impact parameters. The observations are detailed in Table \ref{table:groups_26677} and Figure \ref{fig:26677_detailed}.

QSO-A features two absorption components at this redshift, with the redder absorber featuring \ion{O}{vi} detected at a significant level. No significant absorption is detected in B or C, but some weak absorption in C could be hidden by molecular lines from the sub-DLA. Both of the absorbers are at very similar velocities to the two galaxies, but a simple halo model requires a steep density profile (much steeper than $r^{-2}$), otherwise it would produce absorption in B that is inconsistent with observations. QSO-A also lies close to the minor axis of 26677 and the major axis of 29214. A model consisting of an outflow and disk around the two galaxies respectively can approximately produce the results seen in the observations. 

These galaxies are reasonably well-separated, lying outside of each others' virial radii, so it is not surprising that we see no clear sign of interaction in the absorption. The absorption at impact parameters $\approx$ 250 kpc from both galaxies lies outside the virial radius, but within the impact parameter range that exhibits a bimodal position angle distribution. We suggest in Paper 1 that this is likely due to disk and outflow structures similar to the models used here. This combination of models is also supported by the detection of \ion{O}{vi} at a redshift matching the redder Ly$\alpha$ component that we identify as a likely outflow. The other absorber has \ion{O}{vi} to \ion{H}{i} ratio less than 1/5 of this, better fitting accretion from the IGM. We also note that neither a disk around galaxy 26677 nor an outflow around 29214 would intersect any of the lines of sight at small distances, so both of these structures may also exist around these galaxies.

\begin{table*}
\begin{center}
\caption{Summary of galaxy and absorber properties for group G-238. Any additional galaxies and metal absorbers have velocities shown relative to the first galaxy (26677). with columns as follows: (1) Group redshift; (2) Galaxy ID; (3) Galaxy luminosity (as a multiple of $L_{\star}$; (4) Galaxy inclination; (5) Line-of-sight identifier; (6) Impact parameter between galaxy and line-of-sight at the group redshift; (7) Azimuthal angle between galaxy major axis and the line-of-sight; (8) Absorber column density; (9) Absorber Doppler parameter; (10) Velocity offset between galaxy and absorber (Any additional galaxies and metal absorbers have velocities shown relative to galaxy 26677); (11) Any detected metal ions at the same redshift as this \ion{H}{i} absorber.}
\label{table:groups_26677}
\begin{tabular}{| c| c| c| c| c| c| c| c| c| c| c|}
\hline 
z & Galaxy & Lum ($L_{\star}$) & Inc & LOS & Imp (kpc) & Azimuth & log(N \ion{H}{i}) & b (\kms) & $\Delta$v (\kms) & Other ions \\
(1) & (2) & (3) & (4) & (5) & (6) & (7) & (8) & (9) & (10) & (11) \\ \hline 

0.238 & 26677 & 0.22 & $67^{\circ}$ $\pm$ $1^{\circ}$ & A & 262 & $81^{\circ}$ $\pm$ $1^{\circ}$ & 13.99 $\pm$ 0.20 & 43 $\pm$ 7 & -10 $\pm$ 100 &  \\
 &  &  &  & A & 262 & $81^{\circ}$ $\pm$ $1^{\circ}$ & 13.75 $\pm$ 0.35 & 54 $\pm$ 21 & +60 $\pm$ 100 & \ion{O}{vi} \\
 & & & & B & 502 & $69^{\circ}$ $\pm$ $1^{\circ}$ & (None, limit $\sim$12.8) &  &  &  \\
  & & & & C & 934 & $81^{\circ}$ $\pm$ $1^{\circ}$ & (None, limit $\sim$12.9) &  &  &  \\
 \\
 & 29214 & 0.11 & $43^{\circ}$ $\pm$ $3^{\circ}$ & A & 236 & $8^{\circ}$ $\pm$ $5^{\circ}$ & & & (+70)& \\
 & & & & B & 422 & $52^{\circ}$ $\pm$ $5^{\circ}$ & & & & \\
 & & & & C & 654 & $84^{\circ}$ $\pm$ $5^{\circ}$ & & & & \\
 \\
 & (22676) & 0.32 & & A & 859 & & & & (-180)& \\
 & & & & B & 1069 & & & & & \\
 & & & & C & 1519 & & & & & \\
  \\
 & (33195) & 0.04 & & A & 919 & & & & (+440)& \\
 & & & & B & 616 & & & & & \\
 & & & & C & 495 & & & & & \\
  \hline
\end{tabular}

\end{center}
\end{table*}

\subsection{G-399}

\begin{figure*}
\includegraphics[width=0.930\textwidth]{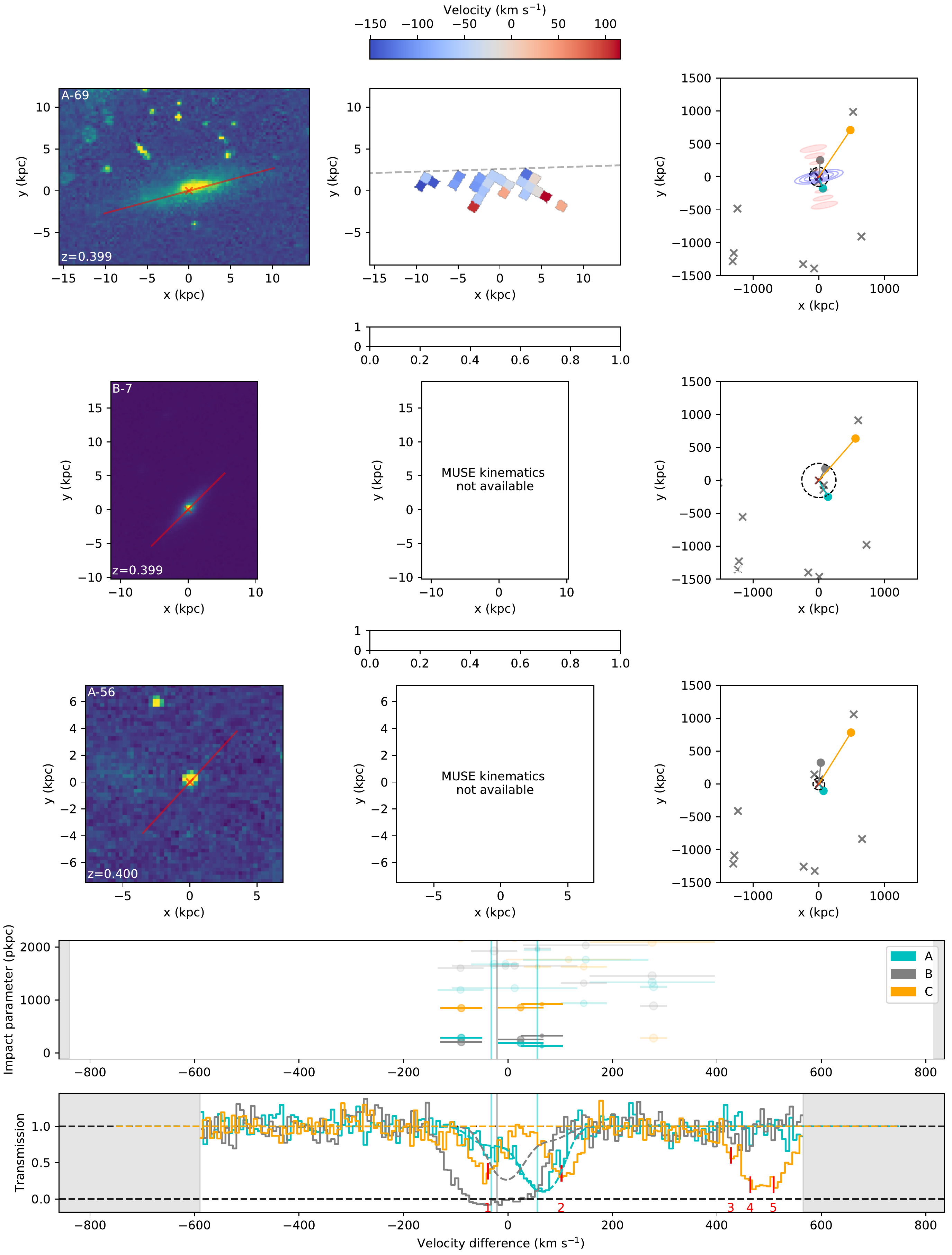}
\caption{Details of the absorption and galaxy environment around group G-399. The layout is identical to that shown in Figure \ref{fig:26677_detailed}, with kinematics measured from the H$\alpha$ emission line seen in the MUSE data, and the model shown in the lower panel combines an outflow with $25^{\circ}$ half-opening angle and 160 \kms{ }velocity with a disk with 150 \kms{ }rotation and 50 \kms{ }infall, both around galaxy A-69. All five absorption components labelled with red ticks are molecular lines associated with the sub-DLA at z$\approx$0.56. Note that galaxy A-69 lies at the edge of the MUSE field, so the velocity map is truncated approximately along the dashed grey line.\label{fig:a69_detailed}}
\end{figure*}

A-56, A-69 and B-7 are the three galaxies within the HST image at z $\approx$ 0.399. The state of the gas causing absorption in the sightlines at this redshift is modelled in \citet{anshul2021}; both sightlines feature transitions from multiple metal ions. The details are given in Table \ref{table:groups_a56} and illustrated in Figure \ref{fig:a69_detailed}. Note that the absorption features in QSO-C are identified with transitions from different redshifts.

B-7 is the largest galaxy in this group, and the closest to QSO-B, but is non-star-forming. The other two galaxies within the HST field are star-forming galaxies but also less massive. A-69 lies on the edge of the MUSE field, with the edge of the field running approximately along the major axis. There may be a velocity gradient across the galaxy, but this is not clear. A-56 lacks a well-determined orientation, as it is indistinguishable from a point source in the HST image.

An outflow around A-69 is capable of producing two absorption components in A, but not the factor of $\approx$10 difference in column density between the two components and the much stronger absorption in B. However, combining this outflow with a disk can reproduce the observed absorption in A whilst remaining consistent with B and C. An outflow with $30^{\circ}$ half-opening angle and 160 \kms{ }velocity, alongside an extended \ion{H}{i} disk with 160 \kms{ }rotation returns an approximate match. An outflow around B-7 is ruled out, as the large velocity offset required to match the absorption in A alongside the opening angle required to cover LOS-A would produce a substantially broader absorption feature.

No disk or outflow around any of the galaxies could produce the absorption in B without substantially exceeding the observed levels of absorption in A or C. The high column density in B is metal-enriched, exhibiting absorption from a range of metal ions, and has a line-of-sight velocity between the two larger galaxies in this group. This may suggest that this material has been stripped from one of the galaxies, or that an outflow from one of the galaxies has been distorted by interaction with the CGM of the other such that it can no longer be fit by our toy models. 

\citet{anshul2021} find that the absorption in QSO-A is consistent with solar metallicity in both components, and use a two-phase model with a low-ionization phase traced by \ion{H}{i} and \ion{C}{iii} (among other ions) and a higher-ionization phase traced primarily by \ion{O}{vi}. The low-ionization phase is consistent with photoionization in the stronger component, and the \ion{O}{vi} in both components is consistent with collisionally ionized, T $\gtrsim 10^{5}$ K gas. This would appear to be consistent with the \ion{O}{vi} resulting primarily from outflowing material in both components. They note that the material observed would likely have been ejected from the central galaxy $>$ 600 Myr ago. It therefore seems possible that the material we have modelled as a rotating disk has been recycled from this outflow, but only gas that has cooled efficiently is seen in this disk. In this scenario, the stronger absorption component consists of both the proposed cool disk and a warm outflow; the cool disk dominates this absorption component for the low-ionization phase, but less so for the \ion{O}{vi}. This would explain the similar metallicity of both phases and components. Stripped material near the minor axis could also reproduce the observations in place of this possible outflow.

\citet{anshul2021} also find multiple phases in the absorption in B, with a broad Ly$\alpha$ component alongside the \ion{O}{vi} and a narrower component matching the \ion{C}{iii} in a lower-ionization phase. They find that the low-ionization phase is consistent with photoionized gas at $\sim$1/10th solar metallicity, and that the \ion{O}{vi} could be produced either by diffuse hot gas at similar metallicity, or by cooler collisionally-ionized gas with near-solar metallicity. They suggest that these could respectively trace either the diffuse, hot CGM or intra-group medium, or the interface between a low-ionization cloud and the hot-CGM (that cloud possibly originating from an outflow). 

If this absorption in B is due to a hot and diffuse CGM or intra-group medium, it must either be distorted or patchy, as a spherical distribution cannot match the ratio of absorption strengths in the two sightlines. Similarly, our biconical outflow models cannot reproduce this ratio, suggesting (if an outflow exists) either a substantial change in outflow rate with time, a very patchy medium, or distortion due to the interaction between the two galaxies.

\begin{table*}
\begin{center}
\caption{Summary of galaxy--absorber group at z $\sim$ 0.399. Any additional galaxies and metal absorbers have velocities shown relative to the first galaxy (A-56). Columns are identical to those in Table \ref{table:groups_26677}.}
\label{table:groups_a56}
\begin{tabular}{| c| c| c| c| c| c| c| c| c| c| c|}
\hline 
z & Galaxy & Lum ($L_{\star}$) & Inc & LOS & Imp (kpc) & Azimuth & log(N \ion{H}{i}) & b (\kms) & $\Delta$v (\kms) & Other ions \\
(1) & (2) & (3) & (4) & (5) & (6) & (7) & (8) & (9) & (10) & (11) \\\hline 

0.399 & A-56 & 0.01 & $35^{\circ}$ $\pm$ $35^{\circ}$ & A & 122 & $76^{\circ}$ $\pm$ $45^{\circ}$ & 13.15 $\pm$ 0.09 & 15 $\pm$ 6 & -100 $\pm$ 40 & \ion{C}{iii}, \ion{O}{vi} \\
 &  &  &  & A & 122 & $76^{\circ}$ $\pm$ $45^{\circ}$ & 14.14 $\pm$ 0.03 & 35 $\pm$ 2 & -10 $\pm$ 40 & \ion{C}{iii}, \ion{O}{vi} \\
 & & & & B & 325 & $38^{\circ}$ $\pm$ $45^{\circ}$ & 16.77 $\pm$ 0.02 & 24.9 $\pm$ 0.5 & -90 $\pm$ 40 & \ion{C}{i}, \ion{N}{iii}, \ion{Si}{iii}, \ion{O}{iii}, \ion{O}{vi}  \\
   & & & & C & 921 & $11^{\circ}$ $\pm$ $45^{\circ}$ & (None, limit $\approx$13.0) & & &  \\
 \\
 & A-69 & 0.19 & $77^{\circ}$ $\pm$ $2^{\circ}$ & A & 186 & $82^{\circ}$ $\pm$ $2^{\circ}$ & & & (-40) & \\
 & & & & B & 254 & $74^{\circ}$ $\pm$ $2^{\circ}$ & & & & \\
 & & & & C & 857 & $44^{\circ}$ $\pm$ $2^{\circ}$ & & & & \\
  \\
 & B-7 & 0.36 & $82^{\circ}$ $\pm$ $1^{\circ}$ & A & 285 & $70^{\circ}$ $\pm$ $1^{\circ}$ & & & (-150) & \\
 & & & & B & 205 & $12^{\circ}$ $\pm$ $1^{\circ}$ & & & & \\
 & & & & C & 846 & $1^{\circ}$ $\pm$ $1^{\circ}$ & & & & \\
  \\
 & (26721) & 0.2 & & A & 937 & & & & (+80) & \\
 & & & & B & 1318 & & & & & \\
 & & & & C & 1626 & & & & & \\
  \\
 & (34572) & 1.2 & & A & 1251 & & & & (+210) & \\
 & & & & B & 889 & & & & & \\
 & & & & C & 280 & & & & & \\
  \hline
\end{tabular}

\end{center}
\end{table*}

\subsection{G-876}

\begin{figure*}
\includegraphics[width=\textwidth]{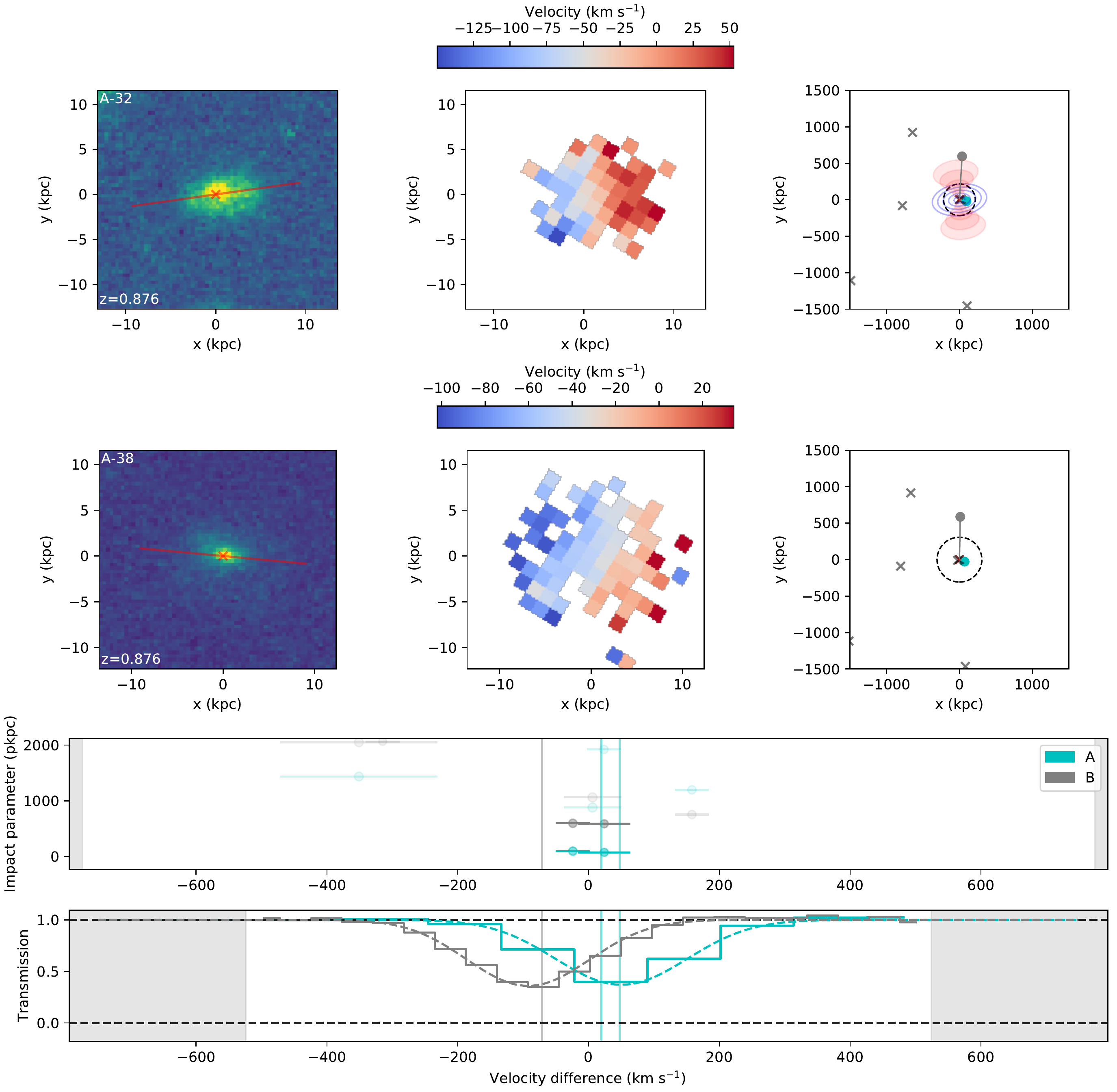}
\caption{Details of the absorption and galaxy environment around group G-876. The layout is identical to that shown in Figure \ref{fig:26677_detailed}, with kinematics measured from the [\ion{O}{ii}] emission line seen in the MUSE data, and the model shown in the lower panel combines an outflow with velocity 160 \kms and opening angle $40^{\circ}$ and a disk with rotation velocity 100 \kms. \label{fig:a32_detailed}}
\end{figure*}

A-32 is an $\approx 0.3 L_{*}$ galaxy at z $\sim 0.88$, inclined at $\sim 30^{\circ}$. It is paired with A-38, a galaxy of similar magnitude that is less than 30 kpc away. Neither galaxy shows any signs of morphological distortion in the HST image or any kinematic signatures of interaction between the two galaxies; both galaxies show a velocity gradient along their major axis that is likely due to rotation (with velocity $\approx$ 100 \kms). This system is detailed in Table \ref{table:groups_a32} and Figure \ref{fig:a32_detailed}.

A-38 is redshifted by $\sim$50 \kms{ }relative to A-32, and is therefore at the same redshift as the absorption in sightline A, whilst sightline B is blueshifted by 100 \kms{ }relative to this galaxy. This also exhibits a $\sim$ 100 \kms{ }velocity gradient. 

Sightline A lies at a distance of $\sim$100 kpc along the major axis of A-32 and 75 kpc at $\sim 20^{\circ}$ to the major axis of A-38, with sightline B at $\sim$600 kpc along the minor axis. Both show absorption, with \ion{H}{i} column densities $\sim 10^{15.8}$ and $10^{15.4}$ $\textrm{cm}^{-2}$ and Doppler parameters 30 and 20 \kms{ }respectively. Sightline A features \ion{O}{iii} and \ion{O}{iv} absorption at this redshift, whilst sightline B does not show any metal absorption. However, the detection limit for these ions allows for the gas seen in LOS-B to have similar \ion{O}{iii} and \ion{O}{iv} column densities to that in A. Note that as both Ly$\alpha$ and Ly$\beta$ lie in the FOS gratings at this redshift, the Doppler parameters are not resolved; higher-order lines appearing in COS constrain the Doppler widths but reveal no additional structure. These galaxies lie beyond the redshift of QSO-C, so no absorption can be detected in the third sightline.

The relatively similar column densities alongside a large difference in impact parameter prevents any of our models from simultaneously matching both absorbers. That the absorption in A is close to the major axis of both galaxies, whilst B is near the minor axis of both galaxies, would suggest that these could be a disk and outflow respectively.

If the absorption in sightline A is associated with A-32, it is co-rotating and could be part of an extended disk with rotation velocity $\approx$100 \kms{ }(depending on any infall component). If associated with A-38, the absorption in A must have comparable rotation and infall velocities, in order to produce absorption with no clear line-of-sight velocity offset. Alternatively, a power-law halo around A-38 could reproduce this lack of velocity offset.

An outflow from either galaxy can also reproduce the absorption in B. In order to reproduce the velocity offset, these putative outflows would require velocities of $\approx$140 \kms{ }(from A-32) or $\approx$210 \kms{ }(from A-38). We note that A-38 cannot produce both a disk matching A and an outflow matching B, as the need for the disk velocity offsets to `cancel' fixes the galaxy orientation, whilst an outflow matching B requires the opposite orientation. 

This still leaves several possible combinations of models that can reproduce the observations. A reasonable fit is shown in Figure \ref{fig:a32_detailed}, and combines an outflow with velocity 140 \kms and opening angle $40^{\circ}$ and a disk with rotation velocity 100 \kms, both originating from A-32.

We also note that galaxy 26501 is substantially brighter than either A-32 or A-38, so may be contributing to the absorption in LOS-B. Furthermore, the velocity difference between the two galaxies and the location of each galaxy near the major axis of the other suggest orbital angular momentum with similar alignment to the rotation of both galaxies. Therefore the absorption that could be identified as a disk around one galaxy may also be tidal material resulting from their interaction, or larger-scale accretion into the group, rather than associated with one of the galaxies.


\begin{table*}
\begin{center}
\caption{Summary of galaxy--absorber group G-876. Any additional galaxies and metal absorbers have velocities shown relative to the first galaxy (A-32). Note that this group is beyond the redshift of QSO-C, so no absorption could be detected. Columns are identical to those in Table \ref{table:groups_26677}.}
\label{table:groups_a32}
\begin{tabular}{| c| c| c| c| c| c| c| c| c| c| c|}
\hline 
z & Galaxy & Lum ($L_{\star}$) & Inc & LOS & Imp (kpc) & Azimuth & log(N \ion{H}{i}) & b (\kms) & $\Delta$v (\kms) & Other ions \\
(1) & (2) & (3) & (4) & (5) & (6) & (7) & (8) & (9) & (10) & (11) \\\hline 

0.876 & A-32 & 0.30 & $55^{\circ}$ $\pm$ $3^{\circ}$ & A & 97 & $19^{\circ}$ $\pm$ $4^{\circ}$ & 15.78 $\pm$ 0.30 & 30 $\pm$ 5 & 40 $\pm$ 20 & \ion{O}{iv} (0), \ion{O}{iii} (-10)\\
 & & & & B & 597 & $79^{\circ}$ $\pm$ $4^{\circ}$ & 15.45 $\pm$ 0.47 & 22 $\pm$ 3 & -50 $\pm$ 20 &  \\
  \\
 & A-38 & 0.24 & $54^{\circ}$ $\pm$ $5^{\circ}$ & A & 75 & $16^{\circ}$ $\pm$ $5^{\circ}$ & & & (+50) & \\
 & & & & B & 588 & $86^{\circ}$ $\pm$ $5^{\circ}$ & & & & \\
 \\
 & (26501) & 0.80 & & A & 879 & & & & (+30)& \\
 & & & & B & 1062 & & & & & \\
  \\
 & (29799) & 0.38 & & A & 1197 & & & & (+180)& \\
 & & & & B & 755 & & & & & \\
    \hline
\end{tabular}

\end{center}
\end{table*}

\section{Discussion} \label{sec:discussion}

\begin{table*}
\begin{center}
\caption{Summary of galaxies in groups and the best-fitting toy models (ordered by redshift). Only galaxies with estimated position angles are shown in this table. Columns are as follows: (1) Group identifier; (2)-(5) Galaxy IDs, luminosities, star-formation class derived from template fitting and presence of kinematics, as given in Table \ref{table:summary_galaxies}; (6) Brief description of model combinations found to produce a reasonable fit. }
\label{table:summary_groups}
\begin{tabular}{| c| c| c| c| c| l|}
\hline 
Group & Galaxy & Lum ($L_{\star}$) & SF-Class & Kinematics & Model(s)  \\ 
(1) & (2) & (3) & (4) & (5) & (6) \\\hline 

G-202 & B-22 & 0.02 & SF & Poor & (4/5 absorbers) B-22 outflow, $\theta$=$15^{\circ}$, v=250 \kms, extent > 100 kpc \\
& 25962 & 0.07 & non-SF & No & AND 31787 outflow, $\theta$=$20^{\circ}$, v=50 \kms, extent > 400 kpc \\
& 31704 & 0.03 & SF & No & AND 32778 disk, height ratio $\approx$10, $v_{\phi}$= 120 \kms, $v_{r}$= 0 \kms, extent > 400 kpc \\
& 31787 & 0.08 & SF & No & (other options possible, e.g. 25962 disk, but less likely) \\
& 32497 & 1.1 & non-SF & No &  \\
& 32778 & 0.4 & SF & No &  \\
\\
G-238 & 26677 & 0.2 & SF & No & (2/2 absorbers) 26677 outflow, $\theta$=$20^{\circ}$, v=120 \kms, extent > 260 kpc  \\
& 29214 & 0.1 & SF & No & AND 29214 disk, height ratio $\approx$40, $v_{\phi}$= 100 \kms, $v_{r}$= 30 \kms, extent > 240 kpc \\
 & & & & & (29214 disk and halo possible instead, with similar disk parameters) \\
\\
G-383 & A-48 & 0.03 & non-SF & Poor & (3/3 absorbers) A-48 outflow, $\theta$=$10^{\circ}$, v=600 \kms, extent > 40 kpc \\
& A-49 & 0.01 & SF & Poor & AND A-49 disk, height ratio $\approx$100, $v_{\phi}$= 150 \kms, $v_{r}$= 40 \kms, extent > 400 kpc \\
& & & & & (A-48 disk with $v_{\phi}$= 320 \kms, $v_{r}$= 0 \kms also possible)  \\
& & & & & (some parameters very uncertain due to edge-on galaxy pair) \\
\\
G-399 & A-56 & 0.01 & SF & No & (2/3 absorbers) A-69 outflow, $\theta$=$25^{\circ}$, v=160 \kms, extent > 180 kpc  \\
& A-69 & 0.19 & SF & Poor & AND A-69 disk, height ratio $\approx$100, $v_{\phi}$= 150 \kms, $v_{r}$= 50 \kms, extent > 200 kpc \\
& B-7 & 0.36 & non-SF & No & (strongest absorber not fit by any toy model) \\
\\
G-517 & B-34 & 0.12 & SF & No & (No detected absorption) \\
& B-43 & 0.6 & SF & Yes & (Upper limit logN $\approx$13.3) \\
\\
G-536 & A-36 & 0.8 & SF & Yes & (4/7 absorbers) A-36 outflow, $\theta$=$55^{\circ}$, v=140 \kms, extent > 120 kpc  \\
& A-37 & 0.3 & SF & Poor & AND A-40 disk, height ratio $\approx$100, $v_{\phi}$= 110 \kms, $v_{r}$= 60 \kms, extent > 400 kpc \\
& A-40 & 0.02 & SF & No & (other absorption may be due to stripping in galaxy interactions, A-37 disk also possible) \\
\\
G-558 & A-72 & 0.10 & SF & Yes & (2/3 absorbers) A-72 outflow, $\theta$=$40^{\circ}$, v=150 \kms, extent > 500 kpc  \\
& A-75 & 0.03 & SF & Yes & OR A-75 outflow, $\theta$=$50^{\circ}$, v=70 \kms, extent > 350 kpc    \\
& A-77 & 0.04 & SF & No & AND \\
& & & & & A-75 disk, height ratio $\approx$70, $v_{\phi}$= 90 \kms, $v_{r}$= 50 \kms, extent > 200 kpc \\
& & & & & OR A-75 halo, $v_{\delta} \approx$ 0, $\alpha \gtrsim 2$\\
& & & & & (sub-DLA not fit, absorber is likely due to a galaxy that is not detected) \\
\\
G-876 & A-32 & 0.30 & SF & Yes & (2/2 absorbers) A-32 disk, height ratio $\approx$20, $v_{\phi}$= 100 \kms, $v_{r}$= 0 \kms, extent > 100 kpc \\
& A-38 & 0.24 & SF & Yes & OR A-38 disk, height ratio $\approx$10, $v_{\phi}$= 100 \kms, $v_{r}$= 100 \kms, extent > 80 kpc \\
& & & & & AND \\
& & & & & A-32 outflow, $\theta$=$40^{\circ}$, v=160 \kms, extent > 600 kpc \\
& & & & & OR A-38 outflow, $\theta$=$35^{\circ}$, v=210 \kms, extent > 600 kpc \\
\\
G-907 & A-16 & 0.17 & SF & Yes & (2/2 absorbers) A-16 disk, height ratio $\approx$20, $v_{\phi}$= 100 \kms, $v_{r}$= 50 \kms, extent > 200 kpc \\
& B-19 & 0.10 & SF & Poor & AND B-19 outflow, $\theta$=$60^{\circ}$, v=80 \kms, extent > 200 kpc \\
& & & & & OR \\
& & & & & A-16 outflow, $\theta$=$60^{\circ}$, v=150 \kms, extent > 180 kpc \\
& & & & & AND B-19 disk, height ratio $\approx$20, $v_{\phi}$= 130 \kms, $v_{r}$= 40 \kms, extent > 180 kpc \\
\\

\hline
\end{tabular}

\end{center}
\end{table*}

The model fits shown above can now be discussed in the context of our previous works, especially the sample of isolated galaxies considered in Paper 2, as well as results from the literature. We summarize the best-fitting results for each galaxy group in Table \ref{table:summary_groups}.

\subsection{Model success}

Firstly, our disk and outflow models provide a plausible fit for 21 of the 28 detected \ion{H}{i} components within 500 \kms{ }and 500 kpc of our sample galaxies. This 75\% success rate is somewhat higher than the $\approx$50-60\% found for the isolated galaxy sample in Paper 2 (13 of 26 within 500 \kms{ } and 12 of 20 within 300 \kms), a $\approx$2$\sigma$ difference under binomial statistics. We note that our `success rate' represents an upper limit on the fraction of detected \ion{H}{i} absorbers that originate in the disk, halo and outflow structures described by our models. A successful fit to our models only tentatively identifies an absorber as a possible disk/halo/outflow. Several alternative origins for apparently successful fits are discussed below.

It may be expected that for group galaxies with gravitational interactions, such forces would distort any structure in the CGM and reduce the rate at which these models can reproduce the absorption at large scales. This is seen in, for example, the M81 group, where 21 cm emission is seen tracing both a disk-like and biconical outflow structure around M82 on small scales, but distorted in the direction of M81 on larger scales \citep[e.g.][]{sorgho2019}. Such distortions would be expected to reduce the success rate of our models in galaxy groups, although in some circumstances our disk and outflow models could appear successful even with these distortions present.

Our higher success rate for group galaxies over isolated galaxies could suggest that galaxy interactions are not having a significant impact on our sample. However, there are several effects that could counter any impact of group interactions. 

Firstly, the larger number of galaxies near to each absorber increases the number of free parameters available for our models, and therefore the likelihood of obtaining a reasonable fit even if the models do not reflect the true physical state of the gas (or reflect the state of the gas poorly due to effects such as intermittent outflows and changes in ionization state, as well as distortions due to group interactions). However, the groups for which some \ion{H}{i} absorption cannot be fit by our models appear to be those with the largest number of galaxies, whereas for those with 2-4 galaxies we have been able to find a plausible fit to all identified absorbers (a difference of $\approx2\sigma$ in mean N$_{gals}$). This suggests that the increased number of free parameters, largest for groups with many galaxies, is not the primary reason for our high success rate.

We also search for other differences between the properties of the groups for which all absorbers were fit, and those with some absorption that could not be fit. No significant differences were found in their redshifts, total stellar masses, impact parameters to the nearest or most massive galaxy, ratios of stellar masses or luminosities of the brightest two galaxies, or the projected separations of the tightest pair of galaxies in each group. These tests therefore provide little indication of the likely origins of unexplained absorbers. We do find that individual absorbers that were not consistent with any of our models have higher \ion{H}{i} column densities than those we could successfully fit, and discuss these in more detail in Sections \ref{sec:discuss_interactions} and \ref{sec:discuss_metals}.

Second, the larger number of galaxies, and resulting smaller impact parameters between the lines-of-sight and the nearest galaxy, may also contribute to the success of our models. With the known correlation between impact parameter and column density \citep[e.g.][]{werk2014, wilde2021}, this contributes to generally higher \ion{H}{i} column densities in our groups (median \ion{H}{i} column densities of $10^{14.5}$ and $10^{13.8} \textrm{cm}^{-2}$ in absorbers associated with this work and our Paper 2 sample respectively). These higher column densities also make successful fits more likely, as weak absorption produced by the toy models can more easily be `hidden' under other absorbers, and column densities are less constrained once absorption begins to saturate. The smaller impact parameters also increase the likelihood of probing any inflowing or outflowing structures closer to the galaxy than any distortions caused by interactions within the group. 

Third, it is possible that some absorption resulting from group interactions could be fit by our models regardless. This could include material along the plane of a galaxy interaction masquerading as an accreting disk, and stripped material near the galaxy minor axis appearing consistent with an outflow. These false fits are unlikely to occur where multiple absorbers are reproduced by a single model, due to the extra constraints, but may contribute to some of our model fits based on a single absorber.



Studies of cool CGM gas using \ion{Mg}{ii} absorption have come to differing conclusions on the origin of the more extended absorption in galaxy groups. \citet{bordoloi2011} suggest that a superposition of CGM absorption from the constituent galaxies can reproduce the observed results, whereas \citet{nielsen2018} prefer a model in which absorbers are generally associated with the intra-group medium rather than any individual galaxy. Using samples extending to slightly higher redshifts, \citet{fossati2019} suggest that this is material originating primarily from tidal interactions between group galaxies, whilst \citet{dutta2020} support a model with contributions from individual halos and tidal interactions.

Our models cannot directly test for tidal material (although we do discuss this briefly in Section \ref{sec:discuss_interactions}), but do produce a superposition of possible structures sometimes found in the CGM of isolated galaxies. However, this differs from the superposition model tested in \citet{bordoloi2011}, \citet{nielsen2018} and \citet{dutta2020} in that the covering fraction of these CGM structures in groups is not constrained to be the same as that around isolated galaxies.

Our high model success rate provides some support for a superposition model for groups using our definition and sample selection, but does not rule out a contribution from intra-group and tidal material. The apparent difference between this conclusion and that of \citet{nielsen2018} may in part result from our larger linking lengths, so our group sample contains some galaxy pairs with larger separations, which are less likely to be affected by group interactions. Our sample also features galaxies at large impact parameters that may explain weak absorbers seen in group environments, which are not satisfactorily explained under their superposition model. \citet{dutta2020} found a similar result, ruling out superposition as the origin for their strongest and weakest absorbers. \citet{nielsen2018} found that combining the expected galaxy-absorber velocity offsets (as observed around isolated galaxies) with the galaxy-galaxy offsets seen in groups produced a model absorber-absorber velocity distribution significantly wider than observed. However, possibly due to including faint MUSE galaxies near the lines-of-sight, our groups exhibit smaller galaxy-galaxy velocity differences than their sample, reducing this inconsistency.

We also note that our sample includes galaxies fainter than those included in the \citet{bordoloi2011} sample. Their lack of faint galaxies likely leads to some interacting galaxies being classified as isolated, therefore making the absorption profiles of group and isolated galaxies more similar.

Our high success rate at reproducing absorption in our sample of galaxy groups suggests that this superposition of disk and outflow structures, similar to those sometimes found around isolated galaxies, may explain a substantial fraction of absorption found around these groups. However, the increase in parameter space due to the larger number of galaxies, effects due to the higher column densities and smaller impact parameters, and absorption from other sources (e.g. intra-group and tidal material) mis-identified as disk/outflow material, are all likely to contribute to the higher success rate found fitting these models to gas around groups than around isolated galaxies.

\subsection{Model parameters}

We briefly discuss the model parameters that produce the best fit for the absorption near these galaxy groups. In most cases these are similar to those found near isolated galaxies in Paper 2, where we discuss the parameters in more detail, but some differences are highlighted here.

\begin{table*}
\begin{center}
\caption{Model outflow properties around galaxies for which outflows can reproduce some of the observed absorption components. Column descriptions: (1)-(5) are from Table \ref{table:summary_galaxies}; (6) specific star-formation rate; (7) maximum impact parameter at which outflow is detected; (8) maximum extent at which absorption is detected (at the point of highest H I density along the sightline with the largest impact parameter); (9) galaxy virial radius; (10) model half-opening angle; (11) model outflow velocity; (12) escape velocity from the location of the sightline at the maximum observed extent. (Note that the starred escape velocities denote $r/r_{vir}$ > 2, so escape velocity is calculated assuming the outflowing material lies beyond the galaxy halo, rather than within an isothermal halo.) }
\label{table:summary_outflows}
\begin{tabular}{| c| c| c| c| c| c| c| c| c| c| c| c| c|}
\hline 
Galaxy & z &  $\textrm{M}_{\star}$ & $\textrm{M}_{h}$ & SFR & sSFR & $\textrm{b}_{max}$ & Extent & $\textrm{r}_{vir}$ & $\theta_{out}$ & $\textrm{v}_{out}$ & $\textrm{v}_{esc}$ \\
 & & $\textrm{log}_{10}(\textrm{M}_{\odot})$ & $\textrm{log}_{10}(\textrm{M}_{\odot})$ & ($\textrm{M}_{\odot}$/yr) & $\textrm{Gyr}^{-1}$ & (kpc) & (kpc) & (kpc) & ($^{\circ}$) & (\kms) & (\kms) \\ 
(1) & (2) & (3) & (4) & (5) & (6) & (7) & (8) & (9) & (10) & (11) & (12) \\ \hline 

B-22 & 0.202 & 8.4 $\pm$ 0.2  & 10.9 $\pm$ 0.3 & $0.06 \pm 0.04$ & $0.22^{+0.19}_{-0.10}$ & 100 & 110 $\pm$ 10 & 90 $\pm$ 20 & 15 & 250 & 110 $\pm$ 30 \\
31787 & 0.202 & 9.7 $\pm$ 0.1 & 11.5 $\pm$ 0.2 & 0.5 $\pm$ 0.2 & 0.08 $\pm$ 0.04 & 390 & 460 $\pm$ 10 & 130 $\pm$ 30 & 20 & 50 & 60* $\pm$ 30 \\
A-48 & 0.383 & 9.6 $\pm$ 0.2 & 11.5 $\pm$ 0.3 & < 0.04 & < 0.01 & 40 & 40 $\pm$ 10 & 120 $\pm$ 30 & 10 & 600 & 210 $\pm$ 50 \\
A-69 & 0.399 & 9.8 $\pm$ 0.2 & 11.6 $\pm$ 0.3 & 0.4 $\pm$ 0.2 & 0.06 $\pm$ 0.03 & 190 & 200 $\pm$ 10 & 130 $\pm$ 30 & 25 & 160 & 130 $\pm$ 40 \\
A-36 & 0.536 & 10.9 $\pm$ 0.2 & 12.9 $\pm$ 0.7 & 6 $\pm$ 3 & 0.08 $\pm$ 0.05 & 120 & 140 $\pm$ 10 & 360 $\pm$ 190 & 55 & 140 & 650 $\pm$ 200 \\
A-72 & 0.558 & 9.7 $\pm$ 0.5 & 11.6 $\pm$ 0.4 & 0.7 $\pm$ 0.2 & $0.13^{+0.20}_{-0.08}$ & 430 & 530 $\pm$ 20 & 120 $\pm$ 60 & 40 & 150 & 60* $\pm$ 30 \\
A-75 & 0.558 & 8.3 $\pm$ 0.4 & 10.9 $\pm$ 0.4 & 0.5 $\pm$ 0.1 & 2.3 $\pm$ 1.3 & 370 & 410 $\pm$ 20 & 80 $\pm$ 30 & 50 & 70 & 30* $\pm$ 10 \\
A-32 & 0.876 & 10.6 $\pm$ 0.2 & 12.4 $\pm$ 0.6 & 7 $\pm$ 2 & 0.16 $\pm$ 0.11 & 600 & 740 $\pm$ 30 & 210 $\pm$ 100 & 40 & 160 & 50 $\pm$ 30 \\
A-38 & 0.876 & 10.8 $\pm$ 0.3 & 12.6 $\pm$ 0.8 & 4.7 $\pm$ 1.1 & 0.08 $\pm$ 0.05 & 590 & 730 $\pm$ 40 & 250 $\pm$ 160 & 35 & 210 & 160 $\pm$ 100 \\
B-19 & 0.907 & 10.4 $\pm$ 0.5 & 12.2 $\pm$ 0.8 & 1.1 $\pm$ 0.3 & $0.05^{+0.10}_{-0.02}$  & 170 & 240 $\pm$ 20 & 170 $\pm$ 110 & 60 & 80 & 250 $\pm$ 160 \\
A-16 & 0.907 & 10.4 $\pm$ 0.5 & 12.2 $\pm$ 0.9 & 2.3 $\pm$ 0.6 & $0.09^{+0.18}_{-0.05}$ & 180 & 230 $\pm$ 30 & 170 $\pm$ 120 & 60 & 150 & 250 $\pm$ 170 \\
\hline

\end{tabular}

\end{center}
\end{table*}

\subsubsection{Outflows}

Table \ref{table:summary_outflows} lists the possible outflows that can reproduce some of the Ly$\alpha$ absorption seen near these groups of galaxies. 

The possible outflows have model parameters that are generally consistent with our results from Paper 2, with a range of half-opening angles extending to $\approx 60^{\circ}$ and velocities mostly between 50 and 250 \kms{ }(excepting the proposed A-48 outflow, for which our best estimate is 600 \kms{ }but much lower velocities of $\sim$200 \kms{ }lie within the 1$\sigma$ range, as the galaxy is very close to edge-on). The larger number of possible outflows over our Paper 2 result is due to a larger number of galaxies in the vicinity of each absorber, so each absorber is more likely to be consistent with at least one possible outflow model.

There is also very little difference in the extents of these outflows (median 200 kpc in groups, 180 kpc isolated; mean 280 and 270 kpc), whether or not these are normalized to the virial radius (mean extent $\approx$1.9 $r_{vir}$ and median $\approx$1.5 $r_{vir}$ for both samples). The median impact parameter to our putative outflows is similar to the median impact parameter between detected \ion{H}{i} and the nearest galaxy (190 kpc). This supports the existence of outflows extending to at least the virial radius from isolated and group galaxies, more similar to those in the FIRE simulations with cosmic rays included \citep[e.g.][]{hopkins2021}, than other models suggesting outflows rarely extend past the virial radius.


Using our best-fit results, seven of the eleven possible outflows would exceed escape velocity were the galaxy isolated, comparable to the two of four outflows exceeding escape velocity in Paper 2. Although the very small sample in Paper 2 provided some suggestion of a correlation between sSFR and $v/v_{esc}$, we do not find a significant correlation using the larger sample here ($\approx 1 \sigma$). 
\citet{schroetter2019} note that the outflows they find using \ion{Mg}{ii} absorption tend to only exceed escape velocity from low-mass galaxies ($\textrm{M}_{\star}/\textrm{M}_{\odot} \lesssim 10^{9.6}$). Whilst our two model outflows from galaxies with smaller masses than this do indeed appear to exceed escape velocity, several of our higher-mass galaxies also exhibit outflows exceeding their escape velocities. These outflows are probed at scales larger than the virial radius, whereas our two outflows around large galaxies probed on scales $< r_{vir}$ do not achieve escape velocity. This appears to support the continued acceleration at large radii seen in the simulations by \citet{hopkins2021} with cosmic rays included, although, as they discuss, only the material near or above the escape velocity would be capable of reaching these scales in the absence of additional acceleration.

In Paper 2, we noted that the model outflows in \citet{schroetter2019} did not feature opening angles as wide as the 45-$60^{\circ}$ found in our isolated sample, possibly due to their stellar masses tending to be larger and resulting in an increased collimation effect. Our group sample includes a larger number of high-mass galaxies, especially at high redshifts, and our model half-opening angles remain large for these galaxies. However, our models suggest narrower outflows provide a better fit to the galaxies at lower redshift, which also tend towards lower masses, opposite to the trend expected due to collimation. This trend may be weaker than it appears, as some of the higher-redshift objects (in the A-16 and A-32 groups) 
are unresolved and the absorber width does not help to constrain the outflow opening angle.

If those relying on unresolved absorption are removed, our total (group and isolated) sample of outflows has opening angles much more similar to those seen in \citet{schroetter2019}, but still shows a trend of wider outflows at higher masses and redshifts. It is unclear whether the redshift or mass difference is contributing to the difference in opening angle, but there are observations suggesting that outflows at higher redshifts are closer to isotropic with respect to the galaxy \citep[e.g.][]{chen2021a}.

\begin{table*}
\begin{center}
\caption{Model disk properties around galaxies for which disks can reproduce some of the observed absorption components. Column descriptions: (1)-(4) are from Table \ref{table:summary_galaxies}; (5) model scale height ratio (i.e. relative disk thickness); (6) maximum impact parameter at which this structure is detected (7) maximum observed disk extent (the 3-d distance from galaxy to the point on the line-of-sight where it intersects the disk plane, for the sightline with largest impact parameter with detected absorption); (8) virial radius; (9) virial velocity; (10) model circular velocity; (11) model radial/infall velocity.}
\label{table:summary_disks}
\begin{tabular}{| c| c| c| c| c| c| c| c| c| c| c| c| }
\hline 
Galaxy & z &  $\textrm{M}_{\star}$ & $\textrm{M}_{h}$ & $\textrm{h}_{r}/\textrm{h}_{z}$ & $\textrm{b}_{max}$ & Extent & $\textrm{r}_{vir}$ & $\textrm{v}_{vir}$ & $\textrm{v}_{\phi}$ & $\textrm{v}_{r}$ \\
 & & $\textrm{log}_{10}(\textrm{M}_{\odot})$ & $\textrm{log}_{10}(\textrm{M}_{\odot})$ & & (kpc) & (kpc)& (kpc) & (\kms) & (\kms) & (\kms)  \\
 (1) & (2) & (3) & (4) & (5) & (6) & (7) & (8) & (9) & (10) & (11) \\  \hline 
32778 & 0.202 & 10.2 $\pm$ 0.1 & 11.8 $\pm$ 0.3 & 20 & 410 & 420 $\pm$ 10 & 170 $\pm$ 60 & 130 $\pm$ 50 & 150 & < 20 \\
29214 & 0.238 & 8.9 $\pm$ 0.1 & 11.1 $\pm$ 0.3 & 40 & 240 & 240 $\pm$ 10  & 100 $\pm$ 20 & 70 $\pm$ 20 & 100 & 30 \\
A-49 & 0.383 & 8.7 $\pm$ 0.6 & 11.1 $\pm$ 0.4 & 200 & 410 & 410 $\pm$ 10 & 90 $\pm$ 50 & 80 $\pm$ 40 & 170 & < 20 \\
A-69 & 0.399 & 9.8 $\pm$ 0.2 & 11.6 $\pm$ 0.3 & 100 & 190 & 730 $\pm$ 300 & 130 $\pm$ 30 & 110 $\pm$ 30 & 150 & 40 \\
A-40 & 0.536 & 9.4 $\pm$ 0.4 & 11.4 $\pm$ 0.3 & 100 & 400 & 460 $\pm$ 60 & 110 $\pm$ 40 & 100 $\pm$ 30 & 110 & 70 \\
A-75 & 0.558 & 8.3 $\pm$ 0.4 & 10.9 $\pm$ 0.4 & 70 & 180 & 290 $\pm$ 60 & 80 $\pm$ 30 & 70 $\pm$ 20 & 90 & 50 \\
A-32 & 0.876 & 10.6 $\pm$ 0.2 & 12.4 $\pm$ 0.6 & 20 & 100 & 110 $\pm$ 10 & 210 $\pm$ 100 & 230 $\pm$ 130 & 100 & < 20 \\
A-38 & 0.876 & 10.8 $\pm$ 0.3 & 12.6 $\pm$ 0.8 & 10 & 80 & 90 $\pm$ 10 & 250 $\pm$ 160 & 250 $\pm$ 160 & 80 & 90 \\
B-19 & 0.907 & 10.4 $\pm$ 0.5 & 12.2 $\pm$ 0.8 & 20 & 170 & 250 $\pm$ 50 & 170 $\pm$ 110 & 190 $\pm$ 120 & 130 & 40 \\
A-16 & 0.907 & 10.4 $\pm$ 0.5 & 12.2 $\pm$ 0.9 & 20 & 180 & 290 $\pm$ 70 & 170 $\pm$ 120 & 200 $\pm$ 130 & 100 & 50 \\ 
\hline

\end{tabular}

\end{center}
\end{table*}

\subsubsection{Disks}

We show the best-fitting parameters for absorption fit by our disk model in Table \ref{table:summary_disks}.

The successful disk-like structures also appear similar to our results from Paper 2, with scale height ratios from 10 to $\sim$ 300, and extents ranging from less than 100 kpc to more than 1 Mpc. The average extents ($\approx$ 300 kpc) are also similar. However, we find substantially smaller circular velocities than around isolated galaxies (by $\approx$ 270 to 120 \kms median, or 3 $v_{vir}$ to 1), and also produce a better fit using smaller infall velocities for our group models. These infall velocities are often better constrained than for isolated galaxies (for which we generally assumed an infall velocity of 0.6$v_{vir}$), as with more high-column density absorption in groups it is more likely that multiple absorbers can be fit by a single model.

Given the small sample size, especially around isolated galaxies, this $\approx1.5\sigma$ difference in circular velocity may simply be due to the galaxy properties and sightline configurations that feature in the sample (although the extents and impact parameters are similar). The assumed infall velocity for isolated galaxies, and its effect on the rotational velocities required to fit the observed absorption, may also contribute, but changing this assumption will increase the rotation velocity for some models and decrease it for others.

A real difference between circular velocities of accreting material onto isolated and group galaxies may be detected here, but it is difficult to demonstrate when using absorption at only one or two locations around each galaxy. For example it is possible that, rather than seeing absorption due to gas falling onto one of the individual galaxies within our groups, we are seeing material falling into the group halo, which would likely extend further than the halo of an isolated galaxy of the same stellar mass as one of the group galaxies. Interaction between infalling material and group gas could slow its apparent velocity and contribute to the observed difference (as a similar interaction between the CGM and galaxy ISM appears to cause velocity lag in the ISM, \citealt{bizyaev2022}). Alignment between the group galaxies and large-scale structure \citep[e.g.][]{tempel2013, zhang2015, hirv2017} could lead to accretion at larger scales looking similar to accretion onto an individual galaxy (possibly supported by several of our groups featuring multiple galaxies with similar position angles, most noticeably G-383 and G-876). Accretion from large-scale structure into a galaxy group may also have been observed by \citet{bielby2017a}.

\subsubsection{Azimuthal angle dependency} \label{sec:angles}

\begin{figure}
\includegraphics[width=\columnwidth]{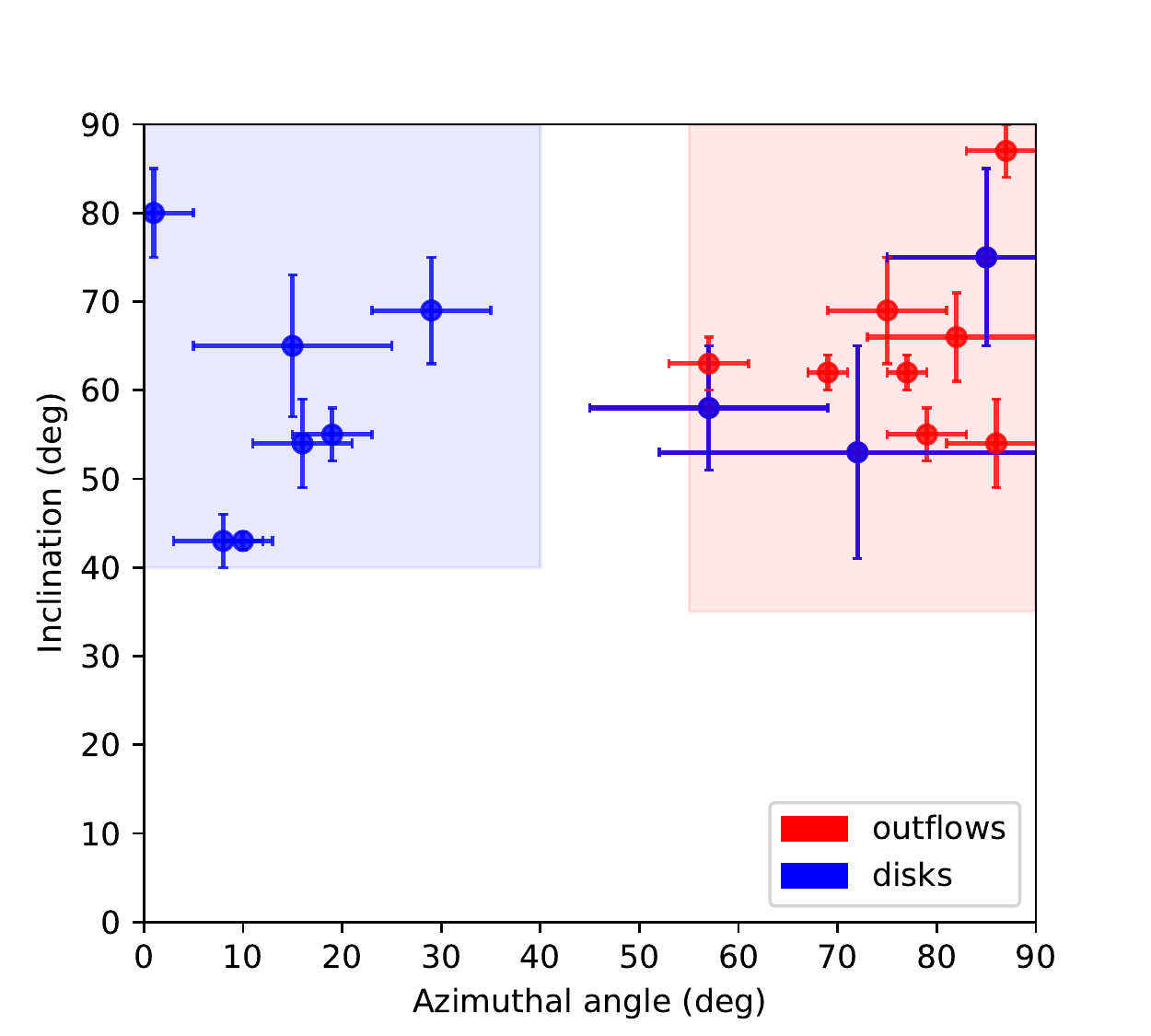}
\caption{Azimuthal angle against inclination for galaxy-absorber pairs in our sample. Absorbers identified as probing a possible outflow are shown in red, with disks in blue. We also shade the regions used to identify the `primary' disk and outflow subsamples in MEGAFLOW in blue and red respectively \citep{zabl2019, schroetter2019}. \label{fig:PA_inc_groups}}
\end{figure}

We also briefly note that attempting to identify disks and outflows purely by cutting in position angle and inclination \citep[as in e.g.][]{bordoloi2011, zabl2019, schroetter2019} produced results inconsistent with our models when applied to our Paper 2 sample of isolated galaxies, although our models were more likely to support disks and outflows near the major and minor axes respectively. We apply this same cut here, and show the results in Figure \ref{fig:PA_inc_groups}.

All of our model outflows lie in the region that would be identified as such using these geometric cuts, whilst our model disks spread over both regions. These results reinforce our conclusions from Paper 2, showing that these geometric cuts are useful but are unlikely to produce pure disk or outflow samples.

\subsection{Correlations between group size and absorption}

\begin{figure}
\includegraphics[width=\columnwidth]{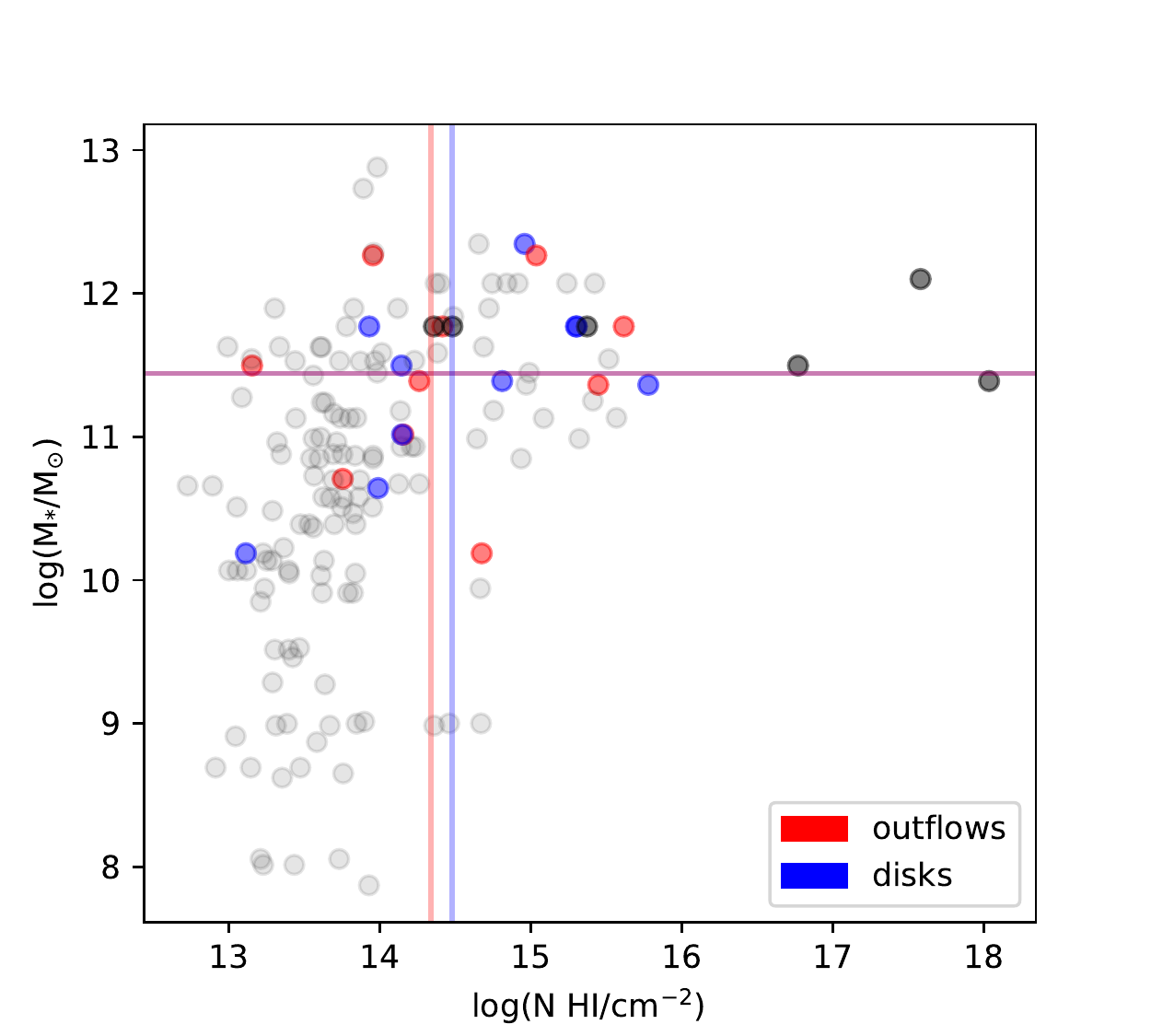}
\caption{Column density against total galaxy stellar mass found within 500 \kms and 1 Mpc of \ion{H}{i} absorbers. Faded points show our full sample of detected \ion{H}{i} absorption, whilst bold points show absorbers within 500 \kms of the galaxy groups considered in this work. The bold points are also coloured by the best-fitting model type, with bold grey points showing absorption that could not reasonably be fit by our models. Faded lines show the median for each model type in both measurements. \label{fig:group_masses}}
\end{figure}

Whether the gas around galaxy groups results primarily from a superposition of the CGM of the individual galaxies, or from stripped material, increasing the number of galaxies in a group is expected to increase the column densities of gas seen around the group. It is therefore unsurprising that we find a correlation between the number of galaxies in a group, and the number of \ion{H}{i} absorption features identified within a 500 \kms{ }window. We also find a strong correlation between the column density of an absorber, and the number or total mass of nearby galaxies (using any combination of a 500 or 300 \kms{ }window in redshift and 300, 500 or 1000 kpc in impact parameter), with Spearman co-efficients suggesting > 95\% confidence in all of these cases. 

These correlations persist if the sample is split to only absorbers in the COS spectra (z $\lesssim$ 0.45) and those in the FOS spectra. This relationship between group size and absorption is therefore present across our 0 $<$ z $<$ 1 redshift range and is robust to the changing galaxy and absorber detection limits across this range.

The total stellar mass of nearby galaxies to each \ion{H}{i} absorber (using a 500 \kms velocity window and 1000 kpc impact parameter cut) is shown in Figure \ref{fig:group_masses}. There is no significant difference between the number or total mass of nearby galaxies surrounding absorbers identified as disks and those identified as outflows using our toy models (using any of the above cuts). As in Paper 2, we find no significant difference between the column densities of disk and outflow absorption. We do find a small difference in Doppler widths (outflows with median $\approx$50 \kms, disks $\approx$40 \kms), but large uncertainties due to unresolved absorption mean that this is also not significant. 

The absorbers within these groups that could not be fit using either model have higher \ion{H}{i} column densities and exist in more massive groups, although the significance is again limited by a small sample size. If these consist of material that has been ejected from galaxies and forms an intra-group medium, this would likely be more prevalent in denser, more massive groups with more interactions between galaxies, as is found here.

\subsection{Galaxy interactions} \label{sec:discuss_interactions}

Tidal interactions between galaxies in groups are expected to strip gas out of the galaxies, which should be detectable \citep[e.g.][]{morris1994}. Streams of stars and gas attributed to tidal effects are seen around nearby galaxies \citep[e.g.][]{majewski1996, ibata2001, foster2014, sorgho2019}, and there are suggestions of observed tidal material around more distant galaxies using emission lines \citep[e.g.][]{epinat2018, johnson2018}. However, identifying this material in absorption is challenging. By considering absorption that could not be reproduced using our toy models, and by looking at close pairs of galaxies most likely to be interacting, we attempt to identify tidal material in our observations.

\begin{figure*}
\includegraphics[width=\textwidth]{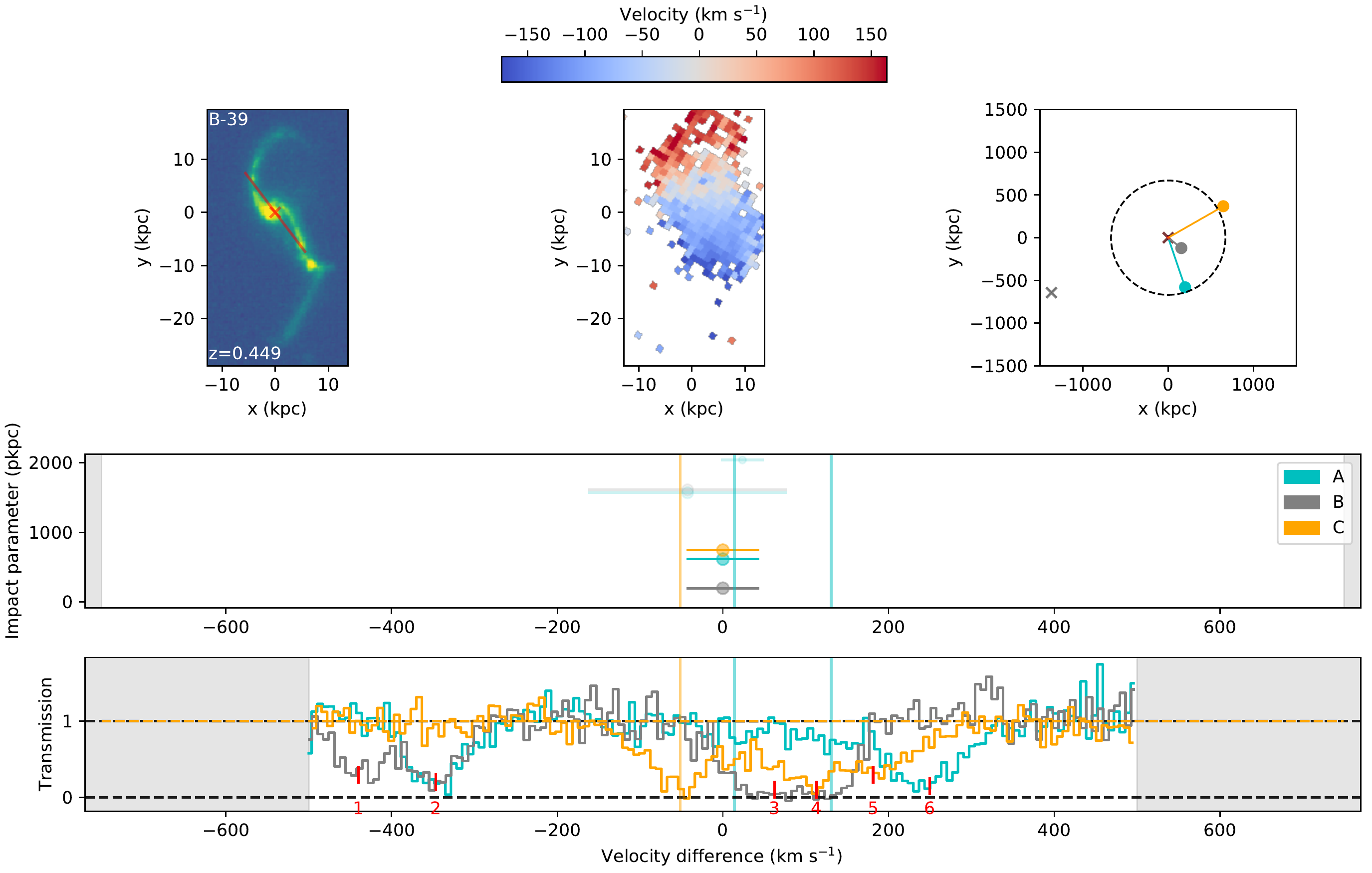}
\caption{Details of the absorption and galaxy environment around the galaxy merger B-39 at z $\sim 0.45$. The layout is identical to that shown in Figure \ref{fig:26677_detailed}, but no model is shown as the position angle and inclination of these galaxies could not be measured. The galaxy kinematics are measured from the [\ion{O}{iii}] emission line seen in the MUSE data. Tick marks (1) and (2) show the Lyman n=6 transition from z$\approx$0.88; whilst (3)-(6) show Ly$\beta$ from z$\approx$0.72. \label{fig:b39_detailed}}
\end{figure*}

Although not included in the main sample due to the difficulty of fitting position angle and inclination measurements, the Q0107 field features a clear ongoing galaxy merger, denoted B-39 and illustrated in Figure \ref{fig:b39_detailed}. We do not attempt to fit our disk/outflow models to this merger. Additionally, the absorption at this redshift is blended with Ly$\beta$ from z$\approx$0.72, making it difficult to identify associated Ly$\alpha$ components.

The VPFIT results suggest two weak components in sightline A, one close to the merger systemic redshift and another $\approx$120 \kms{ }redward, and one stronger component in sightline C, with a small blueshift relative to the systemic redshift. Based on the directions and velocities of the stellar arms/streams seen in the Figure, absorption from an extension of these streams would have the opposite velocity offset to that observed in both of these lines-of-sight.

However, given the lack of other nearby galaxies, it seems likely that at least some of the absorption observed can be attributed to the interaction between these galaxies, possibly on a previous pass between these galaxies such that their orbits have changed the line-of-sight velocity relative to the large-scale absorption. Simulations suggest stripped gas may be found opposing the direction of motion, depending on the orbital configuration \citep{rodriguez2022}, so tidal material cannot be ruled out.

Several absorbers around the galaxy groups we consider in this work can not be reasonably approximated by our disk/outflow/halo toy models. If a large proportion of these absorbers are due to tidal material, these would likely be found more frequently near to close galaxy pairs. However, we do not find any significant difference between the closest pair separations of groups with unexplained absorption, and those for which all absorbers can be fit by our models. 

Despite the lack of evidence for absorption due to interactions between galaxies across the sample as a whole, we can still look individually at these unexplained absorbers and consider whether tidally-stripped material can provide a reasonable explanation.

The unexplained absorption in G-202 has a velocity offset close to our cut-off of 500 \kms, and is therefore likely associated with other galaxies (its high column density making a physical association with at least one galaxy likely, see e.g. T14 and Paper 1). However, B-22 does appear distorted, with a tail of extended emission visible in the direction of LOS-B. Tidal material is therefore a possible explanation for some of the absorption in this sightline, but this absorption is reasonably approximated by our toy models. 

The sub-DLA near G-558 could not be fit using our models, and likely has an undetected host galaxy much closer to the line-of-sight than the detected group of galaxies. 

Strong, metal-enriched absorption in G-399 lies at a velocity between the two large galaxies in this group. As discussed in \citet{anshul2021}, this contains a warm component as well as a cooler gas phase. This is consistent with intra-group material, lying within $r_{vir}$ of the largest group galaxy, and could be tidal in origin.

The galaxies A-36 and A-37 are close together (< 10 kpc, with A-40 a more distant companion), and appear distorted in the HST image, suggesting that they are interacting. The unexplained absorption components all lie blueward of all three galaxies, but have a much larger velocity offset than the velocity difference between the galaxies. This would suggest that this absorption does not identify material ejected by the interactions between these two galaxies. This absorption may be associated with the more massive, but distant, galaxies with a smaller velocity offset.

Therefore, in the four galaxy groups with unexplained absorption, stripped material appears a reasonable hypothesis for two, only one of which contains a nearby pair of galaxies that is likely interacting. 

Stronger evidence for tidal material could come from absorption in the direction of visible stellar streams, or material that lies between the interacting galaxies in velocity and in projection on the sky. This is not the case for the groups in this study, so whilst tidally-stripped gas may be the most likely explanation for a small fraction of absorbers, we do not have good evidence of its detection in this work.

\subsection{Metal absorption}\label{sec:discuss_metals}

Around the twelve isolated galaxies in Paper 2, only two of the \ion{H}{i} absorbers also featured metal absorption around the same galaxy. However, many of the absorbers detected here in galaxy groups do feature metal lines, most commonly \ion{C}{iii} and \ion{O}{vi}, despite similar coverage and signal-to-noise ratio. (No other ions are detected more than three times near our group redshifts, so we primarily discuss \ion{C}{iii} and \ion{O}{vi} here.) This is consistent with our results from Paper 1, in which \ion{O}{vi} detections only appear at impact parameters < 350 kpc from non-group galaxies\footnote{Note that this uses the friends-of-friends algorithm to discriminate between group and non-group galaxies as described in Paper 1. This differs from the definitions of `isolated' and `group' galaxies used in Paper 2 and this work in that the linking lengths used in Paper 1 vary with redshift in order to correct for the changing detection limit with redshift, and exclude MUSE galaxies in order to achieve consistent detection limits across the field. The average linking lengths are comparable to the 500 \kms{ } and 500 kpc used here.}, but are often more extended in group environments \citep[e.g.][]{johnson2015, tchernyshyov2022}. 

\begin{figure*}
\includegraphics[width=\textwidth]{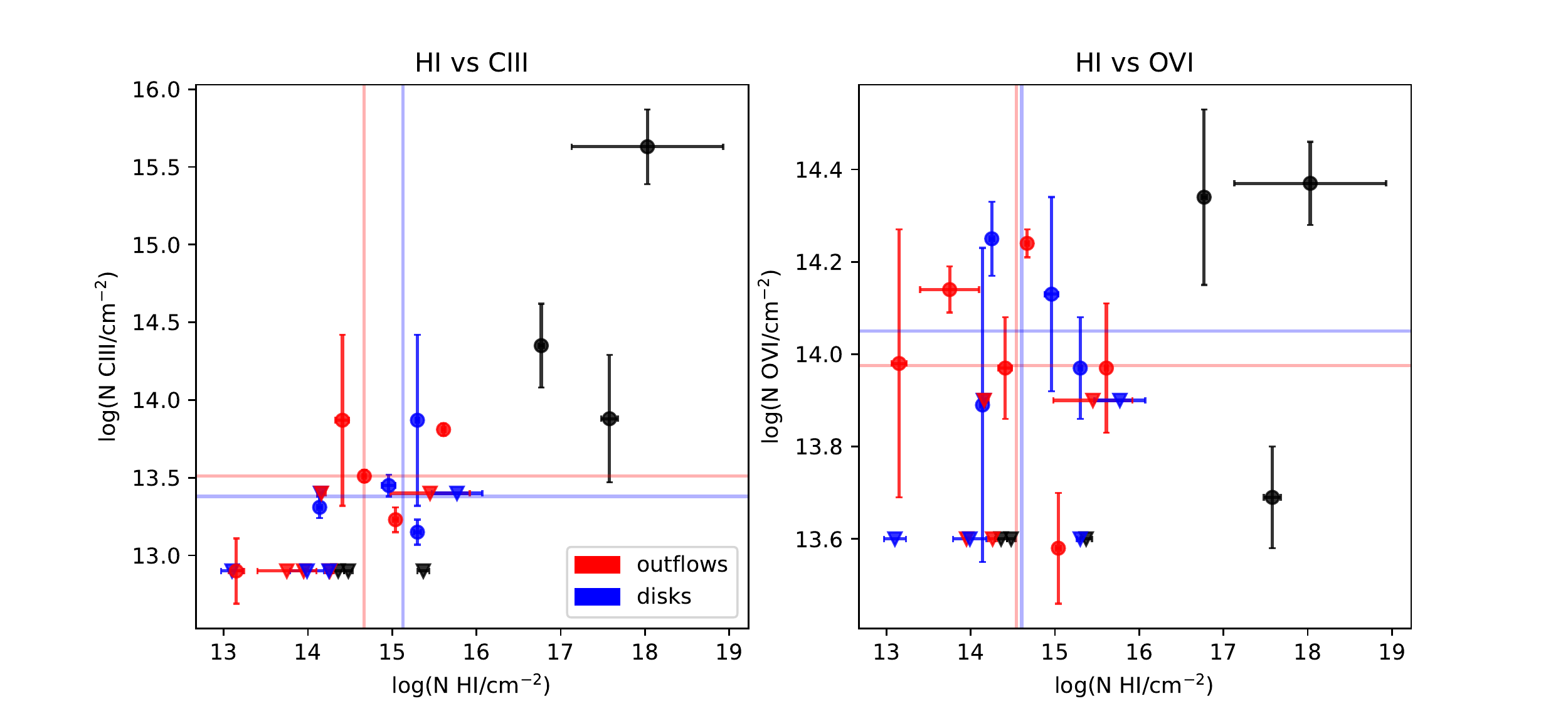}
\caption{Column densities of \ion{H}{i}, \ion{C}{iii} and \ion{O}{vi} absorption found within 500 \kms of our group galaxies. Only metals within 30 \kms of detected \ion{H}{i} are included. Upper limits where metals are not detected are shown by downward-pointing triangles. Points are coloured by best-fitting model type (disk, outflow or none). The median column densities of absorbers with detected metals are shown along each axis. \label{fig:metal_content}}
\end{figure*}

Figure \ref{fig:metal_content} shows the column densities of \ion{C}{iii} and \ion{O}{vi} associated with \ion{H}{i} absorbers around our galaxy groups, coloured by the type of model that provides the best fit to that absorber. In both cases there is no significant difference in either the \ion{H}{i} or metal column densities between the two models. This could suggest that, if our simplified models are indeed indicative of outflow cones and accreting disks around these galaxies, the accreting material does not have significantly different metal content than the outflowing material. However, it is possible that both the metal content and ionization state differ substantially between disks and outflows, such that the \ion{C}{iii} and \ion{O}{vi} column densities appear similar. The similar \ion{H}{i} Doppler parameters of the disk and outflow absorbers suggest similar gas temperatures, but the metal ions likely trace different phases. We would therefore need a more detailed analysis of more metal species in order to confidently determine the metal content of these structures.

With its lower ionization stage, it is unsurprising that \ion{C}{iii} correlates more strongly with \ion{H}{i} than \ion{O}{vi}. We find with > 99\% confidence (using a Spearman rank with non-detections included or excluded) that the column densities of \ion{H}{i} and \ion{C}{iii} are correlated, although we note that this correlation is not significant if absorbers that could not be fit by our toy models are excluded. There is not a significant correlation with \ion{O}{vi} column density. The CGM/IGM is known to be multi-phase \citep[e.g.][]{tripp2008, werk2014, concas2019, anshul2021, nateghi2021}, and hence \ion{O}{vi} likely probes a warmer phase of material than \ion{H}{i} and \ion{C}{iii}, with warm outflows and halos often proposed \citep[e.g][]{werk2016, oppenheimer2016, ng2019}.

In isolated systems, accreting material would be expected to have a lower metallicity than that in outflows, due to enrichment of outflows through stellar evolution within the galaxy \citep[e.g.][]{peroux2020}. However, observations have often failed to detect a relationship between azimuthal angle and metallicity \citep[e.g.][]{peroux2016, kacprzak2019}, possibly in part due to the mixing by azimuthal angle discussed in Paper 2 (and briefly in Section \ref{sec:angles} of this work). Our sample from Paper 2 has too few metal detections to test this effect, but the numbers of \ion{O}{vi} absorbers found along the major and minor axes using our full galaxy sample (Section 4.2 of Paper 1) do suggest higher metal content for minor-axis absorbers. It seems possible that, due to interactions between galaxies, metal enriched gas is more common around group galaxies, and therefore accreting material in group environments has metal content more similar to outflowing material than gas accreting onto isolated galaxies. However, other studies have found suggestions that low-metallicity gas can be masked by the presence of higher-metallicity gas in the line-of-sight \citep[e.g.][]{pointon2019}, which would allow our \ion{H}{i} to be dominated by low-metallicity accretion despite strong metal detections with a different origin (e.g. warm halo, stripped material, intra-group gas).

We also note several high-column-density, metal-enriched absorbers that are not identified as disks or outflows. However, one of these is the sub-DLA in LOS-C, which likely has an undetected host galaxy and may not be strongly affected by the larger-scale environment. Another is close to 500 \kms{ }offset from the galaxy group considered, and is therefore more likely to be associated with other galaxies at smaller velocity offsets. The third lies within the velocity ranged spanned by the nearby group, and could plausibly form part of an intra-group medium.

\section{Summary \& Conclusions}\label{sec:conclusions}

In this study we use multiple lines-of-sight to constrain the origins of \ion{H}{i} absorption seen around 9 pairs and groups of galaxies at z$<$1. By fitting simple disk, halo and outflow models around these galaxies we attempt to determine to what extent the absorption features seen in galaxy groups could be attributed to a superposition of the CGM around the individual galaxies, an extended intra-group medium, or material stripped from the CGM of the galaxies by interactions within the group. We model the absorption around nine galaxy groups in the Q0107 field, probed by three lines-of-sight, and find that:

\begin{enumerate}
    \item Our disk and outflow models reproduce a slightly larger fraction of the identified \ion{H}{i} absorbers where multiple galaxies are present ($\approx$75\%) than around the isolated galaxies we consider in Paper 2 ($\approx$60\%). This supports a model for which a superposition of absorption from the group galaxies results in the observed spectra. Sample variance and a larger parameter space (due to multiple nearby galaxies) appear to be plausible contributors to this higher success rate. 
    
    \item Our `group' sample includes systems with larger separations and lower masses than many previous works considering absorption around group and isolated galaxies. The larger number of galaxies also leads to a smaller impact parameter between absorption and the closest galaxy. Both effects may also contribute to our high success rate and resulting preference towards the superposition model. 
    
    \item The best-fitting parameters for our model outflows are generally similar to those seen for isolated galaxies in our previous work (Paper 2, submitted), including several that appear to extend beyond the virial radius. However, a larger number of putative outflows have smaller opening angles; these galaxies preferentially have lower masses and redshifts than those with apparent wider outflows. This means our overall sample has parameters more similar to \citet{schroetter2019}, who use similar models for \ion{Mg}{ii} absorption.
    
    \item Our model disks have smaller rotation velocities than those around isolated galaxies (120 to 270 \kms, $\approx1.5\sigma$), despite similar impact parameters. This may be due to interactions between infalling material and the extended group halo, but could also be due to the galaxy/sightline configurations sampled in this field. In order to constrain the rotation velocities, we have also had to assume an infall velocity in some cases, which may be distorting the results.
    
    \item  We do not detect any difference in the column densities of absorbers identified as likely disks and outflows, and their Doppler widths do not differ significantly. This matches our results from isolated galaxies in Paper 2. There does not appear to be a difference in the number or total stellar mass of galaxies in groups hosting these possible disks and outflows, although high-column-density absorbers do tend to be seen in denser galaxy environments (both in galaxy number and mass).
    
    \item Absorption that cannot be fit by our models does not occur more frequently in groups with a close pair of galaxies, and therefore does not provide evidence of material stripped by galaxy interactions. Tidally-stripped material is a plausible explanation for two of the four groups hosting these unexplained absorbers, but is not supported by additional evidence such as the absorption matching the velocity offsets between galaxy pairs or lying in the directions of possible tidal stellar features seen in the HST images.  
    
    \item There is no significant difference between the \ion{C}{iii} or \ion{O}{vi} contents of absorbers identified as disks and outflows. Whilst outflowing material is generally expected to be warmer and more metal-enriched, material accreting onto a group galaxy may have been recently stripped or ejected from another nearby galaxy, and thus be more enriched than accretion from the IGM onto an isolated galaxy. We do not find a significant correlation between \ion{H}{i} and \ion{O}{vi} column densities, which fits expectations of these ions tracing different gas phases.

\end{enumerate}

This paper, alongside our previous works using the galaxy and QSO data covering the Q0107 field, demonstrates some advantages of using multiple lines-of-sight probing the gas around galaxies. In addition to being a highly-efficient way to build large samples of galaxy-absorber pairs, which we utilized to detect both a bimodality in the azimuthal angle of these pairs and a preference for aligned angular momentum between gas and galaxies in Paper 1, these sightlines also provide some constraints on the possible CGM structures seen in absorption, and therefore the origins of absorbing material. 

Through kinematic and column density considerations, we are able to rule out simple disk and outflow models for some absorption seen in the three sightlines, and tentatively identify some absorption as likely originating from disk or outflow structures. However, the large number of model parameters available when considering possible models in galaxy groups makes this identification yet more uncertain.

Searching for significant correlations between model parameters and galaxy/group properties will require a larger sample of such observations, probing a wider space of sightline configurations and group properties. Constraining more complex models encompassing a wider range of physical processes (entrainment of gas, changing ionization states, etc.) and therefore with a larger range of free parameters, will require a larger number of sightlines for each galaxy or group.

Whilst relatively few configurations of sightlines can be used to probe the CGM/IGM in this way using current instruments, and such sensitive Ly$\alpha$ measurements at low redshift are only possible using the HST/COS instrument, future instruments will enable these techniques to be used across a much larger number of systems, especially at higher redshifts. CGM tomography at higher redshifts is included in the science cases for several instruments on the ELT \citep[e.g.][]{maiolino2013, evans2015, marconi2021}, which will be able to use much fainter objects as background sources at these higher redshifts, and BlueMUSE \citep{richard2019} will provide efficient observations of galaxies surrounding such sightlines whilst extending Ly$\alpha$ coverage to lower redshifts than most other ground-based observations.

Our work studying the Q0107 system therefore serves as a useful test case for techniques that can be applied to much larger samples (especially at higher redshifts) in the near future.

\section*{Acknowledgements}

Firstly, we thank the anonymous referee for their insightful comments that have improved the quality of this paper.

This work is based on observations made with the NASA/ESA Hubble Space Telescope, obtained from the data archive at the Space Telescope Science Institute. STScI is operated by the Association of Universities for Research in Astronomy, Inc. under NASA contract NAS 5-26555. We also make use of observations collected at the European Southern Observatory under ESO programmes 086.A-0970, 087.A-0857 and 094.A-0131; at the W.M. Keck Observatory under programme A290D; and at the Gemini Observatory under programme GS-2008B-Q-50.

We thank Matteo Fossati for providing the MARZ templates used for estimating redshifts and their uncertainties. We also thank Jill Bechtold for leading the effort to obtain the Keck/DEIMOS data.

We thank the contributors to  SCIPY\footnote{\url{http://www.scipy.org/}}, MATPLOTLIB\footnote{\url{https://matplotlib.org/}}, ASTROPY\footnote{\url{https://www.astropy.org/}}, and the PYTHON programming language, the free and open-source community and the NASA Astrophysics Data system\footnote{\url{https://ui.adsabs.harvard.edu/}} for software and services.

This work also made use of the DiRAC system at Durham University, operated by the Institute for Computational Cosmology on behalf of the STFC DiRAC HPC Facility\footnote{\url{http://www.dirac.ac.uk/}}.

AB acknowledges the support of a UK Science and Technology Facilities Council (STFC) PhD studentship through grant ST/S505365/1. SLM acknowledges the support of STFC grant ST/T000244/1.


SC gratefully acknowledges support from Swiss National Science Foundation grants PP00P2\_163824 and PP00P2\_190092, and from the European Research Council (ERC) under the European Union's Horizon 2020 research and innovation programme grant agreement No 864361. MF also acknowledges funding from the ERC under the Horizon 2020 programme (grant agreement No 757535). This work has been supported by Fondazione Cariplo, grant No 2018-2329.

\section*{Data Availability}

The raw data from the Hubble Space Telescope may be accessed from the MAST archive\footnote{\url{http://archive.stsci.edu/}}, and that from the WM Keck Observatory from the Keck Observatory Archive\footnote{\url{https://koa.ipac.caltech.edu/cgi-bin/KOA/nph-KOAlogin}}. That from the European Southern Observatory may be accessed from the ESO Archive\footnote{\url{http://archive.eso.org/eso/eso_archive_main.html}}, and that from the Gemini Observatory may be accessed from the Gemini Science Archive\footnote{\url{http://www.cadc-ccda.hia-iha.nrc-cnrc.gc.ca/en/gsa/}}. The relevant program IDs are given in Section \ref{sec:data} of this paper. The derived data generated in this research will be shared on reasonable request to the corresponding author.

\addcontentsline{toc}{section}{Acknowledgements}




\bibliographystyle{mnras}
\bibliography{zotero_230220}



\appendix
\section{Description of additional groups}\label{sec:more_groups}

\subsection{G-202}

G-202 is the largest group of galaxies in this field (both in number of galaxies and total stellar mass), at z $\approx$ 0.2. B-22, the only galaxy in the group that lies in the MUSE field, is a small galaxy with a distorted shape, possibly due to tidal interactions within this group. Our estimate of the galaxy's position angle may be slightly affected, but we have taken the orientation of the bright central component (having a larger uncertainty but less likely to be distorted). An additional 5 galaxies at this redshift have estimated position angles, and an further 19 lie within 1 Mpc of at least one of the lines-of-sight.

\begin{figure*}
\includegraphics[width=0.902\textwidth]{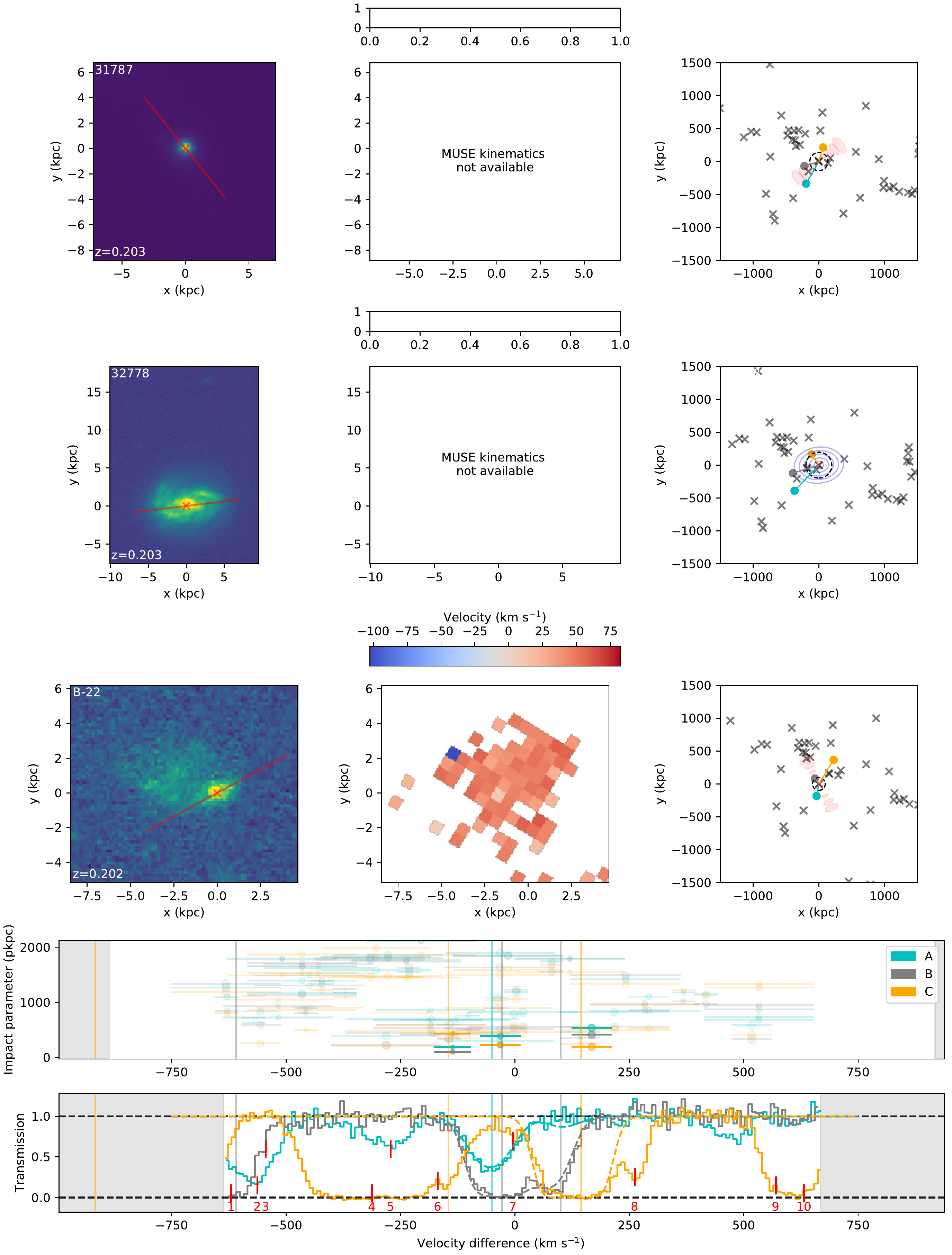}
\caption{Details of the absorption and galaxy environment around group G-202. Features shown are identical to those in Figure \ref{fig:26677_detailed}, with kinematics measured from the H$\alpha$ emission line seen in the MUSE data. Three components make up the model shown in the lower panel: an outflow with half-opening angle $\approx 15^{\circ}$ and velocity $\approx$ 250 \kms from galaxy B-22; an outflow with half-opening angle $\approx 20^{\circ}$ and velocity $\approx$ 50 \kms from galaxy 31787; and a disk with rotation velocity $\approx$ 150 \kms and infall of $\lesssim$ 20 \kms around galaxy 32778. Additional absorbers other than Ly$\alpha$ are marked with red ticks and identified as follows: (1) and (2) as the Lyman n=5-0 transition from z=0.536 in LOS-B and LOS-A respectively; (3) as Ly$\delta$ from z=0.500 in LOS-B; (4) and (5) as the Lyman n=6 transition from z=0.558 in LOS-C (where it is associated with the sub-DLA at z=0.558) and LOS-A; (6) as Ly$\beta$ from z=0.425; and (7)-(10) as molecular lines from the sub-DLA.  \label{fig:b22_detailed}}
\end{figure*}

Absorption is visible in all three lines-of-sight, although several of the absorption components visible in Figure \ref{fig:b22_detailed} are due to features other than Ly$\alpha$, including Lyman lines and molecular hydrogen lines from the sub-DLA at z=0.558. The five absorption components we have attempted to fit are listed in Table \ref{table:groups_b22}, alongside another strong absorber in LOS-B which is close to 500 \kms{ }blueward of the galaxy. This large velocity offset makes it unlikely to be associated with any of the galaxies in the HST field, and more likely associated with one of the galaxies with slightly lower redshift.

Despite the relatively dense galaxy environment, a combination of disks and outflows provides a good approximation to four of the five components, and there may be a contribution to the fifth from a molecular line. The model shown in the Figure is an outflow with half-opening angle $\approx 15^{\circ}$ and velocity $\approx$250 \kms{ }from galaxy B-22 (matching the bluer component in B), an outflow with half-opening angle $\approx 20^{\circ}$ and velocity $\approx$50 \kms{ }from galaxy 31787 (matching the absorption in A), and a disk with rotation of 150 \kms{ }and negligible infall around galaxy 32778 (matching the strong absorption in C and redder component of absorption in B).


Stripped material is expected in groups of interacting galaxies \citep[e.g][]{morris1994}, and the distortion in B-22 may be due to this. However, a `tail' of gas extending out along the path of this apparent stream would not intersect any of the three lines-of-sight. All absorbers identified have widths $\lesssim$ 50 \kms, so we also do not clearly identify any hot intra-group gas. 

Two components of absorption in LOS-B are also seen in \ion{C}{iii} and \ion{O}{vi}, with redshifts matching the two Ly$\alpha$ components. No metals are detected in the A or C sightlines, imposing 1$\sigma$ upper limits of $10^{12.9}$ and $10^{13.6}$ $\textrm{cm}^{-2}$ on \ion{C}{iii} and \ion{O}{vi} respectively in sightline A, whilst these lines could not be detected in LOS-C due to the sub-DLA. With the metals in C unconstrained, the absorbers in B and C could have similar \ion{C}{iii}/Ly$\alpha$ and \ion{O}{vi}/Ly$\alpha$ ratios, consistent with both probing rotating/accreting material associated with the same galaxy. The \ion{C}{iii}/Ly$\alpha$ and \ion{O}{vi}/Ly$\alpha$ ratios of the bluer absorption component in B (identified as an outflow in our best model result) are lower than that of the redder component (identified as accretion), but these ions likely trace different ionization phases, so their ratios may not reflect a difference in the metal content of the gas. 

If the accreting material does indeed have a higher metal content, it could have been stripped or ejected from other group galaxies, rather than accreted from the IGM, but we cannot distinguish this from a scenario in which the carbon and oxygen in the accreting material are ionized such that \ion{C}{iii} and \ion{O}{vi} represent a larger fraction.

We also note that B-22, with a proposed strong outflow, has a lower SFR (and sSFR) than most galaxies in this group. Therefore this putative outflow is either sustained by a very low star-formation rate, or the galaxy has quenched in the past $\approx$ 300 Myr. The disk model also requires angular momentum aligned on scales of $\approx$ 400 kpc, larger than seen in most models. Despite these difficulties, we cannot rule out a superposition of disk/outflow models for this group.

\begin{table*}
\begin{center}
\caption{Summary of galaxy and absorber properties for G-202. Columns are labelled as in Table \ref{table:groups_26677}, although details of galaxies outside the HST field are shown more briefly.}
\label{table:groups_b22}
\begin{tabular}{| c| c| c| c| c| c| c| c| c| c| c|}
\hline 
z & Galaxy & Lum ($L_{\star}$) & Inc & LOS & Imp (kpc) & Azimuth & log(N \ion{H}{i}) & b (\kms) & $\Delta$v (\kms) & Other ions \\ 
(1) & (2) & (3) & (4) & (5) & (6) & (7) & (8) & (9) & (10) & (11) \\ \hline 

0.202 & B-22 & 0.02 & $66^{\circ}$ $\pm$ $5^{\circ}$ & A & 186 & $51^{\circ}$ $\pm$ $9^{\circ}$ & 13.95 $\pm$ 0.02 & 52 $\pm$ 3 & 90 $\pm$ 40 & None \\
 & & & & B & 103 & $82^{\circ}$ $\pm$ $9^{\circ}$ & 17.58 $\pm$ 0.10 & 16 $\pm$ 1 & -470 $\pm$ 40 & \ion{C}{ii-iii}, \ion{O}{vi}, \ion{Si}{ii-iii} \\
  & & & & B &  &  & 15.03 $\pm$ 0.03 & 43 $\pm$ 2 & +110 $\pm$ 40 & \ion{C}{iii}, \ion{O}{vi} \\
   & & & & B &  &  & 14.96 $\pm$ 0.08 & 18 $\pm$ 1 & +240 $\pm$ 40 & \ion{C}{iii}, \ion{O}{vi} \\
  & & & & C & 431 & $30^{\circ}$ $\pm$ $9^{\circ}$ & 13.96 $\pm$ 0.05 & 43 $\pm$ 6 & -10 $\pm$ 40 & None \\
  & & & & C &  &  & 14.65 $\pm$ 0.05 & 64 $\pm$ 3 & +280 $\pm$ 40 & None \\
   \\
 & 25962 & 0.07 & $81^{\circ}$ $\pm$ $1^{\circ}$ & A & 295 & $65^{\circ}$ $\pm$ $1^{\circ}$ & & & (+100)& \\
 & & & & B & 516 & $42^{\circ}$ $\pm$ $1^{\circ}$ & & & & \\
 & & & & C & 895 & $53^{\circ}$ $\pm$ $1^{\circ}$ & & & & \\
   \\
 & 31704 & 0.03 & $41^{\circ}$ $\pm$ $4^{\circ}$ & A & 457 & $16^{\circ}$ $\pm$ $8^{\circ}$ & & & (+280)& \\
 & & & & B & 360 & $20^{\circ}$ $\pm$ $8^{\circ}$ & & & & \\
 & & & & C & 246 & $79^{\circ}$ $\pm$ $8^{\circ}$ & & & & \\
 \\
 & 31787 & 0.08 & $62^{\circ}$ $\pm$ $2^{\circ}$ & A & 387 & $69^{\circ}$ $\pm$ $2^{\circ}$ & & & (+100)& \\
 & & & & B & 230 & $69^{\circ}$ $\pm$ $2^{\circ}$ & & & & \\
 & & & & C & 226 & $56^{\circ}$ $\pm$ $2^{\circ}$ & & & & \\
  \\
 & 32497 & 1.1 & $85^{\circ}$ $\pm$ $2^{\circ}$ & A & 394 & $62^{\circ}$ $\pm$ $1^{\circ}$ & & & (-150)& \\
 & & & & B & 226 & $77^{\circ}$ $\pm$ $1^{\circ}$ & & & & \\
 & & & & C & 215 & $54^{\circ}$ $\pm$ $1^{\circ}$ & & & & \\
 \\ 
 & 32778 & 0.4 & $43^{\circ}$ $\pm$ $1^{\circ}$ & A & 535 & $40^{\circ}$ $\pm$ $2^{\circ}$ & & & (+300)& \\
 & & & & B & 412 & $10^{\circ}$ $\pm$ $2^{\circ}$ & & & & \\
 & & & & C & 195 & $63^{\circ}$ $\pm$ $2^{\circ}$ & & & & \\
 \hline
 & & & & \multicolumn{3}{c}{Impact params (kpc)} & & & & \\
 & & & & A & B & C & & & & \\ \hline
 & (24419) & 0.01 & & 628 & 719 & 1118 & & & (-430) & \\
 & (27581) & 0.03 & & 724 & 930 & 1048 & & & (-380) & \\
 & (28913) & 0.02 &  & 679 & 536 & 814 & & & (-50) & \\
 & (30488) & 0.1 &  & 848 & 973 & 949 & & & (+420) & \\
 & (31439) & 1.0 &  & 1082 & 889 & 1036  & & & (+100) & \\
 & (31978) & 0.1 &  & 697 & 439 & 478 & & & (-50) & \\
 & (32021) & 0.1 &  & 678 & 415 & 438 & & & (0) & \\
 & (32308) & 0.8 &  & 597 & 331 & 355 & & & (-10) & \\

 & (32591) & 0.03 &  & 856 & 598 & 586 & & & (-50) & \\
  & (32755) & 0.7 &  & 784 & 531 & 572 & & & (+130) & \\
 & (33166) & 0.03 &  & 759 & 493 & 344  & & & (+30) & \\
 & (33268) & 0.9 &  & 820 & 556 & 505 & & & (+350) & \\
 & (33496) & 0.1 &  & 1100 & 844 & 799 & & & (-490) & \\
 & (34292) & 0.03 &  & 895 & 809 & 500 & & & (-320) & \\
  & (34719) & 0.3 &  & 834 & 590 & 259 & & & (-420) & \\
 & (34992) & 2.9 &  & 818 & 551 & 456 & & & (+150) & \\
 & (35102) & 0.5 &  & 1163 & 1132 & 861 & & & (-330) & \\
  & (36229) & 0.1 &  & 1106 & 856 & 528 & & & (+430) & \\
 & (38978) & 0.1 &  & 1488 & 1305 & 903 & & & (-310) & \\
     \hline
\end{tabular}

\end{center}
\end{table*}

\subsection{G-383}

\begin{figure*}
\includegraphics[width=\textwidth]{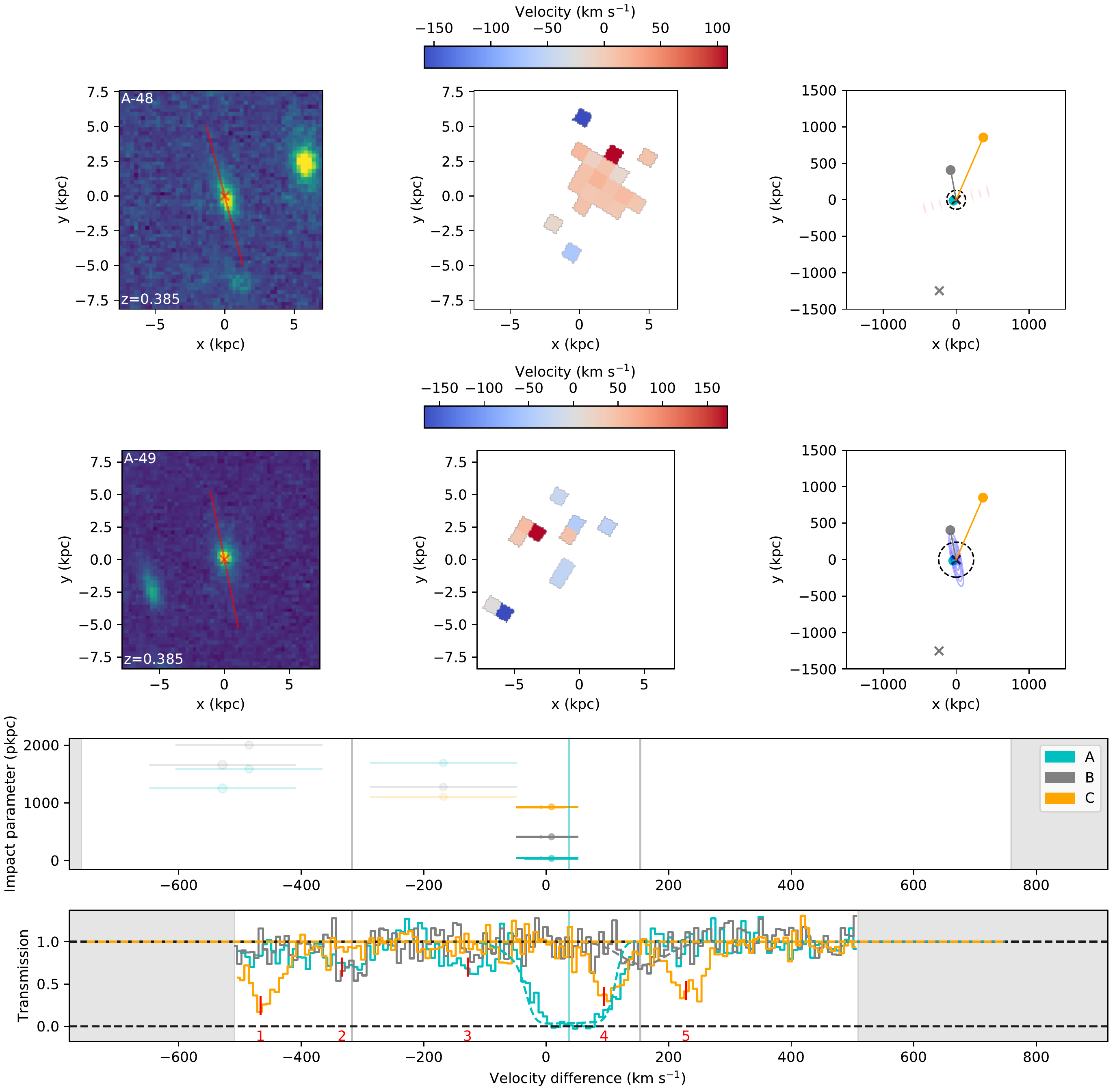}
\caption{Details of the absorption and galaxy environment around group G-383, a galaxy pair at z $\sim 0.39$, with kinematics taken from the H$\alpha$ emission line seen in the MUSE data. The layout is identical to that shown in Figure \ref{fig:26677_detailed}, and the model shown in the lower panel combines an outflow with 600 \kms velocity and $10^{\circ}$ half-opening angle, and a rotating disk around A-49 with rotation velocity 180 \kms. Additional absorbers indicated by red ticks are identified as follows: (1), (4) and (5) are H2 molecular lines from the sub-DLA at z $\approx$ 0.56; (2) is Ly$\delta$ from z=0.73; and (3) is Ly$\beta$ from z=0.64. Note that the close separation between the galaxies in both velocity and projection makes them difficult to distinguish in the central panel. \label{fig:a48_detailed}}
\end{figure*}

A-48 and A-49 are a close pair of small, star-forming galaxies near to the line-of-sight to QSO-A, at z $\approx$ 0.38. They are separated by $\approx$ 5 kpc and aligned such that they lie approximately along each other's minor axis. A-48 is larger and closer to the QSO. They are illustrated in Figure \ref{fig:a48_detailed} and detailed in Table \ref{table:groups_a48}. Other larger galaxies exist 1-1.5 Mpc from the sightlines, but outside the HST image so do not have measured position angles. Whilst emission-line velocities are measurable from the MUSE data for both galaxies, neither show a clear velocity gradient.

Strong absorption exists in LOS-A, near the minor axis of both galaxies and well within the virial radius, featuring \ion{C}{iii} and \ion{O}{vi} absorption. Both indicate warm, metal-enriched material that would be consistent with a galactic wind (although, as the Ly$\alpha$ column densities of the other absorbers are small, their \ion{C}{iii} and \ion{O}{vi} fractions are not constrained, and could be higher). As both galaxies are very close to edge on, the required outflow velocity in order to produce the 20 or 40 \kms{ }line-of-sight velocity difference is massively affected by both the uncertainty in the velocity offset itself and the galaxy inclination. Using the measured values of both requires a fast outflow of $\sim$ 500 \kms{ }to reproduce the observations, but any value of outflow velocity $\lesssim$ 1000 \kms{ }can approximate the observations within the 1$\sigma$ region of $i$ and $\Delta \textrm{v}$. 

Weak absorption in QSO-B is apparent at $\approx$ 320 \kms{ }blueward of the galaxy pair, as well as 150 \kms{ }redward. Possible absorbers in A at -320 and -150 are attributed to higher-order Lyman lines from higher redshift gas (Ly-$\beta$ from z $\approx$ 0.64, and Ly$\gamma$ from z $\approx$ 0.73). All apparent transitions in C are attributed to molecular lines from the sub-DLA at z $\approx$ 0.56. 

Neither of the absorbers in B can be reproduced using a reasonable outflow, as it would need a very large opening angle to get close to the major axis, and this would produce a very large velocity spread on the absorption in A. As the absorbers lie near the major axis, a disk-like structure could be considered. This could be invoked for either or both absorbers in B, with potentially the two galaxies rotating in different directions producing the -320 and +150 \kms{ }absorbers. Such a disk could not fit the absorption in A in addition to that in B, as a disk thick enough to produce strong absorption along the minor axis would also cover a wide range of line-of-sight velocities along LOS-B, and therefore wider absorption than is observed.

Given the uncertainties in inclination and velocity offset, an outflow fitting the absorption in A alongside a disk from either galaxy matching the redward component of absorption in B seems plausible. The large velocity offset required to match the blue component in B would require a larger disk velocity than the virial velocity of either galaxy. It is therefore more likely to be associated with the more distant galaxies, or not physically associated with any of the galaxies in this group.

\begin{table*}
\begin{center}
\caption{Summary of galaxy and absorber properties of group G-383. Any additional galaxies and metal absorbers have velocities shown relative to the first galaxy (A-48). Columns are identical to those in Table \ref{table:groups_26677}.}
\label{table:groups_a48}
\begin{tabular}{| c| c| c| c| c| c| c| c| c| c| c|}
\hline 
z & Galaxy & Lum ($L_{\star}$) & Inc & LOS & Imp (kpc) & Azimuth & log(N \ion{H}{i}) & b (\kms) & $\Delta$v (\kms) & Other ions \\
(1) & (2) & (3) & (4) & (5) & (6) & (7) & (8) & (9) & (10) & (11) \\\hline 

0.383 & A-48 & 0.03 & $87^{\circ}$ $\pm$ $3^{\circ}$ & A & 38 & $87^{\circ}$ $\pm$ $4^{\circ}$ & 14.67 $\pm$ 0.02 & 53 $\pm$ 2 & +40 $\pm$ 40 & \ion{C}{iii}, \ion{O}{vi}\\
 & & & & B & 414 & $4^{\circ}$ $\pm$ $4^{\circ}$ & 13.22 $\pm$ 0.09 & 20 $\pm$ 7 & -320 $\pm$ 40 & (None)  \\
 & & & & B & 414 & $4^{\circ}$ $\pm$ $4^{\circ}$ & 13.11 $\pm$ 0.13 & 24 $\pm$ 11 & +150 $\pm$ 40 & (None) \\
  & & & & C & 932 & $38^{\circ}$ $\pm$ $4^{\circ}$ & (None, limit $\sim13.0$) & & &  \\
 \\
 & A-49 & 0.02 & $80^{\circ}$ $\pm$ $5^{\circ}$ & A & 43 & $81^{\circ}$ $\pm$ $4^{\circ}$ & & & (-20) & \\
 & & & & B & 411 & $0^{\circ}$ $\pm$ $4^{\circ}$ & & & & \\
 & & & & C & 927 & $35^{\circ}$ $\pm$ $4^{\circ}$ & & & & \\
  \hline
\end{tabular}

\end{center}
\end{table*}

\subsection{G-517}

\begin{figure*}
\includegraphics[width=\textwidth]{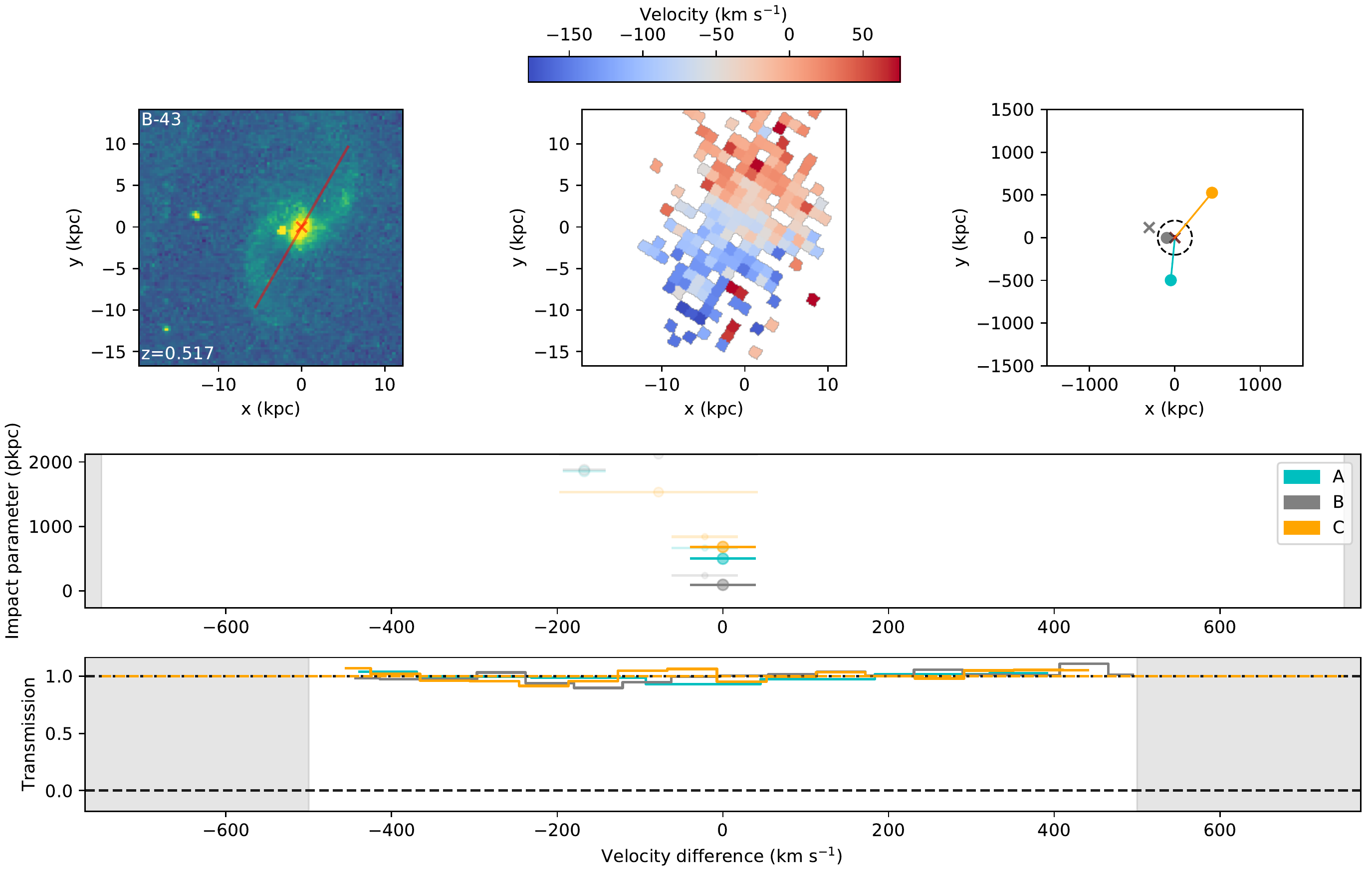}
\caption{Details of the absorption and galaxy environment of group G-517, with kinematics from [\ion{O}{iii}] as seen in the MUSE data. The layout is identical to that shown in Figure \ref{fig:26677_detailed}, except that no model is shown.  \label{fig:b43_detailed}}
\end{figure*}

Group G-517 is made up of only two galaxies. B-43 is a $\approx0.6 L_{*}$ spiral galaxy at z$\approx$0.52, with a far smaller ($\approx0.02 L_{*}$) companion galaxy B-34, illustrated in Table \ref{table:groups_b43} and Figure \ref{fig:b43_detailed}. A large group exists at the same redshift, but this is more than 1.5 Mpc from the lines-of-sight. B-34 is sufficiently faint that it is difficult to obtain a position angle. The uncertainty on the B-43 position angle may also be underestimated as the flux in the spiral arms dominates over the disk that we use to measure the position angle. 

There is no significant absorption in any of the lines-of-sight. This is somewhat surprising for LOS-B, as it lies near the minor axis at a small impact parameter to a fairly large star-forming galaxy. This either indicates that such outflows are not ubiquitous around SF galaxies, that these outflows are `patchy' (with covering factor $<$1), or that any outflow around this galaxy is heating the gas enough that no \ion{H}{i} absorption is visible. Coverage of several metal species including \ion{C}{iii} and \ion{O}{vi} is provided by our observations at this redshift, but no metals are detected.

\begin{table*}
\begin{center}
\caption{Summary of galaxy and absorber properties for group G-517. Any additional galaxies and metal absorbers have velocities shown relative to the first galaxy (B-43). Columns are identical to those in Table \ref{table:groups_26677}.}
\label{table:groups_b43}
\begin{tabular}{| c| c| c| c| c| c| c| c| c| c| c|}
\hline 
z & Galaxy & Lum ($L_{\star}$) & Inc & LOS & Imp (kpc) & Azimuth & log(N \ion{H}{i}) & b (\kms) & $\Delta$v (\kms) & Other ions \\
(1) & (2) & (3) & (4) & (5) & (6) & (7) & (8) & (9) & (10) & (11) \\\hline 

0.517 & B-43 & 0.6 & $66^{\circ}$ $\pm$ $1^{\circ}$ & A & 501 & $3^{\circ}$ $\pm$ $6^{\circ}$ & (None, limit $\approx$13.3) &  &  & \\
 & & & & B & 95 & $86^{\circ}$ $\pm$ $6^{\circ}$ & (None, limit $\approx$13.3) &  &  &  \\
  & & & & C & 683 & $37^{\circ}$ $\pm$ $6^{\circ}$ & (None, limit $\approx$13.4) & & &  \\
 \\
 & (B-34) & 0.02 & & A & 666 & & & & (-20)& \\
 & & & & B & 238 & & & & & \\
 & & & & C & 841 & & & & & \\
 
  \hline
\end{tabular}

\end{center}
\end{table*}

\subsection{G-536}

\begin{figure*}
\includegraphics[width=0.933\textwidth]{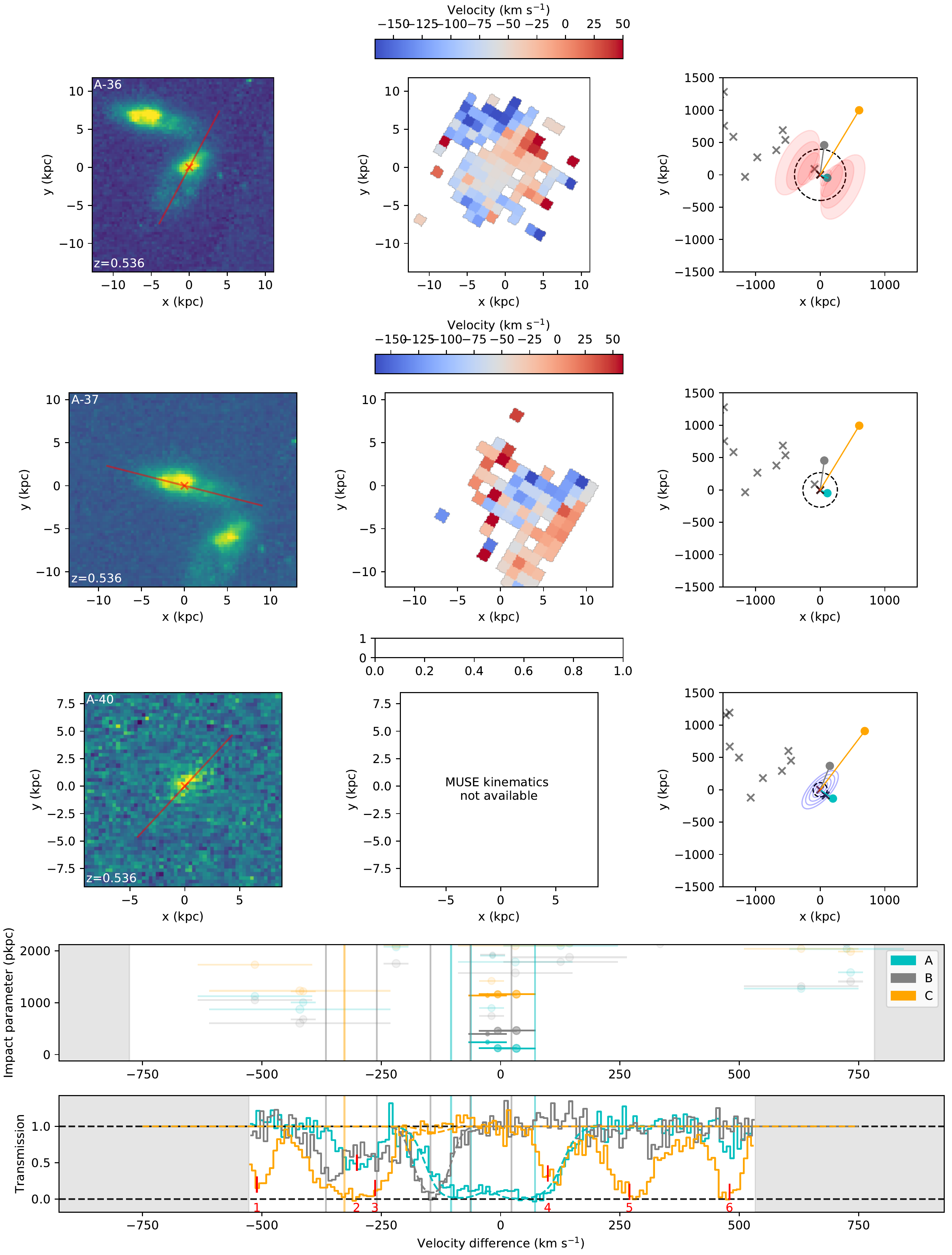}
\caption{Details of the absorption and galaxy environment around G-536. The layout is identical to that shown in Figure \ref{fig:26677_detailed}, although the lower panel shows the relevant Ly$\beta$ absorption, and kinematics shown are from the [\ion{O}{ii}] emission line seen in the MUSE data. The model shown in the lower panel combines an outflow with velocity 140 \kms and half-opening angle $55^{\circ}$ around A-36 and a disk with rotation velocity 110 \kms around A-40. Absorption originating from another redshift is labelled with red ticks. Component (2) is Ly$\alpha$ from z $\approx$ 0.295, whilst all of the other labelled components are molecular lines originating from the sub-DLA in LOS-C. \label{fig:a36_detailed}}
\end{figure*}

A-36 and A-37 are a close pair of galaxies that are likely interacting (given the asymmetry in A-36), with A-40 a more distant third galaxy. Details are given in Table \ref{table:groups_a36} and Figure \ref{fig:a36_detailed}. A substantial number of galaxies exist at similar redshifts with larger impact parameters. All three MUSE galaxies are star-forming, and the interacting pair have measurable kinematics. Strong absorption is visible along all three lines of sight. In galaxy A-36, the bright bulge appears to be off-centre with respect to the disk, likely indicating interaction with A-37. However, GALFIT is able to fit both galaxies reasonably well using a simple disk that is not strongly distorted.

In Figure \ref{fig:a36_detailed}, we show the Ly$\beta$ transition in absorption at the redshift of this group. This is due to the Ly$\alpha$ lying in the lower-resolution FOS spectra, whilst Ly$\beta$ produces strong absorption in the COS spectra, so more structure in the Ly$\beta$ can be discerned.

Given several galaxies, there are multiple ways of generating the absorption features in the observed spectra. A wide outflow around A-36 can approximately reproduce the saturated Ly$\beta$ absorption in LOS-A, or a narrower outflow could explain the redmost component of this absorption. An outflow from galaxy A-37 with sufficient velocity to cover the galaxy-absorber offset seen to the absorbing components in LOS-B produces absorption that is too broad and therefore inconsistent with the observations. Depending on the orientation of the galaxy, an outflow around A-40 can also reasonably match either the blue or red end of the broad saturated absorption in LOS-A.

Disk models may also contribute. A disk around A-36 will intersect both the A and B sightlines, and the galaxy velocity gradient acts in the direction such that any co-rotating disk cannot match the absorption found blueward of the galaxy. Given the saturation in A, such a disk cannot reproduce any substantial fraction of this absorption without also being detected in LOS-B redward of the galaxy. The observations therefore rule out this scenario. A disk around A-37 can fit a part of the absorption in A, most likely the blue end based on the galaxy kinematics, whilst a disk around A-40 can reproduce the strong absorption in B and a portion of the absorption in A.

The model shown in Figure \ref{fig:a36_detailed} includes this disk around A-40, with a rotation velocity of $\approx$ 110 \kms. An outflow from galaxy A-36 produces the other two components in A. However, the absorption in A could instead be produced by a disk-like structure around A-37 or an outflow around A-40 in combination with a weaker outflow from A-36. 

Several metal features are detected at this redshift in sightlines A and B. LOS-A exhibits \ion{C}{iii}, \ion{O}{iv} and \ion{O}{vi}, whilst LOS-B features \ion{O}{ii} alongside \ion{C}{ii}, \ion{C}{iii} and \ion{C}{iv}. The metals in A appear to form two components, rather than the three components detected in \ion{H}{i}, with similar column densities in the three detected ions. This is consistent with the two components being part of the same structure, an outflow from A-36 being our best model fit. The absorption in LOS-B has a lower \ion{C}{iii} content and a wider range of low ions detected, which also seems consistent with our models suggesting that this is primarily accretion onto A-40 (although the presence of some metals requires that this accretion is not pristine).

We therefore cannot rule out a superposition of disks and outflows in this group, although some combinations are ruled out. Given their close proximity and the distortions visible in the HST image, we would expect substantial interactions in the CGM of these galaxies, likely including tidal stripping, but cannot show this using the current observations of these galaxies. The absorption that is blueshifted by 300-500 \kms{ }is not fit by our models, and may originate from these galaxies or some of the more distant galaxies at this redshift.

\begin{table*}
\begin{center}
\caption{Summary of galaxy--absorber group G-536. Non-MUSE galaxies and metal absorbers have velocities shown relative to the first galaxy (A-36). Columns are identical to those in Table \ref{table:groups_26677}.}
\label{table:groups_a36}
\begin{tabular}{| c| c| c| c| c| c| c| c| c| c| c|}
\hline 
z & Galaxy & Lum ($L_{\star}$) & Inc & LOS & Imp (kpc) & Azimuth & log(N \ion{H}{i}) & b (\kms) & $\Delta$v (km/s) & Other ions \\
(1) & (2) & (3) & (4) & (5) & (6) & (7) & (8) & (9) & (10) & (11) \\\hline 

0.536 & A-36 & 0.77 & $62^{\circ}$ $\pm$ $2^{\circ}$ & A & 120 & $77^{\circ}$ $\pm$ $2^{\circ}$ &  15.61 $\pm$ 0.02 & 39 $\pm$ 2 & 40 $\pm$ 60 & \ion{O}{iv} \& \ion{C}{iii}(40 \& -110)   \\
 & & & & A & 120 & $77^{\circ}$ $\pm$ $2^{\circ}$ & 15.30 $\pm$ 0.02 & 67 $\pm$ 4 & -90 $\pm$ 60 & \ion{O}{vi} (0 \& -120) \\
 & & & & A & 120 & $77^{\circ}$ $\pm$ $2^{\circ}$ & 14.41 $\pm$ 0.08 & 290 $\pm$ 50 & -140 $\pm$ 60 &  \\
 & & & & B & 463 & $28^{\circ}$ $\pm$ $2^{\circ}$ & 15.30 $\pm$ 0.04 & 24 $\pm$ 2 & -180 $\pm$ 60 & \ion{C}{ii}(-150, -230), \ion{O}{ii} (-140) \\
 & & & & B & 463 & $28^{\circ}$ $\pm$ $2^{\circ}$ & 14.36 $\pm$ 0.07 & 39 $\pm$ 7 & -290 $\pm$ 60 & \ion{C}{iii}(-180), \ion{C}{iv} (0) \\
 & & & & B & 463 & $28^{\circ}$ $\pm$ $2^{\circ}$ & 14.48 $\pm$ 0.05 & 28 $\pm$ 3 & -400 $\pm$ 60 & \\
 & & & & C & 1166 & $5^{\circ}$ $\pm$ $2^{\circ}$ & 15.37 $\pm$ 0.07 & 53 $\pm$ 5 & -360 $\pm$ 40 & (No metals)\\
 \\
 & A-37 & 0.31 & $72^{\circ}$ $\pm$ $3^{\circ}$ & A & 123 & $31^{\circ}$ $\pm$ $2^{\circ}$ &  & & (-40)  &   \\
 &  &  &  & B & 459 & $75^{\circ}$ $\pm$ $2^{\circ}$ & &  &  &   \\
 &  &  &  & C & 1163 & $52^{\circ}$ $\pm$ $2^{\circ}$ & &  &  &   \\
 \\ 
 & A-40 & 0.02 & $65^{\circ}$ $\pm$ $8^{\circ}$ & A & 240 & $63^{\circ}$ $\pm$ $10^{\circ}$ &  & & (-60)  &   \\
 &  &  &  & B & 398 & $15^{\circ}$ $\pm$ $10^{\circ}$ & &  &  &   \\
 &  &  &  & C & 1140 & $30^{\circ}$ $\pm$ $10^{\circ}$ & &  &  &   \\
 \\
 & (28106) & 0.3 &  & A & 897 &  & &  & (-50)  &   \\
 &  &  &  & B & 744 &  & &  &  &   \\
 &  &  &  & C & 1421 &  & &  &  &   \\
 \\
 & (29364) & 1.1 &  & A & 874 &  & &  & (-450)  &   \\
 &  &  &  & B & 606 &  & &  &  &   \\
 &  &  &  & C & 1229 & & &  &  &   \\
  \\
 & (29521) & 0.4 & & A & 1007 &  &  & & (-450)  &   \\
 &  &  &  & B & 681 &  & &  &  &   \\
 &  &  &  & C & 1220 &  & &  &  &   \\
  
  \hline
\end{tabular}

\end{center}
\end{table*}

\subsection{G-558}

\begin{figure*}
\includegraphics[width=\textwidth]{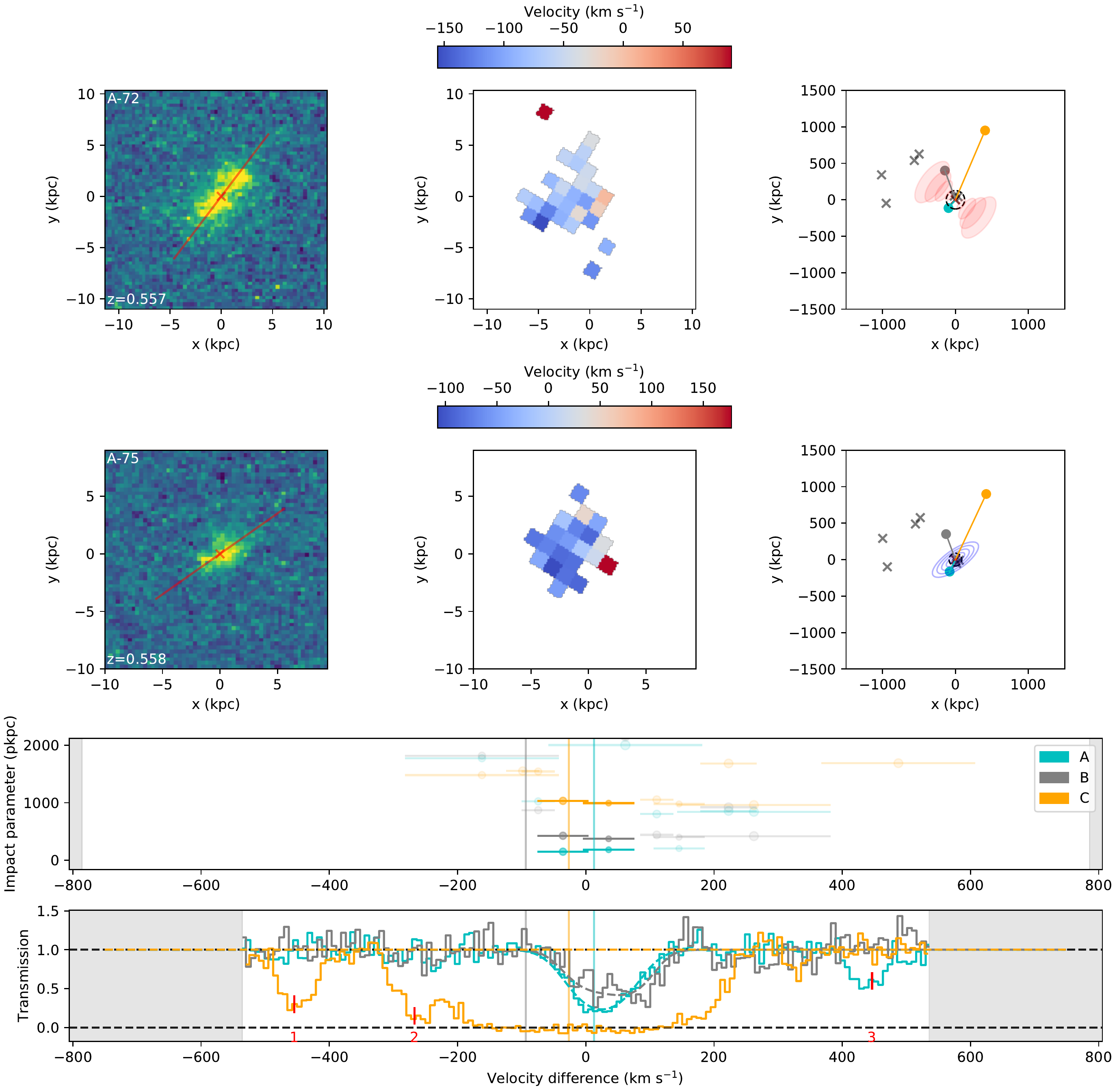}
\caption{Details of the Ly$\beta$ absorption and galaxy environment around group G-558 at the redshift of the sub-DLA. The layout is identical to that shown in Figure \ref{fig:26677_detailed}. The model shown is an outflow around galaxy A-72 (with half-opening angle $\approx 40^{\circ}$ and velocity 150 \kms) and a disk around A-75 (with a rotational velocity of 100 \kms and an infall component of $\approx$ 50 \kms). Additional features labelled (1) and (2) are H2 molecular lines at the same redshift as the sub-DLA (z$\approx$0.558), whilst (3) is the Lyman 7-0 transition from z=0.718. \label{fig:a75_detailed}}
\end{figure*}

This group of galaxies lies at the redshift of the sub-DLA at z=0.558 in the spectrum of QSO-C. Whilst the three galaxies lying in the MUSE field (A-72, A-75 and A-77) are the closest to QSO-A, there are several other galaxies at this redshift at similar impact parameters to QSOs B and C. As neither the high-resolution imaging nor the IFU data cover QSO-C, no host galaxy has been identified for the sub-DLA. A-77 is a very low-surface-brightness galaxy, making it difficult for GALFIT to fit, so the position angle and inclination are not well-determined. The details are given in Table \ref{table:groups_a72} and the galaxies and absorption at this redshift are shown in Figure \ref{fig:a75_detailed}. The three closest galaxies are separated by $\approx$ 50 kpc. There may be some distortion visible in the light from these galaxies (they do not appear to be smooth disks), but this cannot be clearly attributed to interactions between the galaxies.

We use Ly-$\beta$ for the fitting, as unlike Ly-$\alpha$ it lies in the higher-resolution COS wavelength range, and can therefore better constrain the models. The absorption in C is saturated across a large range of wavelengths, even in Ly-$\beta$, but absorption is also visible in both the A and B sightlines. 

We note that the absorption in B at this redshift in both Ly$\alpha$ and Ly$\beta$ is a blend of \ion{H}{i} absorption lines from other redshifts, making it difficult to identify which of the absorption features seen in Figure \ref{fig:a75_detailed} are Ly$\beta$ and therefore should be considered in our fitting. In particular, the absorption component identified at $\approx$ -100 \kms{ }in LOS-B and marked with a vertical line in the figure is apparent in Ly$\alpha$ but not clearly visible in Ly$\beta$. The absorption in LOS-B that is clearly visible at the wavelength of Ly$\beta$ was not clearly identified as such, and appears to be a blend of Ly$\alpha$ from z $\approx$ 0.32, Ly$\beta$ from this redshift, and Ly$\delta$ from z $\approx$ 0.64. However, the Ly$\beta$ does contribute, as the column densities estimated for the other \ion{H}{i} systems making up this blend are not sufficient to produce the observed absorption.

Note that the velocity offset to galaxy 30142 is $\approx$ 300 \kms, such that any outflow with sufficient velocity to match this would also produce a larger range of line-of-sight velocities than seen in the observed absorption. This also applies to galaxy 29812, despite a somewhat smaller velocity offset (as A and B lie nearer the minor axis). A disk around 30142 also fails to reproduce the column density ratio between the two lines-of-sight, as the smaller impact parameter to LOS-B leads to absorption stronger than observed. Although a disk around 29812 is possible and can reproduce the absorption in LOS-B, such a disk would need a radius of $\approx$ 1 Mpc in order to reach the sightline. The absorption near this group of galaxies is therefore likely due to the trio of galaxies closest to the sightlines. A-77 does not have position angle and inclination measurements, so we can only model these structures around A-72 and A-75.

Neither a disk nor an outflow can simultaneously match both of the clear absorption components seen in sightlines A and B. For both galaxies with measured position angles, A lies at a smaller impact parameter and nearer to the major axis, so our disk models produce much larger column densities at A than at B (even accounting for the other contributors to the blended absorption in B). Any outflow with large enough opening angle to cover LOS-A and fast enough velocity to match the galaxy-absorber offset will also produce much wider absorption in B than is observed. A combination of models is therefore required.

We find that the absorption in A can be approximately reproduced by a halo or disk around A-75, whilst the absorption in B could be attributed to an outflow from A-72 or A-75 (or a disk from 29812, but this is less likely due to the larger extent required). We do detect \ion{O}{vi} absorption in LOS-A but not in LOS-B. However, our detection limits allow the absorption in B to have a similar \ion{O}{vi} to \ion{H}{i} ratio to that in A. We therefore do not consider this non-detection to make an outflow origin for the absorption in B any less plausible. 

The orientation (which side of the disk is inclined away from the observer, denoted by the $S_{Nr}$ parameter) required of A-75 for an outflow to reach the absorption in B is not consistent with the disk model that fits A, so the combination of A-75 disk and outflow is not able to fit the observations. As A-72 is the most luminous and has the highest SFR of the two galaxies, we prefer the combination of outflow from A-72, and a disk or halo from A-75. A combination of outflow and halo from A-75 is also consistent with the observations and, as we note throughout this work, stripped material is difficult to rule out using these observations.

The model shown in Figure \ref{fig:a75_detailed} consists of an outflow around galaxy A-72 and a disk around A-75. The outflow has half-opening angle $\approx 40^{\circ}$ and velocity 150 \kms, whilst the disk has rotational velocity of 100 \kms and an infall component of $\approx$ 50 \kms. 

\begin{table*}
\begin{center}
\caption{Summary of galaxy--absorber group G-558. Any additional galaxies and metal absorbers have velocities shown relative to the first galaxy (A-72). Columns are identical to those in Table \ref{table:groups_26677}.}
\label{table:groups_a72}
\begin{tabular}{| c| c| c| c| c| c| c| c| c| c| c|}
\hline 
z & Galaxy & Lum ($L_{\star}$) & Inc & LOS & Imp (kpc) & Azimuth & log(N \ion{H}{i}) & b (\kms) & $\Delta$v (\kms) & Other ions \\ 
(1) & (2) & (3) & (4) & (5) & (6) & (7) & (8) & (9) & (10) & (11) \\\hline 

0.558 & A-72 & 0.10 & $63^{\circ}$ $\pm$ $3^{\circ}$ & A & 148 & $3^{\circ}$ $\pm$ $4^{\circ}$ & 14.81 $\pm$ 0.03 & 37 $\pm$ 2 & +50 $\pm$ 40 & \ion{O}{vi} \\
 & & & & B & 426 & $57^{\circ}$ $\pm$ $4^{\circ}$ & 14.26 $\pm$ 0.05 & 260 $\pm$ 40 & -60 $\pm$ 40 & None \\
  & & & & C & 1033 & $14^{\circ}$ $\pm$ $4^{\circ}$ & 18.0 $\pm$ 0.9 & 64 $\pm$ 12 & +10 $\pm$ 40 & \ion{C}{ii-iii}, \ion{N}{iii}, \ion{O}{vi}, \ion{Si}{ii-iii}  \\
 \\
 & A-75 & 0.03 & $69^{\circ}$ $\pm$ $6^{\circ}$ & A & 183 & $29^{\circ}$ $\pm$ $6^{\circ}$ & & & (+70)& \\
 & & & & B & 373 & $75^{\circ}$ $\pm$ $6^{\circ}$ & & & & \\
 & & & & C & 994 & $30^{\circ}$ $\pm$ $6^{\circ}$ & & & & \\
  \\
 & A-77 & 0.04 &  & A & 205 &  & & & (+180)& \\
 & & & & B & 401 &  & & & & \\
 & & & & C & 980 &  & & & & \\
  \\
 & (26644) & 0.5 &  & A & 857 &  & & & (+260)& \\
 & & & & B & 922 &  & & & & \\
 & & & & C & 1684 &  & & & & \\
   \\
 & (27894) & 0.1 &  & A & 1025 &  & & & (-40)& \\
 & & & & B & 871 &  & & & & \\
 & & & & C & 1544 &  & & & & \\
 \\
 & (29812) & 0.2 & $78^{\circ}$ $\pm$ $12^{\circ}$ & A & 804 & $83^{\circ}$ $\pm$ $8^{\circ}$ & & & (+150)& \\
 & & & & B & 442 & $47^{\circ}$ $\pm$ $8^{\circ}$ & & & & \\
 & & & & C & 1054 & $6^{\circ}$ $\pm$ $8^{\circ}$ & & & & \\
 \\
 & (30142) & 0.6 & $74^{\circ}$ $\pm$ $2^{\circ}$ & A & 842 & $6^{\circ}$ $\pm$ $2^{\circ}$ & & & (+300)& \\
 & & & & B & 419 & $35^{\circ}$ $\pm$ $2^{\circ}$ & & & & \\
 & & & & C & 960 & $87^{\circ}$ $\pm$ $2^{\circ}$ & & & & \\
 
  \hline
\end{tabular}

\end{center}
\end{table*}

\subsection{G-907} \label{sec:a16_text}

\begin{figure*}
\includegraphics[width=\textwidth]{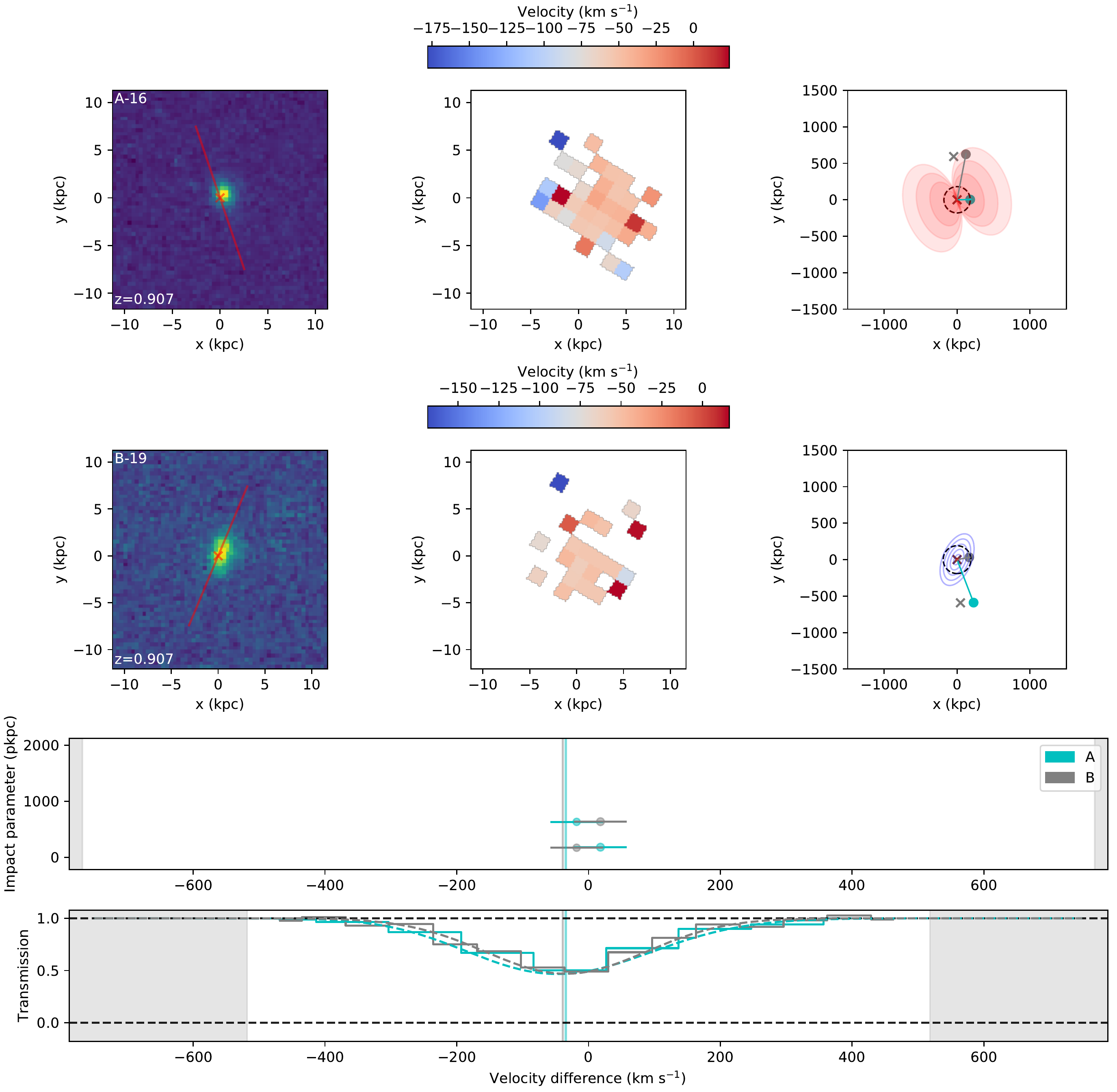}
\caption{Details of the absorption and galaxy environment around G-907, a galaxy pair at z $\sim 0.91$. The layout is identical to that shown in Figure \ref{fig:26677_detailed}, with kinematics from the [\ion{O}{ii}] emission line seen in the MUSE data, and the model shown in the lower panel combines an outflow with velocity 150 \kms and half-opening angle $60^{\circ}$ around A-16 and a disk with rotation velocity 130 \kms around B-19. \label{fig:a16_detailed}}
\end{figure*}

A-16 and B-19 are both small, star-forming galaxies at z $\sim$0.9, illustrated in Table \ref{table:groups_a16} and Figure \ref{fig:a16_detailed}. Both lines-of-sight show moderately strong absorption slightly blueward of both galaxies. The galaxies have sufficient signal-to-noise for kinematics to be measured, but are not well-resolved by MUSE so do not show a clear velocity gradient. They are separated by $\approx$600 kpc, so it is unsurprising that there is no sign of interaction in the HST image.

There are no clear constraints from the kinematics, and the absorption features in the FOS spectra are unresolved. Ly$\beta$ also lies in FOS at this redshift, so cannot be used to resolve these features. Each galaxy lies $\approx$ 200 kpc from one sightline, and $\approx$ 700 kpc from the other. Our models therefore cannot produce both absorption features with near-identical strengths and velocities using a disk and outflow from the same galaxy. However, the parameters can easily be tuned such that a disk or outflow from each galaxy can reproduce the absorption in the nearest line-of-sight.

In Figure \ref{fig:a16_detailed} we show an outflow with velocity 150 \kms{ }and half-opening angle $60^{\circ}$ from A-16, and a disk with scale height ratio 20:1 with circular and inward velocities of 130 and 30 \kms{ }respectively around B-19. However, an outflow from B-19 and disk around A-16 can similarly reproduce the observations. In this case, the outflow would require only 80 \kms{ }velocity, and a similar opening angle, and the disk would require $v_{\phi} \approx 150$ \kms{ }and $v_{r} \approx 50$ \kms. This scenario also produces more model absorption in the more distant sightline than the A-16 outflow, which better constrains the two disk velocity components and the outflow opening angle. Either scenario is possible with parameters within the bounds suggested by Paper 2 and other works. 

Although our models cannot easily discriminate between these two scenarios, we find that the absorption around these galaxies is consistent with plausible CGM structures around the individual galaxies. Tidal material is not ruled out, but given a separation between the galaxies that is substantially larger than the virial radius, this is unlikely to dominate the observed absorption.

\begin{table*}
\begin{center}
\caption{Summary of galaxy--absorber group G-907. Any additional galaxies and metal absorbers have velocities shown relative to the first galaxy (A-16). Note that this group is beyond the redshift of QSO-C, so no absorption could be detected. Columns are identical to those in Table \ref{table:groups_26677}.}
\label{table:groups_a16}
\begin{tabular}{| c| c| c| c| c| c| c| c| c| c| c|}
\hline 
 z & Galaxy & Lum ($L_{\star}$) & Inc & LOS & Imp (kpc) & Azimuth & log(N \ion{H}{i}) & b (\kms) & $\Delta$v (\kms) & Other ions \\ 
(1) & (2) & (3) & (4) & (5) & (6) & (7) & (8) & (9) & (10) & (11) \\\hline 

0.907 & A-16 & 0.17 & $53^{\circ}$ $\pm$ $12^{\circ}$ & A & 180 & $72^{\circ}$ $\pm$ $20^{\circ}$ & 14.16 $\pm$ 0.05 & 170 $\pm$ 20 & -50 $\pm$ 40 & \\
 & & & & B & 635 & $30^{\circ}$ $\pm$ $20^{\circ}$ & 14.15 $\pm$ 0.03 & 157 $\pm$ 11 & -60 $\pm$ 40 &  \\
  \\
 & B-19 & 0.10 & $58^{\circ}$ $\pm$ $7^{\circ}$ & A & 631 & $44^{\circ}$ $\pm$ $12^{\circ}$ & & & (-40) & \\
 & & & & B & 170 & $57^{\circ}$ $\pm$ $12^{\circ}$ & & & & \\
 
    \hline
\end{tabular}

\end{center}
\end{table*}

\bsp	
\label{lastpage}
\end{document}